\documentclass[sort&compress,11pt]{article}
\pdfoutput=1

\usepackage{amssymb,amsmath,amsthm,mathtools}
\usepackage[toc,page]{appendix}
\usepackage{graphicx}
\usepackage{slashed}
\usepackage[]{hyperref}
\usepackage{subfigure}
\usepackage[utf8]{inputenc}
\usepackage{color}
\usepackage{cite}

\newcommand{\change}[1]{{#1}}

\usepackage{bbm}
\makeatletter
\makeatletter
\newcommand\footnoteref[1]{\protected@xdef\@thefnmark{\ref{#1}}\@footnotemark}
\makeatother

\if@twoside \oddsidemargin -17pt \evensidemargin 00pt
\else \oddsidemargin 00pt \evensidemargin 00pt
\fi
\expandafter\ifx\csname mathrm\endcsname\relax\def\mathrm#1{{\rm #1}}\fi

\makeatother

\def\refeq#1{\mbox{(\ref{#1})}}
\def\reffi#1{\mbox{Fig.~\ref{#1}}}

\def\refta#1{\mbox{Table~\ref{#1}}}
\def\reftas#1{\mbox{Tables~\ref{#1}}}
\def\refse#1{\mbox{Section~\ref{#1}}}
\def\refses#1{\mbox{Sections~\ref{#1}}}
\def\refapp#1{\mbox{App.~\ref{#1}}}
\def\refapps#1{\mbox{Apps.~\ref{#1}}}
\def\citere#1{\mbox{Ref.~\cite{#1}}}
\def\citeres#1{\mbox{Refs.~\cite{#1}}}

\def\ie{i.e.\ }
\def\eg{e.g.\ }

\allowdisplaybreaks[3]   

\newcommand{\red}[1]{\textcolor{red}{#1}}
\newcommand{\green}[1]{\textcolor{green}{#1}}
\newcommand{\blue}[1]{\textcolor{blue}{#1}}
\definecolor{orange}{rgb}{1,0.5,0}
\newcommand{\orange}[1]{\textcolor{orange}{#1}}

\newcommand{\cmm}[1]{\ensuremath{#1}\ifmmode\else{}\fi}
\newcommand{\nmc}[2]{\newcommand{#1}{\cmm{#2}}}

\newcommand{\ga}[1]{\Gamma_{#1 #1}}

\nmc{\gs}{{\fontfamily{pzc}\fontsize{13pt}{12pt}\selectfont GraphShot}}

\newcommand{\GeV}{\unskip\,\mathrm{GeV}}

\newcommand{\TeV}{\unskip\,\mathrm{TeV}}

\nmc{\Pp}{\mathrm{p}}
\nmc{\PW}{\mathrm{W}}
\nmc{\PZ}{\mathrm{Z}}
\nmc{\PH}{\mathrm{H}}
\nmc{\Pb}{\mathrm{b}}

\nmc{\g}{g}
\nmc{\gy}{{g^\prime}}
\nmc{\Gf}{G_\mathrm{F}}

\nmc{\B}{{\rm B}}
\nmc{\R}{{\rm R}}

\nmc{\im}{\mathop{\mathrm{Im}}\nolimits}

\nmc{\qq}{q}
\nmc{\slq}{\slashed{\qq}}
\nmc{\pp}{p}
\nmc{\slp}{\slashed{\pp}}
\nmc{\kk}{k}
\nmc{\slk}{\slashed{\kk}}
\nmc{\Prl}{P_\mathrm{L}}
\nmc{\Prr}{P_\mathrm{R}}

\nmc{\ii}{\mathrm{i}}

\nmc{\rd}{\mathrm{d}}

\nmc{\ft}{\mathrm{F.T.}}

\nmc{\Az}{A_0}

\nmc{\h}{h}
\nmc{\Hl}{H_\mathrm{l}}
\nmc{\Hh}{H_\mathrm{h}}
\nmc{\Hnov}{\widetilde{H}}
\nmc{\GZ}{G_0}
\nmc{\Ha}{H_\mathrm{a}}
\nmc{\Hpm}{H^\pm}
\nmc{\Gpm}{G^\pm}
\nmc{\Gp}{G^+}
\nmc{\Gm}{G^-}
\nmc{\Hmp}{H^\mp}
\nmc{\Gmp}{G^\mp}
\nmc{\Phin}{\Phi_n}
\nmc{\Phii}{\Phi_i}
\nmc{\Phione}{\Phi_1}
\nmc{\Phitwo}{\Phi_2}
\nmc{\phionepm}{\phi_1^\pm}
\nmc{\phitwopm}{\phi_2^\pm}
\nmc{\phinp}{\phi_n^+}
\nmc{\phiip}{\phi_i^+}
\nmc{\phii}{\varphi_i}
\nmc{\phij}{\varphi_j}
\nmc{\phik}{\varphi_k}
\nmc{\phin}{\varphi_n}
\nmc{\rhoi}{\rho_i}
\nmc{\rhon}{\rho_n}
\nmc{\rhoone}{\rho_1}
\nmc{\rhotwo}{\rho_2}
\nmc{\etan}{\eta_n}
\nmc{\etai}{\eta_i}
\nmc{\etaone}{\eta_1}
\nmc{\etatwo}{\eta_2}

\nmc{\vth}{\tilde v_\h}
\nmc{\tilh}{\tilde\h}
\nmc{\al}{\alpha}
\nmc{\be}{\beta}
\nmc{\vone}{v_1}
\nmc{\vtone}{\left(\vbone+\dvone\right)}
\nmc{\vtwo}{v_2}
\nmc{\vttwo}{\left(\vbtwo+\dvtwo\right)}
\nmc{\vn}{v_n}
\nmc{\vi}{v_i}
\nmc{\vtn}{\tilde v_n}
\nmc{\vti}{\tilde v_i}
\nmc{\tb}{t_\be}
\nmc{\tanb}{\tan\be}
\nmc{\ca}{\cos\al}
\nmc{\catwo}{\cos^2\al}
\nmc{\satwo}{\sin^2\al}
\nmc{\sa}{\sin\al}
\nmc{\stwoa}{\sin2\al}
\nmc{\cbe}{\cos\be}
\nmc{\ctwobe}{\cos2\be}
\nmc{\cbeptwo}{\cos^2\be}
\nmc{\sbe}{\sin\be}
\nmc{\sbeptwo}{\sin^2\be}
\nmc{\cab}{c_{\al\be}}
\newcommand{\sab}{\ensuremath{s_{\alpha\beta}}}

\newcommand{\ct}{\cw}
\nmc{\mw}{M_\mathrm{W}}
\nmc{\sw}{s_\mathrm{w}}
\nmc{\cw}{c_\mathrm{w}}
\nmc{\mv}{M_V}
\nmc{\ms}{M_S}
\nmc{\mscale}{{M^{*}}}
\nmc{\msp}{M_{S'}}
\newcommand{\mz}{M_\mathrm{Z}}
\nmc{\mf}{m_f}
\nmc{\mfL}{\mf^\mathrm{l}}
\nmc{\mfU}{\mf^\mathrm{u}}
\nmc{\mfD}{\mf^\mathrm{d}}
\nmc{\mfb}{m_{f,\mathrm{B}}}
\newcommand{\mt}{m_\mathrm{t}}
\newcommand{\mh}{M_{\mathrm{h}}}
\nmc{\mhl}{M_{\Hl}}
\nmc{\mhh}{M_{\Hh}}
\nmc{\Msb}{M_{\rm sb}}
\nmc{\mhc}{M_{\Hpm}}
\nmc{\mha}{M_{\Ha}}
\nmc{\tth}{t_{\h}}
\nmc{\Th}{T_{\h}}
\nmc{\hTh}{\hat{T}_{\h}}
\nmc{\hThone}{\hat{T}_{\h}^{(1)}}
\nmc{\thone}{\tth^{(1)}}
\nmc{\Thone}{{T}_{\h}^{(1)}}
\nmc{\Thtwo}{{T}_{\h}^{(2)}}
\nmc{\Tn}{{T}_n}
\nmc{\tn}{t_n}
\nmc{\Ti}{{T}_i}
\nmc{\ti}{t_i}
\nmc{\Tl}{{T}_l}
\nmc{\tl}{t_l}
\nmc{\Tiol}{{T}_i^{(1)}}
\nmc{\Titl}{{T}_i^{(2)}}
\nmc{\Tnol}{{T}_n^{(1)}}
\nmc{\Tntl}{{T}_n^{(2)}}
\nmc{\Thl}{T_{\Hl}}
\nmc{\thlone}{\thl^{(1)}}
\nmc{\Thlone}{\Thl^{(1)}}
\nmc{\hThlone}{\hat{{T}}_{\Hl}^{(1)}}
\nmc{\Thh}{T_{\Hh}}
\nmc{\thhone}{\thh^{(1)}}
\nmc{\Thhone}{\Thh^{(1)}}
\nmc{\hThhone}{\hat{{T}}_{\Hh}^{(1)}}
\nmc{\gh}{c}
\nmc{\xiw}{\xi_W}
\nmc{\xiz}{\xi_Z}
\nmc{\xia}{\xi_{A}}
\nmc{\xiaz}{\xi_{AZ}}
\nmc{\xiphiz}{\xi_{\GZ}}
\nmc{\xihh}{\xi_{\Hh}}
\nmc{\rxi}{\mathrm{R}_\xi}
\nmc{\xib}{\xi_\beta}
\nmc{\dxib}{\delta_{\xib}}
\nmc{\ci}{\gh^i}
\nmc{\cj}{\gh^j}
\nmc{\aci}{\overline{\gh}^i}
\nmc{\acp}{\overline{\gh}^+}
\nmc{\acm}{\overline{\gh}^-}
\nmc{\acz}{\overline{\gh}^z}
\nmc{\acg}{\overline{\gh}^{\gamma}}
\nmc{\dbrs}{\delta_{\rm BRST}\,}
\nmc{\dBRS}{s}
\nmc{\hb}{h_{\B}}
\nmc{\dhb}{\Delta\hb}
\nmc{\mhb}{M_{\mathrm{h},\B}}
\nmc{\mhr}{M_{\mathrm{h},\R}}
\nmc{\lb}{\lambda_{\B}}
\nmc{\vb}{v_{\B}}
\nmc{\vbh}{\vb}
\nmc{\vh}{v_{\h}}
\nmc{\dvh}{\dv}
\nmc{\dvhone}{\dvh^{(1)}}
\nmc{\dvhtwo}{\dvh^{(2)}}
\nmc{\muB}{\mu_{\B}}
\nmc{\mwb}{M_{\mathrm{W},\B}}
\nmc{\mzb}{M_{\mathrm{Z},\B}}
\nmc{\gb}{g_{\mathrm{B}}}
\nmc{\vbi}{v_{i,\B}}
\nmc{\vbone}{v_{1,\B}}
\nmc{\vbtwo}{v_{2,\B}}
\nmc{\vbn}{v_{n,\B}}
\nmc{\vbl}{v_{l,\B}}
\nmc{\cbone}{c_{1,\B}}
\nmc{\cbi}{c_{i,\B}}
\nmc{\cbj}{c_{j,\B}}
\nmc{\cbm}{c_{m,\B}}
\nmc{\cbk}{c_{k,\B}}
\nmc{\phibone}{\varphi_{1,\B}}
\nmc{\Phibone}{\varPhi_{1,\B}}
\nmc{\phibtwo}{\varphi_{2,\B}}
\nmc{\Phibtwo}{\Phi_{2,\B}}
\nmc{\rhobone}{\rho_{1,\B}}
\nmc{\rhobtwo}{\rho_{2,\B}}
\nmc{\phibi}{\varphi_{i,\B}}
\nmc{\Phibi}{\varPhi_{i,\B}}
\nmc{\phibn}{\varphi_{n,\B}}
\nmc{\Phibn}{\Phi_{n,\B}}
\nmc{\phibl}{\varphi_{l,\B}}
\nmc{\Phibl}{\varPhi_{l,\B}}
\nmc{\phibit}{{\phi}^\prime_{i,\B}}
\nmc{\mhlb}{M_{\Hl,\B}}
\nmc{\mhhb}{M_{\Hh,\B}}
\nmc{\mhcb}{M_{\Hpm,\B}}
\nmc{\mhab}{M_{\Ha,\B}}
\newcommand{\moneb}{m_{1,\B}^2}
\newcommand{\mtwob}{m_{2,\B}^2}
\newcommand{\loneb}{\lambda_{1,\B}}
\newcommand{\ltwob}{\lambda_{2,\B}}
\newcommand{\lthrb}{\lambda_{3,\B}}
\newcommand{\lfoub}{\lambda_{4,\B}}
\newcommand{\lfivb}{\lambda_{5,\B}}
\newcommand{\lib}{\lambda_{i,\B}}

\newcommand{\dvone}{\Delta v_1}
\newcommand{\dvtwo}{\Delta v_2}
\nmc{\dv}{\Delta v}
\nmc{\dvn}{\Delta v_n}
\nmc{\dvi}{\Delta v_i}
\nmc{\dvl}{\Delta v_l}

\nmc{\dvoneone}{\dvone^{(1)}}

\nmc{\dvtwoone}{\dvtwo^{(1)}}

\nmc{\Gh}{\ga{\h}}
\nmc{\Gi}{\ga{i}}
\nmc{\onePI}{\Sigma}

\newcommand{\gss}{\onePI_{SS}^\mathrm{1PI}}
\newcommand{\gssp}{\onePI_{SS'}^\mathrm{1PI}}

\newcommand{\gazt}{\onePI_{AZ}^{\mathrm{1PI},\mathrm{T}}}
\newcommand{\gzzt}{\onePI_{ZZ}^{\mathrm{1PI},\mathrm{T}}}
\newcommand{\gaat}{\onePI_{AA}^{\mathrm{1PI},\mathrm{T}}}
\newcommand{\gwwt}{\onePI_{WW}^{\mathrm{1PI},\mathrm{T}}}

\newcommand{\gazth}{\hat{\Sigma}_{AZ}^{\mathrm{T}}}
\newcommand{\gvvt}{\Sigma_{VV}^{\mathrm{1PI},\mathrm{T}}}
\newcommand{\gvvth}{\hat{\Sigma}^{\mathrm{T}}_{VV}}
\newcommand{\gssh}{\hat{\Sigma}_{SS}}
\newcommand{\gvs}{\onePI_{VS}^{\mathrm{1PI},\mu}}
\newcommand{\gvsh}{\hat{\Sigma}_{VS}^{\mu}}
\newcommand{\gssph}{\hat{\Sigma}_{SS'}}

\newcommand{\gffS}{\onePI_{ff}^{\mathrm{1PI},\rm S}}

\newcommand{\gffL}{\onePI_{ff}^{\mathrm{1PI},\rm L}}
\newcommand{\gffR}{\onePI_{ff}^{\mathrm{1PI},\rm R}}
\newcommand{\gffh}{\hat{\Sigma}_{ff}}
\newcommand{\gffSh}{\hat{\Sigma}_{ff}^{\rm S}}

\newcommand{\gffLh}{\hat{\Sigma}_{ff}^{\rm L}}
\newcommand{\gffRh}{\hat{\Sigma}_{ff}^{\rm R}}
\newcommand{\tvv}{t_{VV}}
\newcommand{\tss}{t_{SS}}
\newcommand{\tvs}{t_{VS}}
\newcommand{\tssp}{t_{SS'}}
\newcommand{\dzvv}{\delta Z_{VV}}
\newcommand{\dzss}{\delta Z_{SS}}
\newcommand{\dzssp}{\delta Z_{SS'}}
\newcommand{\dzsps}{\delta Z_{S'S}}
\newcommand{\dzspsp}{\delta Z_{S'S'}}
\newcommand{\dzaz}{\delta Z_{AZ}}
\newcommand{\dzza}{\delta Z_{ZA}}
\newcommand{\thl}{t_{\Hl}}
\newcommand{\thh}{t_{\Hh}}
\newcommand{\thlhl}{t_{\Hl\Hl}}
\newcommand{\thhhh}{t_{\Hh\Hh}}
\newcommand{\thhhl}{t_{\Hh\Hl}}
\newcommand{\thlhh}{t_{\Hl\Hh}}
\newcommand{\tha}{t_{\Ha\Ha}}
\newcommand{\tgha}{t_{\GZ\Ha}}
\newcommand{\thag}{t_{\Ha\GZ}}
\newcommand{\tgzero}{t_{\GZ\GZ}}
\newcommand{\thc}{t_{\Hpm\Hmp}}
\newcommand{\tghc}{t_{\Gpm\Hmp}}
\newcommand{\thcg}{t_{\Hpm\Gmp}}
\newcommand{\tgpm}{t_{\Gpm\Gmp}}
\newcommand{\tw}{t_{W^{\pm}W^{\mp}}^{\mu\nu}}
\newcommand{\tz}{t_{ZZ}^{\mu\nu}}
\newcommand{\twhc}{t_{W^{\pm}\Hmp}^\mu}

\newcommand{\twg}{t_{W^{\pm}\Gmp}^\mu}

\newcommand{\twmgp}{t_{W^{-}\Gp}^\mu}
\newcommand{\tzha}{t_{Z\Ha}^\mu}

\newcommand{\tzg}{t_{Z \GZ}^\mu}

\newcommand{\tff}{t_{ff}}
\newcommand{\tfp}{t_{f,1}}
\newcommand{\tfm}{t_{f,2}}

\nmc{\kone}{\left(\thl\sa  - \thh\ca\right)}
\nmc{\ktwo}{\left(\thl\ca + \thh\sa \right)}
\nmc{\dzh}{\delta Z_{\h}}
\nmc{\dzmh}{\delta Z_{\mh}}
\nmc{\dmh}{\delta \mh}
\nmc{\dmf}{\delta \mf}
\nmc{\dmfU}{\delta \mfU}
\nmc{\dmfL}{\delta \mfL}
\nmc{\dmfD}{\delta \mfD}
\nmc{\dmv}{\delta \mv}
\nmc{\dms}{\delta \ms}
\nmc{\dmsp}{\delta \msp}
\nmc{\dmw}{\delta \mw}
\nmc{\dmz}{\delta \mz}

\newcommand{\dzhhhl}{\delta Z_{\Hh\Hl}}
\newcommand{\dzhlhh}{\delta Z_{\Hl\Hh}}

\newcommand{\dzgha}{\delta Z_{\GZ\Ha}}
\newcommand{\dzhag}{\delta Z_{\Ha\GZ}}

\newcommand{\dzw}{\delta Z_{WW}}
\newcommand{\dzz}{\delta Z_{ZZ}}
\newcommand{\dza}{\delta Z_{AA}}

\newcommand{\dzfl}{\delta Z_{f,\rm L}}
\newcommand{\dzfr}{\delta Z_{f,\rm R}}

\nmc{\ftone}{t_{s}}
\nmc{\fttwo}{f_{t,2}}
\nmc{\ftthr}{f_{t,3}}
\nmc{\ftfou}{f_{t,4}}
\nmc{\tab}{t_{\alpha\beta}}
\newcommand{\pics}{Pics}
\newcommand{\SMatrix}{$S$-matrix}
\newcommand{\SM}{SM}
\newcommand{\THDM}{2HDM}
\newcommand{\MSSM}{MSSM}
\nmc{\msbar}{{\overline{\mathrm{MS}}}}
\nmc{\PP}{{\overline{\mathrm{P.P.}}}}
\newcommand{\mL}{\mathcal{L}}
\newcommand{\ts}{{\it FJ~Tadpole Scheme}}
\newcommand{\bfts}{{\bf FJ Tadpole Scheme}}

\def\draftdate{\relax}
\def\mda{\relax}
\def\mua{\relax}
\def\mla{\relax}
\def\draft{
\def\thtystars{******************************}
\def\sixtystars{\thtystars\thtystars}
\typeout{}
\typeout{\sixtystars**}
\typeout{* Draft mode!
         For final version remove \protect\draft\space in source file *}
\typeout{\sixtystars**}
\typeout{}
\def\draftdate{\today}
\def\mua{\marginpar[\boldmath\hfil$\uparrow$]%
                   {\boldmath$\uparrow$\hfil}\color{black}%
                    \typeout{marginpar: $\uparrow$}\ignorespaces}
\def\mda{\color{red}\marginpar[\boldmath\hfil$\downarrow$]%
                   {\boldmath$\downarrow$\hfil}%
                    \typeout{marginpar: $\downarrow$}\ignorespaces}
\def\mla{\marginpar[\boldmath\hfil$\rightarrow$]%
                   {\boldmath$\leftarrow $\hfil}%
                    \typeout{marginpar: $\leftrightarrow$}\ignorespaces}
\def\Mua{\marginpar[\boldmath\hfil$\Uparrow$]%
                   {\boldmath$\Uparrow$\hfil}\color{black}%
                    \typeout{marginpar: $\uparrow$}\ignorespaces}
\def\Mda{\color{red}\marginpar[\boldmath\hfil$\Downarrow$]%
                   {\boldmath$\Downarrow$\hfil}%
                    \typeout{marginpar: $\downarrow$}\ignorespaces}
\def\Mla{\marginpar[\boldmath\hfil\textcolor{red}{$\Rightarrow$}]%
                   {\boldmath\textcolor{red}{$\Leftarrow $}\hfil}%
                    \typeout{marginpar: $\leftrightarrow$}\ignorespaces}
\overfullrule 5pt
\oddsidemargin -15mm
\marginparwidth 29mm
}

\makeatletter
\newcommand{\specialnumber}[1]{%
  \def\tagform@##1{\maketag@@@{(\ignorespaces##1\unskip\@@italiccorr#1)}}%
}
\newcommand{\specialeqref}[2]{\begingroup
  \def\tagform@##1{\maketag@@@{(\ignorespaces##1\unskip\@@italiccorr#2)}}%
  \eqref{#1}\endgroup}
\makeatother

\numberwithin{equation}{section}

\begin{document}

\thispagestyle{empty}
\def\thefootnote{\fnsymbol{footnote}}
\setcounter{footnote}{1}
\null
\hfill
\\
\begin{flushright}
\draftdate
\end{flushright}
\vskip 1.2cm
\begin{center}

{\Large \boldmath{\bf Gauge-independent \msbar~renormalization in the 2HDM}
\par} \vskip 2.5em
{\large
{ Ansgar~Denner, Laura~Jenniches, Jean-Nicolas~Lang, Christian~Sturm}\\[3ex]
{\normalsize \it
Institut f\"ur Theoretische Physik und Astrophysik, \\
Universit\"at W\"urzburg, Emil-Hilb-Weg 22,
D-97074 W\"urzburg, Germany}\\[1ex]
}
\par \vskip 1em
\end{center}\par
\vfill \vskip .0cm \vfill {\bf Abstract:} \par 


We present a consistent renormalization scheme for the CP-conserving
Two-Higgs-Doublet Model based on \msbar{} renormalization of the
mixing angles and the
soft-$Z_2$-symmetry-breaking
scale \Msb\ in the Higgs sector.  This scheme requires to treat
tadpoles fully consistently in all steps of the calculation in order
to provide gauge-independent $S$-matrix elements. We show how bare
physical parameters have to be defined and verify the gauge
independence of physical quantities by explicit calculations in a
general $R_{\xi}$-gauge.  The procedure is straightforward and
applicable
to other models with extended Higgs sectors.  In contrast to the
proposed scheme, the \msbar{} renormalization of the mixing angles
combined with popular on-shell renormalization schemes gives rise to
gauge-dependent results already at the one-loop level.  We present
explicit results for
electroweak NLO corrections to selected processes in the appropriately
renormalized Two-Higgs-Doublet Model and in particular discuss their scale dependence.
\par
\vskip 1cm
\noindent
September 2016
\par
\null
\setcounter{page}{0}
\clearpage
\def\thefootnote{\arabic{footnote}}
\setcounter{footnote}{0}


\section{Introduction\label{sec:Introduction}}
The discovery of a Higgs boson at the Large Hadron Collider
(LHC)~\cite{Chatrchyan:2012xdj,Aad:2012tfa} was a tremendous success
for elementary particle physics.  The Higgs boson is now a central
object of intense research in both experiment and theory in order to
determine its properties precisely. In particular, it is interesting
to investigate whether the Higgs boson belongs to the Standard Model
(SM) or whether it is part of a more general theory. In this context,
models with additional Higgs bosons are of special interest. An
extended Higgs sector can contribute to solve open problems in
particle physics, like for example the question of the origin of the
matter--antimatter asymmetry in the universe or the nature of dark
matter.

At the LHC, detectable differences between the SM and a theory with an
extended Higgs sector can be small. Therefore, accurate theory
predictions are strongly desirable and in turn require the knowledge
of higher-order corrections.  QCD corrections essentially dress the
basic electroweak (EW) interactions of the Higgs boson and do not
fundamentally change by adding additional Higgs bosons to the theory.
Electroweak corrections, on the other hand, can significantly modify
the predictions for physical observables, like cross sections and
partial decay widths, through an extended Higgs sector.  For this
reason, we dedicate special attention to the calculation of
next-to-leading order (NLO) EW corrections to Higgs production.  This
requires a renormalization of the new physical parameters and fields
of the extended Higgs sector.

Within this work, we consider in particular the CP-conserving \THDM{}
of type II with a softly broken $Z_2$ symmetry
\cite{Gunion:2002,Branco:2011iw}. The Higgs sector of this model
depends on four physical mass parameters, the masses $\mhl$ and $\mhh$
of the light and heavy, neutral, scalar Higgs bosons, the mass $\mha$
of the pseudo-scalar Higgs boson, and the mass $\mhc$ of the charged
Higgs boson. In addition, there are two mixing angles, $\al${} and
$\be${}, as well as the soft-$Z_2$-breaking scale $\Msb$. The
renormalization of the \THDM{} of type II has been discussed in the
context of supersymmetry (see
e.g.~\citeres{Pierce:1992hg,Freitas:2002um,Baro:2008bg}) and also in
the general
case~\cite{Santos:1996vt,Kanemura:2004mg,LopezVal:2009qy,Krause:2016oke}.
A reasonable renormalization scheme should fulfil the following three
conditions \cite{Freitas:2002um}: it should lead to gauge-independent
physical counterterms, it should be numerically stable, \ie the size
of the higher-order corrections should be moderate, and it should
preferably be defined in a process-independent way.

The masses of the Higgs bosons, vector bosons and fermions are
naturally renormalized in the classical on-shell scheme, which is
straightforward to apply. In contrast, for the parameters $\alpha$,
$\beta$, and $\Msb$ there is no natural renormalization scheme. As
long as these parameters are unknown, it is difficult to identify
processes that can be measured precisely and that would allow to
extract the values of these parameters accurately and in a stable way.
Moreover, process-dependent renormalization can lead to unnaturally
large corrections for the predictions of other processes (see e.g.\ 
\citere{Krause:2016oke}) or require to artificially split off IR
singularities \cite{Freitas:2002um}.  Motivated by studies on the
renormalization of the quark-mixing matrix in the SM
\cite{Denner:1990yz}, process-independent renormalization of the
mixing angles $\alpha$ and $\beta$ based on the field renormalization
constants of the Higgs fields have been proposed
\cite{Kanemura:2004mg,LopezVal:2009qy,Kanemura:2015mxa}. Taking these
recipes naively leads, however, to gauge-dependent renormalization of
the mixing angles \cite{Gambino:1998ec}. Therefore, it has been
suggested \cite{Yamada:2001px,Krause:2016oke} to render these methods
gauge independent by applying the pinch technique
\cite{Cornwall:1981zr,Cornwall:1989gv}. This, however, merely trades
the gauge dependence for a dependence on the prescription intrinsic in
the pinch technique.

We consider it favourable to stick to a renormalization scheme that
does not depend on particular conventions and propose to use the
\msbar{} renormalization scheme which is by default used in
perturbative QCD.  \msbar{} renormalization is simple to apply and to
implement. In addition, it leads to a residual scale dependence that
helps to estimate the uncertainty caused by unknown higher-order
corrections via a suitable scale variation. If the perturbative
corrections become large along with the scale dependence, this signals
the onset of the non-perturbative regime where the Higgs sector
becomes strongly coupled.  This will be illustrated in this work.

The use of an \msbar{} renormalization scheme, however, requires
particular attention in the proper treatment of the Higgs tadpoles.
Tadpoles have no physical meaning and drop out in properly calculated
\SMatrix{} elements. If all parameters of the theory can be
renormalized on-shell, as in the SM, (some) tadpoles can be
consistently omitted to simplify practical calculations.  However, if
parameters are renormalized in the \msbar{} scheme, it becomes
crucial to define all bare physical parameters and counterterms, like
for instance the mass counterterms of the EW gauge bosons or Higgs
bosons, in a gauge-independent way. This in turn requires a consistent
treatment of tadpoles.  Within the \THDM, we show that a
careless treatment of tadpoles, as often used in the literature, in
combination with \msbar{} renormalization of the mixing angles
$\alpha$ and $\beta$ yields gauge-dependent results already at the
one-loop level.

A consistent treatment of tadpoles has been formulated for the SM by
Fleischer and Jegerlehner in \citere{Fleischer:1980ub}.  In this
paper, we generalize this scheme, which we dub \ts{}, and apply it to
the \THDM\ in order to allow for a consistent gauge-independent
\msbar{} renormalization of the mixing angles $\alpha$ and $\beta$ as
well as the soft-breaking scale $\Msb$.  We stress that the \ts{} as
introduced here provides a consistent universal description of
tadpoles in arbitrary theories with spontaneous symmetry breaking
(SSB), such as general multi-Higgs models.

The outline of this paper is as follows.
In \refse{sec:tadpoles} we discuss the role of the Higgs tadpoles for
the proper definition of gauge-independent, physical parameters and
counterterms for a general Higgs sector and recapitulate the use of
the \ts{} within the SM.
Our notation and conventions for the \THDM{} are introduced in
\refse{sec:THDM}.
Section \ref{sec:renoconditions} is devoted to the extension of the
\ts{} to the \THDM{} and the formulation of the renormalization
conditions.
In \refse{sec:gaugedep} we discuss the gauge dependence of popular
schemes and relate these schemes to the \ts{}.
In \refse{sec:processes} we demonstrate the applicability of the \ts{}
by employing it in NLO calculations of Higgs-boson production
processes in the \THDM{}. In particular, we discuss Higgs-boson
production in gluon fusion as well as Higgs-boson production through
Higgs strahlung.
Finally, in \refse{sec:conclusion} we close with our conclusions.
In the Appendices we present results for the Higgs-tadpole
counterterms in the \THDM, proof the gauge dependence of the \msbar{}
renormalization of $\beta$ in popular tadpole schemes, and illustrate
the tadpole contributions to the two-loop Higgs-boson self-energy.


\section{The role of tadpoles and gauge dependence}\label{sec:tadpoles}

Before discussing the renormalization of the \THDM{}, we examine the
tadpole renormalization in theories with an extended Higgs sector in
general, and we revisit the treatment of tadpoles in the \SM{}. As
tadpole contributions are gauge dependent (see e.g.\ 
\citeres{Gambino:1999ai,Actis:2006} or \refapp{sec:appA}), their
gauge-dependent parts have to cancel in any physical quantity.  This
cancellation is always given when a physical renormalization scheme
such as the on-shell scheme is applied, which is usually the case in
the EW \SM{}.  For Beyond-Standard-Model (BSM) theories, on the other
hand, on-shell renormalization of parameters may introduce a process
dependence and/or lead to unnaturally large NLO corrections.  Hence,
it is preferable to renormalize new parameters in the \msbar{} scheme,
at least until first evidence of the BSM theory allows for a
meaningful definition of a process-dependent renormalization scheme.
In this case, the treatment of tadpole counterterms requires some care
to warrant the gauge independence of \SMatrix{} elements: if \msbar{}
counterterms are used, it is essential to ensure that all counterterms
of physical parameters are gauge independent. This requires a
gauge-independent definition of bare physical parameters which, in
particular, must not employ shifted vacuum expectation values (vevs)
in the presence of SSB. We show how this can be achieved in general
and in a systematic way. In Section \ref{sec:tadpoles-gen}, we discuss
a proper definition of the vevs for a general Higgs sector. The
treatment of tadpoles closely follows the arguments which have been
presented by Fleischer and Jegerlehner in App.~A of
\citere{Fleischer:1980ub} for the \SM{}. We refer to this scheme as
the \ts{} in the following.

The tadpole counterterms are absent at tree level. At higher orders in
perturbation theory, they can be used to remove explicit tadpole
contributions, which simplifies practical calculations.  We show that
the \ts{} is gauge independent regardless of whether tadpoles are
removed by a consistent renormalization or taken into account
explicitly.  This implies that any physical quantity is independent of
the (consistent) renormalization of tadpoles.


\subsection{The \bfts{} for a general Higgs sector}\label{sec:tadpoles-gen}

The bare Higgs Lagrangian is defined in terms of bare fields $\Phibi$
and bare parameters $\cbj$,
\begin{align}
  \mL_{\rm H,\B}\left( \Phibone,\ldots,\Phibl;\cbone,\ldots,\cbk;\ldots \right),\label{eq:LHBbare}
\end{align}
with $i=1,\ldots,l$ and $j=1,\ldots,k$. 
\change{We stress that $\cbj$ are the theory-defining parameters, \ie
  the bare parameters of the original Lagrangian with unbroken gauge symmetry.}
After spontaneous symmetry
breaking, the neutral scalar components $\phibi$ of the Higgs
multiplets obtain vevs $\vbi+\dvi$, and the Lagrangian can be written
as
\begin{align}
  \mL_{\rm H,\B}\left(\phibone + \vbone + \dvone,\ldots,\phibl + \vbl + \dvl;
                  \ldots;\cbone,\ldots,\cbk;\ldots \right),
\label{eq:barelag}
\end{align}
where the shifts of the vevs $\dvi$ are introduced for later
convenience. The $\vbi$ are chosen in such a way that the vevs of the
shifted fields $\phibi$
\begin{align}
  \left\langle \phibi \right \rangle &= 0 + \ti\left( \dvone,\ldots,\dvl \right)
  + \text{higher-order corrections},
  \label{eq:fieldvev}
\end{align}
vanish at tree level, where $\dv=0$.  Here $\dv=0$ is short for
$\dvone=0,\ldots,\dvl=0$.  \change{Since the $\vbi$ are thus defined
  to minimize the bare scalar potential, they are directly given in
  terms of the bare (theory-defining) parameters of the Lagrangian
  [see \refeq{eq:bareSMpar} for the SM].}

The tadpole counterterm $\ti$ is defined by
the expression obtained by taking the derivative of the Lagrangian
with respect to the field $\phii$, setting all fields to zero and
keeping only the linear and higher-order terms in $\dvi$.  This can be
generalized to tadpole counterterms of $n$-point functions. To this
end, we define the tadpole Lagrangian $\Delta \mathcal{L}$ in the
\ts{}, which gives rise to all the tadpoles in the theory, as
\begin{align}
  \Delta \mathcal{L}:= \mathcal{L}-\left.\mathcal{L}\right|_{\dv=0}.
  \label{eq:deltalag}
\end{align}
Then, the tadpole counterterm is defined by the expression
\begin{align}
  \ti\left( \dvone,\ldots,\dvl \right) 
  &\equiv \Delta\mathcal{L}_{i}\left( \dvone,\ldots,\dvl;\ldots \right)
\quad \text{with} \quad \Delta\mathcal{L}_{i}:=
\left.
\frac{\partial\Delta\mathcal{L}}{\partial{\phii}}
\right|_{\varphi=0},
  \label{eq:rentadpoleeq1}
  \intertext{where the field $\phii$ can be any scalar in the theory.
    The tadpole counterterms to two-point functions are given by
    derivatives of the Lagrangian with respect to $\phii$ and $\phij$,
    \ie}
  t_{ij}\left( \dvone,\ldots,\dvl \right) 
  &\equiv \Delta\mathcal{L}_{ij}\left( \dvone,\ldots,\dvl;\ldots \right)
  \quad \text{with} \quad \Delta\mathcal{L}_{ij}:=
  \left.\frac{\partial^2\Delta\mathcal{L}}{\partial{\phii}\partial{\phij}}\right|_{\varphi=0},
  \label{eq:rentadpoleeq2}
  \intertext{where the fields $\phii$ and $\phij$ can be scalars,
    vector bosons or fermions. 
Analogously, we obtain the tadpole counterterm to the interaction of three
fields} t_{ijk}\left(
    \dvone,\ldots,\dvl \right) &\equiv \Delta\mathcal{L}_{ijk}\left(
    \dvone,\ldots,\dvl;\ldots \right) \quad \text{with} \quad
  \Delta\mathcal{L}_{ijk}:= \left.\frac{\partial^3\Delta\mathcal{L}}
    {\partial{\phii}\partial{\phij}\partial{\phik}}\right|_{\varphi=0},
  \label{eq:rentadpoleeq3}
\end{align}
where the fields $\phii$, $\phij$ and $\phik$ can only be scalars or
vector bosons in renormalizable quantum field theories. Tadpole
counterterms to scalars arise from the Higgs potential, while the
tadpole counterterms involving vector bosons and fermions originate
from the kinetic terms and the Yukawa terms, respectively, of the
Lagrangian of the theory.  The one-particle irreducible (1PI) tadpole
loop corrections are given by
\begin{align}
  \Ti &=
  0 + 
  \underbrace{
  \raisebox{-5pt}{\includegraphics{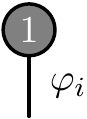}}
  }_{\Tiol}+
  \underbrace{
  \raisebox{-5pt}{\includegraphics{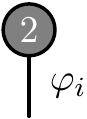}}
  }_{\Titl}+ \ldots
  =:
  \raisebox{-4.8pt}{\includegraphics{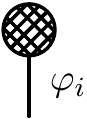}}
  ,
  \label{eq:perturbativeTn}
\end{align}
with the contributions%
\footnote{The tadpole loop contributions \Ti{} should not be confused
  with the tadpole counterterms \ti.} at tree level (0), one loop
(${\Tiol}$), two loops (${\Titl}$), and the hatched graph denoting the
sum of all 1PI tadpole graphs.  Note that ${\Ti^{(N)}}$ contains
$(N-1)$-loop counterterm contributions.

Choosing the shifts of the vevs as $\dv=0$ implies that all tadpole
counterterms $\ti$ vanish, and $\left\langle\phibi\right\rangle=0$
holds at tree level.  Already at one-loop order, the bare fields
receive a non-vanishing vev due to the tadpole loop corrections in
Eq.~\eqref{eq:perturbativeTn}.  In the following, we illustrate the
tadpole renormalization for general two-point functions.  We define
the self-energy ${\Sigma}_{ii}(\qq^2)$ of a field $i$ as the
higher-order (beyond tree-level) contributions to the inverse
connected 2-point function and denote the corresponding 1PI
contributions by $\Sigma^{\mathrm{1PI}}_{ii}$.  We indicate
renormalized functions by a hat.  The renormalized self-energy at
one-loop order can be written in terms of 1PI graphs as follows
\begin{align}
  \hat{\Sigma}_{ii}^{(1)}\left(\qq^2\right) &=
\underbrace{
  \raisebox{-7pt}{\includegraphics{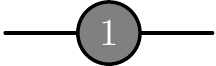}} + 
  \raisebox{-1pt}{\includegraphics{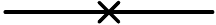}}
}_{\hat{\Sigma}^{(1),\mathrm{1PI}}_{ii}}
  +
  \underbrace{
  \sum_n \bigl(
    \raisebox{-0.5pt}{\includegraphics{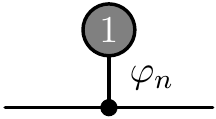}} + 
    \raisebox{-1pt}{\includegraphics{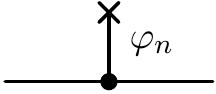}}
  \bigr)}_{\hat{T}^{(1)}_{ii}}
 \label{eq:fullselfenergya}\\
  &=
  \raisebox{-7pt}{\includegraphics{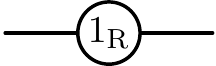}} + 
  \sum_n
  \raisebox{-0.5pt}{\includegraphics{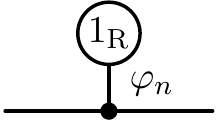}}
  ,
    \label{eq:fullselfenergyb}
\end{align}
where the subscript $\mathrm{R}$ indicates renormalized 1PI graphs.
For the two-loop contributions we obtain
\begin{align}
  \hat{\Sigma}_{ii}^{(2)}\left(\qq^2\right) =&
  \raisebox{-7pt}{\includegraphics{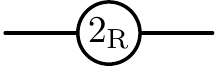}} + 
  \raisebox{-0.5pt}{\includegraphics{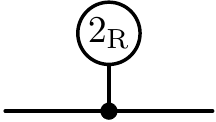}} + 
  \raisebox{-0.5pt}{\includegraphics{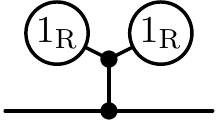}} + 
  \raisebox{-0.5pt}{\includegraphics{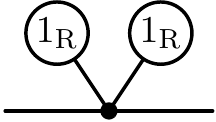}} 
    \notag
  \\ 
  &+\raisebox{-7.5pt}{\includegraphics{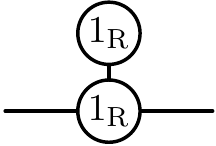}} + 
    \raisebox{-0.5pt}{\includegraphics{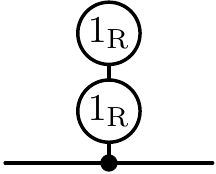}},
    \label{eq:fullselfenergy2loop}
\end{align}
where we suppressed the summation over tadpoles of different fields
for simplicity.  In the \ts{} the counterterm corresponding to the
two-point irreducible part reads
\begin{align}
  \raisebox{-1pt}{\includegraphics{\pics/2pct}} &=
  t_{ii}\left( \dvone,\ldots,\dvl \right) - 
  \delta m_i^2 +
  \left(p^2-m_i^2\right)\delta Z_i,
\end{align}
where the tadpole counterterm $t_{ii}$ is defined in
Eq.~\eqref{eq:rentadpoleeq2}, $\delta m_i^2$ is the mass counterterm,
and $\delta Z_i$ is the field-renormalization counterterm.  In
addition, the renormalized tadpole contribution $\hat{T}_{ii}$ with
the sum over all scalar fields has to be taken into account. The
physical mass is defined as the zero of the
full 2-point function.%
\footnote{In the presence of mixing this requires an appropriate
  renormalization of the mixing energies $\Sigma_{ij}$, $i\ne j$.}
Accordingly in the on-shell scheme the mass counterterm $\delta m_i^2$
is obtained by requiring the renormalized on-shell self-energy
$\hat{\Sigma}_{ii}(m_i^2)$, as defined in Eq.\ 
\eqref{eq:fullselfenergya}, to vanish.  One can show that the
resulting counterterm is gauge independent, for example, by means of
extended BRST symmetry (following the proof for $\PW$~bosons in
\citere{Gambino:1999ai} in the $R_{\xi}$-gauge).  We have verified
that the gauge parameters $\xiz$ and $\xiw$ [see Eq.~\eqref{eq:GFRxi}]
cancel for mass counterterms in the SM and the \THDM\ by explicit
calculations in the $R_{\xi}$ gauge.

As stated above, both $t_{ii}$ and $\ti$ vanish for $\dv=0$.
Therefore, the non-renormalized sum of the tadpole diagrams $T_{ii}$,
\ie the third contribution in Eq.~\eqref{eq:fullselfenergya}, has to
be taken into account. To avoid calculating these contributions, we
can allow $\dv\neq 0$ in Eq.~\eqref{eq:barelag} and relate $\dv$ to
the loop tadpole corrections.  More precisely, we choose the shift in
the vev $\dv$ such that the fields $\varphi$ do not develop a vev to
any order in perturbation theory, which is equivalent to setting all
renormalized tadpole contributions $\hat\Ti$ of the theory to zero.
The shifts $\dv$ are determined by solving the non-linear equations
\begin{align}
  -\Ti &= \ti\left( \dvone,\ldots,\dvl \right),\label{eq:rentadpoleeq4}
  \intertext{with $\ti$ defined in
    Eq.~\eqref{eq:rentadpoleeq1}. This equation is solved order by
    order in perturbation theory upon using the perturbative
expansion of $\dv$,}
  \dvi &= \dvi^{(1)} + \dvi^{(2)} + \ldots,
\label{eq:propervexp}
\end{align}
and the perturbative expansion of the tadpole contributions
\eqref{eq:perturbativeTn}.

As an important consequence of the consistent inclusion of tadpole
contributions, as in Eqs.~\eqref{eq:fullselfenergyb} and
\eqref{eq:fullselfenergy2loop}, connected Green's functions do not
depend on a particular choice of $\dv$. To proof this, we consider the
generating functional of Green's functions $Z[j]$ defined using the
Lagrangian \eqref{eq:barelag} with $\dv=0$ and the generating
functional $Z^\prime[j]$ based on Eq.~\eqref{eq:barelag} with an
arbitrary $\dv\ne0$.  Restricting ourselves for simplicity to the case
with only one Higgs field $\varphi$, the two functionals can be
related by a field redefinition $\varphi \to \varphi - \Delta v$ in
$Z^\prime[j]$ using the invariance of the path integral measure under
a constant shift.  The corresponding generating functionals of
connected Green's functions $W[j]=\log Z[j]$ and $W^\prime[j]=\log
Z'[j]$ are related by
\begin{align}
  W^\prime[j]  = W[j] - \ii \dv \int \mathrm{d}^4 x\;j(x).
\end{align}
 Consequently, it follows for the connected Green's functions
\begin{align}
  \left.\frac{\delta^n W^\prime}{\ii\delta j(x_1) \ldots \ii\delta
      j(x_n)} \right|_{j=0}
&= 
   \left.\frac{\delta^n W}{\ii\delta j(x_1) \ldots \ii\delta j(x_n)} \right|_{j=0}, \quad \text{for}\quad
  n > 1,
  \label{eq:invariantconnGF}
\\
  \left.\frac{\delta W^\prime}{\ii\delta j(x_1)} \right|_{j=0}
&= 
   \left.\frac{\delta W}{\ii\delta j(x_1)} \right|_{j=0} - \dv.
\end{align}
Note that this does not imply that the vertex functions are the same.
The connection between the vertex functions in both formulations is given by
\begin{align}
  \Gamma^\prime[\bar \varphi^\prime(j)] = \Gamma[\bar \varphi(j)] \quad
  \text{with}  \quad
  \bar \varphi(j(x)) := \frac{\delta W}{\ii \delta j(x)} \quad
  \text{and}
  \quad \bar \varphi^\prime = \bar \varphi - \dv.
\end{align}
Treating $\dv$ perturbatively, the $n$-point vertex functions are related by
\begin{align}
  \frac{\delta^n
  \left(\Gamma^\prime[\bar\varphi]-\Gamma[\bar\varphi]\right)}{\delta
  \bar\varphi(x_1) \ldots \bar\varphi(x_n)} &= 
    \frac{\delta^n\left( \Gamma[\bar\varphi + \dv]-
    \Gamma[\bar\varphi]\right)}{\delta
  \bar\varphi(x_1) \ldots \bar\varphi(x_n)}
  =: \frac{\delta^n \Gamma^\Delta[\bar\varphi]}{\delta \bar\varphi(x_1)
  \ldots \bar\varphi(x_n)}
  \label{eq:implicittadpoles}
  = \mathcal{O}\left(\dv\right).
\end{align}
However, Eq.~\eqref{eq:invariantconnGF} states that the tadpole
counterterm dependence originating from $\Gamma^\Delta$ cancels in
connected Green's functions with more than one external leg.

According to Eq.~\eqref{eq:invariantconnGF} the tadpole
renormalization condition ${\hat{T}_{i}}=0$ does not modify connected
Green's functions in the \ts{} since $\dv$ can be freely chosen, in
particular, such that the tadpole equations \eqref{eq:rentadpoleeq4}
are fulfilled.  This has interesting consequences for the
interpretation of the tadpole counterterms. For example, using that
the expression \eqref{eq:fullselfenergya} is independent of $\dv$, we
conclude that the one-loop two-point tadpole counterterm derived from
Eq.~\eqref{eq:deltalag} obeys
\begin{align}
  \underset{t_{ij}}{\raisebox{-1pt}{\includegraphics{\pics/2pct}}} = - \sum_n
  \raisebox{-1pt}{\includegraphics{\pics/2ptct}},
  \label{eq:dvindep2point}
\end{align}
independently of the nature of the external particle(s).  This can
also be seen directly at one-loop order by computing the two-point
tadpole contribution $t_{ij}$ which can be derived from
Eq.~\eqref{eq:implicittadpoles}.  To this end, we assume a typical
scalar potential $V$ in the Lagrangian \eqref{eq:LHBbare} without
derivative interactions. Expanding $t_{ij}$ to first order in $\dv$
and using that $\dv$ acts as a field shift, we obtain
\begin{align}
  \underset{t_{ij}}{\raisebox{-1pt}{\includegraphics{\pics/2pct}}}
  &= \ft \sum_n \dvn \left.\frac{\partial\delta^2 S}{\partial
  \dvn \delta \varphi_i \delta\varphi_j}\right|_{\varphi=0, \Delta v=0} 
  +\mathcal{O}\left((\dv)^2\right), \notag\\
  &= 
  \ft \sum_n \dvn \left.\frac{\delta^3 S}{
  \delta \varphi_n \delta \varphi_i \delta \varphi_j}\right|_{\varphi=0, \Delta v=0}
  +\mathcal{O}\left((\dv)^2\right),
  \label{eq:twopointtadpolegeneric}
\end{align}
where we use that $\Gamma$ is given by the action $S$ at
tree-level\footnote{The contribution \eqref{eq:twopointtadpolegeneric}
  is a higher-order contribution because $\dv$ is identified with a
  higher-order correction.} and $\ft$ denotes the Fourier transform
that translates Green's functions from configuration space to momentum
space.  Defining the mass squared matrix of the scalar fields
\begin{align}
  (M^2)_{ij}:= \left.\frac{\partial^2 V}{\partial \varphi_i
  \partial\varphi_j}\right|_{\varphi=0,\dv=0},
\end{align}
the explicit tadpole counterterm to the two-point function reads
\begin{align}
\sum_n\raisebox{-1pt}{\includegraphics{\pics/2ptct}} =
  \sum_{n,k} 
  \left(\ft \left.\frac{\delta^3 S}{
  \delta \varphi_n \delta \varphi_i \delta\varphi_j}\right|_{\varphi=0, \Delta v=0}\right)
  (M^2)_{nk}^{-1} t_k .
\label{eq:explicit_tadpole_ct}
\end{align}
For the tadpole counterterm, we find
\begin{align}
t_i &= 
\left.
\frac{\partial\Delta\mathcal{L}}{\partial{\phii}}
\right|_{\varphi=0}
= \left.\sum_n\frac{\partial^2\mathcal{L}}{\partial{\phii}\partial\dvn}
\right|_{\varphi=0,\dv=0} \dvn + \mathcal{O}\left((\dv)^2\right) \notag\\
&= -\left.\sum_n\frac{\partial^2 V}{\partial{\phii}\partial\phin}
\right|_{\varphi=0,\dv=0} \dvn + \mathcal{O}\left((\dv)^2\right) 
= -\sum_n \left(M^2\right)_{in} \dvn + \mathcal{O}\left((\dv)^2\right) .
\end{align}
Inserting this into Eq.~\eqref{eq:explicit_tadpole_ct}, 
yields
\begin{align}
\sum_n  \raisebox{-1pt}{\includegraphics{\pics/2ptct}} 
  &=
  \sum_{n,k,l} 
  \ft \left.\frac{\delta^3 S}{
  \delta \varphi_n \delta \varphi_i \delta\varphi_j}\right|_{\varphi=0, \dv=0}
\left(M^2\right)_{nk}^{-1} \Bigl[-\left(M^2\right)_{kl} \dvl\Bigr]
  +\mathcal{O}\left((\dv)^2\right)\notag\\
  &= 
  -\ft \sum_n\left.\frac{\delta^3 S}{
  \delta \varphi_n \delta \varphi_i \delta \varphi_j}\right|_{\varphi=0,
\dv=0}\dvn +\mathcal{O}\left((\dv)^2\right).
\end{align}
Combining this result with Eq.~\eqref{eq:twopointtadpolegeneric}, we
have explicitly verified Eq.~\eqref{eq:dvindep2point} at one-loop
order.

Using the tadpole renormalization condition \eqref{eq:rentadpoleeq4},
the one-loop two-point tadpole counterterm can be expressed as
\begin{align}
  \underset{t_{ij}}{
  \raisebox{-1pt}{\includegraphics{\pics/2pct}}
  } = - 
  \sum_n \raisebox{-1pt}{\includegraphics{\pics/2ptct}} =
  \sum_n \raisebox{-1pt}{\includegraphics{\pics/2pt}},
  \quad \mathrm{if} \quad t_n = - \Tn \quad\forall n.
  \label{eq:tadpoleinterp}
\end{align}
Therefore, the tadpole counterterms mimic the contribution
of tadpole diagrams \Ti, once $\Ti$ is identified with $-\ti$.

We conclude that the \ts{} is equivalent to a scheme where tadpoles
are not renormalized, which corresponds to setting $\dv$ to zero in
Eq.~\eqref{eq:barelag} and computing all tadpole diagrams explicitly.
The consistent use of the \ts{} defined in Eqs.~\eqref{eq:barelag} and
\eqref{eq:rentadpoleeq1}--\eqref{eq:rentadpoleeq3} guarantees the
independence of the chosen tadpole renormalization, meaning that the
value for any physical quantity is independent of the value of $t_i$
and thus $\hat{T}_{i}$.

We stress that the shift $\dvi$ in the vevs is not a parameter of the
theory but can be chosen arbitrarily. By solving the tadpole equation
\eqref{eq:rentadpoleeq4}, which can be done order by order in
perturbation theory, the shift can be expressed as a function of
tadpole counterterms. After spontaneous symmetry breaking, the bare
physical parameters like particle masses can be expressed through the
theory defining parameters, \ie the coupling constants in the Higgs
potential before spontaneous symmetry breaking. In the \ts{}, tadpole
contributions are never absorbed into the definition of bare physical
parameters, which is crucial to assure gauge independence in some
renormalization schemes.

\change{Here, we would like to mention a general consequence of BRST
  invariance \cite{Piguet:1984js}: \SMatrix~elements calculated in
  terms of the bare theory-defining parameters $\cbj$ of
  Eq.~\refeq{eq:LHBbare} are gauge independent. Thus, renormalization
  schemes that fix the $\cbj$ in a gauge-independent way lead to a
  gauge-independent \SMatrix. A possible gauge-dependent definition of
  $\dv$ does not spoil the gauge independence of the \SMatrix, as long
  as it does not enter the renormalization conditions. However, the
  latter requirement is violated in popular schemes as detailed
  below.}

In the following, within the \ts{} we always take advantage of the
tadpole renormalization condition $\hat{T}_{i} = 0$, and explicit
(counterterm-)tadpoles do not show up.  In \refse{sec:tadpolesSM}, we
describe the \ts{} scheme in the \SM{}, and in
\refse{sec:tadpolesTHDM}, it is applied to the \THDM{}.  For both
models, we discuss how the bare physical masses are properly related
to the original parameters of the Lagrangian.


\subsection{The \bfts{} in the \SM{}}\label{sec:tadpolesSM}

In the \SM{}, physical parameters such as the particle masses and the
EW couplings are usually renormalized using on-shell and physical
renormalization conditions leading to
gauge-independent physical observables.%
\footnote{The renormalization of the strong coupling is not influenced
  by tadpoles at the one-loop level.}  Nevertheless, the techniques
discussed in the previous section, which result in gauge-independent
counterterms to physical parameters in arbitrary renormalization
schemes, can already be illustrated in the \SM{}.  Following the
notation in \citere{Denner:1991}, the bare Lagrangian for the Higgs
field $\Phi_{\B}$ defined in Eq.~\eqref{eq:LHBbare} can be written as
\begin{align}
  \mL_{\mathrm{H},\B} &= (D_{\mu}\Phi_{\B})^{\dagger}(D^{\mu}\Phi_{\B})-V_{\B}(\Phi_{\B}).
\label{eq:SMHiggsLagranigian}
        \intertext{The Higgs field couples to the gauge bosons through the covariant derivative $D_{\mu}$. The bare Higgs potential is given by}
        V_{\B}(\Phi_{\B}) &= \frac{\lb}{4}(\Phi_{\B}^{\dagger}\Phi_{\B})^2-\mu_{\B}^2\Phi_{\B}^{\dagger}\Phi_{\B}
        \intertext{with the bare Higgs doublet $\Phi_{\B}$ defined as}
        \Phi_{\B} &= \left(\begin{array}{c}
        \phi_{\B}^+(x)\\
        \frac{1}{\sqrt{2}}\left[\vbh+\dvh+\hb(x)+\ii\chi_{\B}(x)\right]
        \end{array}\right).
\end{align}
We insert the Higgs doublet into the potential and collect the terms
linear, $V_{\B}^{1}$, and quadratic, $V_{\B}^{2}$, in the bare,
neutral, scalar Higgs field $\hb$,
\begin{align}
        \label{eq:Higgs-potential}
        \begin{split}
        V_{\B}(\Phi_{\B}) &\supset \left(\vbh+\dvh\right)\left(\frac{\lb}{4}\left(\vbh+\dvh\right)^2-\mu_{\B}^2\right)\hb(x)+\left(\frac{3\lb}{8}\left(\vbh+\dvh\right)^2-\frac{1}{2}\mu_{\B}^2\right)\hb^2(x)\\
        &\equiv V_{\B}^{1}\left( \dvh,\hb \right) + V_{\B}^{2}\left( \dvh,\hb \right).
        \end{split}
\end{align}
The relation between $\dvh$ and the tadpole counterterm $\tth$ is determined
according to Eq.~\eqref{eq:rentadpoleeq1},
\begin{align}
        \tth \hb(x) &= - V_{\B}^{1}.
\end{align}
Using the tree-level condition, 
$\dvh=0\Leftrightarrow\tth=0$, 
gives the relations 
between the bare parameters
\begin{align}
        \lambda_{\B} &= \frac{4\mu_{\B}^2}{\vbh^2},
\qquad \mu_{\B}^2=\frac{\mhb^2}{2},
\qquad  \vbh=\frac{2\mwb}{\gb},
\label{eq:bareSMpar}
\end{align}
where the last relation defines the bare W-boson mass.  The exact form
of the shift $\dvh$ in terms of the tadpole counterterm $\tth$ can be
obtained from Eq.~\eqref{eq:rentadpoleeq1}, which requires the
knowledge of the linear term of the Higgs potential
\eqref{eq:Higgs-potential} for $\dvh\neq0$,
\begin{align}
  \begin{split}
        V_{\B}^{1}\left( \dvh,\hb \right)
        &= \left(\vbh+\dvh\right)\left(\frac{\lb}{4}\left(\vbh+\dvh\right)^2-\mu_{\B}^2\right)\hb(x)\\
        &= \frac{\mhb^2\dvh}{8 \mwb^2}\left(2 \mwb + \gb \dvh\right)\left(4 \mwb + \gb \dvh\right)\hb(x),\\
        &\overset{!}{=}- \tth\hb(x).
  \end{split}
\end{align}
We can relate $\dvh$ to the tadpole counterterms $\tth$ at every order in
perturbation theory. For $\dvh =\dvhone+ \dvhtwo+\ldots$ and $\tth^{(L)}$ being
the $L$-loop tadpole counterterm corresponding to the \SM{} Higgs boson, we obtain
\begin{align}
        \dvhone = -\frac{\tth^{(1)}}{\mhb^2}
        \label{dvhonesol}
\end{align}
at one-loop order and
\begin{align}
        \dvhtwo = -\frac{\tth^{(2)}}{\mhb^2} - \frac{3\gb \left(\dvhone\right)^2}{4 \mwb}
        \label{eq:dvhtwosol}
\end{align}
at two loops. The tadpole counterterm to the two-point function of the
\SM{} Higgs boson can be obtained from Eq.~\eqref{eq:rentadpoleeq2}
\begin{align}
        V_{\B}^{2}\left( \dvh,\hb \right)&=
  \left(\frac{3\lb}{4}\left(\vbh+\dvh\right)^2-\mu_{\B}^2\right)\frac{\hb^2(x)}{2}\\
        &\overset{!}{=} \frac{\mhb^2-t_{hh}}{2}\hb^2(x),\notag
\end{align}
where $\lb$, $\mu_{\B}^2$, and $\vbh$ are replaced according to
Eq.~\eqref{eq:bareSMpar} as before.  This yields
\begin{align}
t_{hh} \hb^2 &=
  - \left(\frac{3 \gb\,\mhb^2 \dvh}{2 \mwb} + \frac{3\gb\,\mhb^2
    \left(\dvh\right)^2}{8
  \mwb^2}\right) \hb^2.
\label{eq:twopointtadpole}
\end{align}
At one-loop order the bare parameters in
Eq.~\eqref{eq:twopointtadpole} can be replaced by renormalized ones.
Omitting anything beyond one loop, the two-point tadpole counterterm
is given by
\begin{align}
        t_{hh}^{(1)} &= \frac{3 g \thone}{2 \mw}.\label{eq:thh-SM}
\end{align}
Using the on-shell condition $\qq^2=\mhr^2$, where $\mhr$ denotes the
renormalized Higgs-boson mass, and the tadpole renormalization
condition $\hat{T}_{h}=0$, the renormalized on-shell two-point
function of the Higgs boson reads
\begin{align}
    \hat{\Sigma}_{hh}^{(1)}(\mhr^2) &=
    \Sigma_{hh}^{(1),\mathrm{1PI}}(\mhr^2) - \frac{3 g \Thone}{2 \mw} -
    \left(\dmh^{2}\right)^{(1)}.\label{eq:Shh-SM}
\end{align}
This expression can be used to determine the counterterm of the
Higgs-boson mass $\dmh^2$.

We note that at the two-loop order bare parameters need to be
expressed in terms of counterterms and renormalized parameters in
Eqs.~\eqref{dvhonesol} and \eqref{eq:twopointtadpole}, omitting any
terms beyond two loops. The Higgs-boson self-energy at two loops,
focussing on the tadpole dependence, is discussed in
\refapp{sec:2lse}.

We note that additional tadpole counterterms are required in the \SM{} for two-
and three-point functions of scalars, vector bosons and fermions. Tadpole
counterterms for two- and three-point functions involving vector bosons
originate from the kinetic terms of the \SM{} Higgs sector, while tadpole
counterterms to fermion self-energies 
result
from the Yukawa terms of the \SM{} Lagrangian.


\subsection{Gauge independence of physical parameters in the
\SM}\label{sec:gaugeinvsm}
As mentioned at the beginning of \refse{sec:tadpoles}, the physical
parameters of the \SM{} are usually renormalized on shell. In this
case, gauge dependencies introduced by careless treatments of tadpoles
cancel in all renormalized physical quantities. However, when some
parameters are renormalized in the \msbar{} scheme this is not
generally the case.  Such problems can originate from gauge-dependent
counterterms resulting from a gauge-dependent definition of bare
physical parameters. We use the \SM{} to demonstrate potential
problems with gauge dependence in renormalization schemes commonly
used in the literature. In order to illustrate the effect of different
tadpole renormalization schemes on the gauge dependence of the
counterterms of physical parameters, we compare the gauge dependence
of the Higgs-boson mass counterterm $\dmh^2$ in the scheme described
in \citere{Denner:1991} and the $\beta_h$ scheme from
\citere{Actis:2006} to the \ts{}.

The scheme described in \citere{Denner:1991} requires the vev of the bare
Higgs-boson field to vanish at one-loop order
\begin{align}
  \langle\hb\rangle=0,
  \label{eq:vevDS}
\end{align}
such that $\tth$ is fixed via Eq.~\eqref{eq:rentadpoleeq4} and thus
gauge dependent.  At the same time, the bare Higgs-boson mass $\mhb^2$
is defined as the coefficient of the quadratic term in the Higgs
field.  As a consequence, no tadpole counterterm $t_{hh}$ appears in
the two-point function. However, the so-defined bare Higgs-boson mass,
\eg at one loop
\begin{align}
  \mhb^2 = 2\mu_{\B}^2 - t_{hh}
 = 2\mu_{\B}^2 - \frac{3 \gb\tth}{2 \mwb},
  \label{eq:mhbareDS}
\end{align}
depends on the tadpole counterterm $\tth$ 
and thus becomes gauge dependent as well.

The mass counterterm of the Higgs boson, defined as the difference
between the bare mass squared $\mhb^2$ and the renormalized
Higgs-boson mass squared $\mhr^2$,
\begin{align}
  \mhb^2 \equiv \mhr^2 + \dmh^2,
  \label{eq:massctdef}
\end{align}
is determined by requiring that the renormalized self-energy
\eqref{eq:fullselfenergya} vanishes on-shell, \ie for $\qq^2=\mhr^2$.
Since the renormalized tadpole contribution $\hThone$ vanishes [see
Eq.~\eqref{eq:vevDS}], the Higgs-boson mass counterterm is given by
the 1PI contribution
\begin{align}
  \hat{\Sigma}_{hh}^{\mathrm{1PI}}\left(\mhr^2\right)=\Sigma_{hh}^{\mathrm{1PI}}\left(\mhr^2\right)-\dmh^2\overset{!}{=}0.
  \label{eq:massctdefDS}
\end{align}
The gauge dependence of $\Sigma_{hh}^\mathrm{1PI}(\mhr^2)$,
which results in a gauge-dependent mass counterterm $\dmh^2$, can be
shown by means of an explicit calculation as in \citere{Kniehl:2000}.

In the scheme of \citere{Denner:1991} also the bare gauge-boson and
fermion masses become gauge dependent, since they are defined using
the shifted vev $(\vbh+\dvh)$. For instance, the bare W-boson mass is
given by
\begin{align}
\mwb = \frac{1}{2}\gb(\vbh+\dvh)  
= \frac{1}{2}\gb\left(\vbh-\frac{\tth}{\mhb^2}\right) 
\end{align}
at one-loop order.

The gauge dependence of the Higgs-boson mass counterterm can also be
understood from its definition \eqref{eq:massctdef}. As the
renormalized mass parameter is identified with the physical mass in
the on-shell scheme, which has to be gauge independent, the gauge
dependence of a bare parameter must be compensated by the gauge
dependence of the counterterm. Using the short-hand notation
$\partial_{\xi}$ for $\partial/\partial\xi$, \eqref{eq:massctdef}
leads to
\begin{align}
  \partial_{\xi}\mhb^2&=\partial_{\xi}\mhr^2+\partial_{\xi}\dmh^2\label{eq:gauge-dep-mh}
\end{align}
in the $R_{\xi}$-gauge with gauge parameter $\xi$. As
$\partial_{\xi}\mhr^2=0$, the gauge dependence of $\mhb^2$ is directly
related to the gauge dependence of $\dmh^2$.

A similar discussion applies to the $\beta_h$ scheme in
\citere{Actis:2006} which also requires the vev $\langle\hb\rangle$ to
vanish at higher orders and defines the bare masses using the shifted
vev, \eg
\begin{align}
\mhb^2 &= \frac{1}{2}\lb(\vbh+\dvh)^2 
= \frac{1}{2}\lb\vb\left(\vbh-2\frac{\tth}{\mhb^2}\right)  
= \frac{1}{2}\lb\vbh^2 - \frac{\gb\tth}{\mwb} ,
\\
\mwb &= \frac{1}{2}\gb(\vbh+\dvh)  
= \frac{1}{2}\gb\left(\vbh-\frac{\tth}{\mhb^2}\right) 
\end{align}
at one-loop order. In this scheme the parameter $\muB$ is eliminated
from the bare Lagrangian in favour of $\tth$ and $\mhb^2$, while $\lb$
is expressed in terms of $\mhb^2$, $\mwb^2$ and $\gb$.  As a
consequence, all tadpoles to 3-point functions are absorbed into the
definition of the bare physical parameters.  This has the advantage
that tadpole counterterms appear exclusively in one- and two-point
functions for the Higgs and would-be Goldstone bosons.  However, the
bare masses become gauge dependent via the dependence on the tadpole
$\tth$.  The Higgs-boson mass counterterm reads
\begin{align}
  \dmh^2 ={}& \Sigma_{hh}^{\mathrm{1PI}}\left(\mhr^2\right) - \frac{\gb\Thone}{2\mwb}.
\end{align}
This definition differs from the one resulting from
Eq.~\eqref{eq:Shh-SM} upon imposing the on-shell mass renormalization
condition $\hat{\Sigma}_{hh}(\mhr^2)=0$ by gauge-dependent tadpole
terms.
In \citere{Actis:2006}, the $\beta_t$ scheme has been introduced to
cure this problem. There, the bare particle masses are defined in
terms of the bare vev and are therefore gauge independent. 
When tadpoles are renormalized requiring $\hat{T}_{h}=0$, the \ts{} in
the \SM{} is equivalent to the $\beta_t$ scheme.

The theory-defining bare parameters of the original Lagrangian, e.g.\ 
$\lb$ and $\muB$ are gauge independent by definition. When introducing
a new set of bare parameters, these can become gauge dependent if vevs
or tadpoles enter their definition. Bare parameters that are defined
exclusively by the theory-defining bare parameters remain gauge
independent.

In the scheme of \citere{Denner:1991} and the $\beta_h$ scheme of
\citere{Actis:2006} discussed above, all bare particle masses are
gauge dependent.  However, the gauge dependence cancels in physical
quantities since all parameters of the theory are defined by on-shell
renormalization conditions.  This can be seen as follows: the gauge
dependence of the counterterms results from the omission of
gauge-dependent tadpole contributions. Since these are momentum
independent they cancel in renormalized quantities that are defined by
subtracting the same quantity at a fixed point in momentum space. This
holds for on-shell schemes or momentum-subtraction schemes but not for
MS or \msbar{} schemes, where only the divergent parts are subtracted.
Since in the scheme of \citere{Denner:1991} as well as in the
$\beta_h$ scheme of \citere{Actis:2006} all renormalization conditions
are based on complete subtraction of the relevant vertex functions,
the resulting physical quantities and scattering amplitudes are gauge
independent.  This changes when some parameters are renormalized in
the \msbar{} scheme, potentially inducing gauge dependence in the
\SMatrix{} if applied to gauge-dependent bare parameters.

For later convenience we describe a simple way to construct the
different tadpole schemes from the original bare Lagrangian
\eqref{eq:barelag} or \eqref{eq:SMHiggsLagranigian} augmented by
gauge, fermion and Yukawa terms. The starting point is the bare
Lagrangian \change{in terms of the theory-defining parameters, \ie
  $\mu^2_{\B}$, $\lambda_{\B}$, $g_{\B}$, \ldots for the SM, where the
  bare scalar fields have been shifted by an independent parameter $v_{\B}$ (which is not
  yet fixed by a minimum condition) and vanishing $\dv$.}
Then, the tadpole renormalization in the different schemes at
one-loop order can be introduced by shifting the bare parameters as
follows:
\begin{description}
\item [Scheme 1] \strut\\{} 
The bare Lagrangian in the scheme of \citere{Denner:1991} is obtained
upon performing the shifts 
\begin{align}
  \lb &\to \lb + \frac{2\tth}{\vb^3},\qquad
  \mu_{\B}^2 \to \mu_{\B}^2  + \frac{3}{2}\frac{\tth}{\vb}.
 \label{eq:dennerscheme_sm}
\end{align}
\item [Scheme 2] \strut\\{} 
The bare Lagrangian in the $\beta_h$ scheme of \citere{Actis:2006}
results from the shifts 
\begin{align}
  \lb &\to \lb,\qquad
  \mu_{\B}^2 \to \mu_{\B}^2  + \frac{\tth}{\vb}.
 \label{eq:betahscheme_sm}
\end{align}
\item [Scheme 3] \strut\\{} 
Finally, the bare Lagrangian in the \ts{} scheme is obtained via
\begin{align}
 \begin{split}
  \vbh &\to \vbh - \frac{\tth}{\mh^2}.
 \label{eq:tadpolescheme_sm}
\end{split}
\end{align}
\end{description}
\change{Only after these shifts the vev $\vb$ is fixed by minimizing
  the scalar potential for $\tth=0$ and thus related to $\mu_{\B}^2$ and
  $\lb$.}

We have shown that the \ts{} is the natural scheme for dealing with
the tadpoles, as it prevents that tadpoles are absorbed into the
definition of bare parameters. Moreover, the tadpole renormalization
condition $\hat{T}=0$ is very useful since no explicit tadpole loop
contributions have to be computed.  We stress that the presented
tadpole renormalization procedure is general and not restricted to the
SM or the \THDM.


\section{Two-Higgs-doublet model---Lagrangian and fields \label{sec:THDM}}
In this section, we review the definition of the Lagrangian of the
\THDM.  We restrict ourselves to the case of a CP-conserving type-II
\THDM\ with a softly broken $Z_2$ symmetry.  For a comprehensive
introduction to the \THDM\ we refer to \eg
\citeres{Gunion:2002,Branco:2011iw}.


\subsection{Fields and potential in the symmetric basis}

Let \Phii{} denote the $i$-th Higgs doublet with $i=1,2$ defined by
\begin{align}
\Phii=\left(
  \begin{array}{c}
    \phiip\\
    \frac{1}{\sqrt{2}}\left(\vi+\rhoi+\ii\etai\right)
  \end{array}
\right).
\label{eq:higgsdoublets}
\end{align}
The most general potential of the \THDM\ has 3 quadratic and 7 quartic
products of Higgs doublets, each coming with a real or complex
coupling constant. Requiring CP conservation and $Z_2$ symmetry
($\Phione\to-\Phione$,$\Phitwo\to\Phitwo$) simplifies the Lagrangian
resulting in five real couplings $\lambda_1\ldots\lambda_5$ and two
real mass parameters $m^2_1$ and $m^2_2$. Soft breaking of the $Z_2$
symmetry allows for the third mass parameter $m^2_{12}$.  Since the
theory is spontaneously broken, we assign two vacuum expectation
values $v_1$ and $v_2$ which, under the same symmetry restriction, can
be chosen to be real. The most general potential is then given by
\begin{align}
  \begin{split}
        V={}&m_1^2\Phione^{\dagger}\Phione+m_2^2\Phitwo^{\dagger}\Phitwo
     -m_{12}^2\left(\Phione^{\dagger}\Phitwo+\Phitwo^{\dagger}\Phione\right)\\
        &+\frac{\lambda_1}{2}\left(\Phione^{\dagger}\Phione\right)^2
        +\frac{\lambda_2}{2}\left(\Phitwo^{\dagger}\Phitwo\right)^2
        +\lambda_3\left(\Phione^{\dagger}\Phione\right)\left(\Phitwo^{\dagger}\Phitwo\right)
        +\lambda_4\left(\Phione^{\dagger}\Phitwo\right)\left(\Phitwo^{\dagger}\Phione\right)\\
        &+\frac{\lambda_5}{2}\left[\left(\Phione^{\dagger}\Phitwo\right)^2
        +\left(\Phitwo^{\dagger}\Phione\right)^2\right].
  \end{split}
\label{eq:hpgb}
\end{align}


\subsection{Fields and potential in the mass eigenbasis}
After spontaneous symmetry breaking the eight degrees of freedom in
the doublets \eqref{eq:higgsdoublets} split into three would-be
Goldstone bosons $G_0$ and $G^\pm$ and five physical Higgs bosons
$\Hl, \Hh, \Ha, \Hpm$. In order to identify the mass eigenstates, the
part of the Lagrangian quadratic in the fields needs to be
diagonalized.  The mixing angle \be{} is introduced to separate
would-be Goldstone bosons from charged and pseudoscalar physical Higgs
fields, and the angle \al{} is required to diagonalize the neutral
Higgs sector.  With the rotation matrices
\begin{align}
  R(\alpha)=
  \left(\begin{array}{cc}
        \ca & -\sa\\
        \sa & \ca
  \end{array}\right), \qquad
  R(\beta)=
  \left(\begin{array}{cc}
        \cbe & -\sbe\\
        \sbe & \cbe
  \end{array}\right), 
  \label{eq:rotations}
\end{align}
the mass eigenstates of Higgs- and would-be-Goldstone-boson fields are
obtained by the following transformations
\begin{align}
  \left(\begin{array}{c}
    \rhoone\\
    \rhotwo
  \end{array}\right)=
  R(\al)
  \left(\begin{array}{c}
    \Hh\\
    \Hl
  \end{array}\right), \qquad
  \left(\begin{array}{c}
    \phionepm\\
    \phitwopm
  \end{array}\right)={}&
  R(\beta)
  \left(\begin{array}{c}
    G^{\pm}\\
    H^{\pm}
  \end{array}\right), \qquad
  \left(\begin{array}{c}
    \etaone\\
    \etatwo
  \end{array}\right)=
  R(\be)
  \left(\begin{array}{c}
    \GZ\\
    \Ha
  \end{array}\right),
  \label{eq:rotationfields}
\end{align}
for a suitable choice of \al{} and \be{}. 

The Higgs sector is coupled to the gauge sector by means of covariant
derivatives. Identifying the mass eigenstates of the vector bosons,
one obtains the well-known tree-level relations in the \THDM
\begin{align}
  \mw = \frac{1}{2} \g v, \qquad
  \mz = \frac{1}{2} \sqrt{\g^2 + \gy^2}\; v,\qquad
  v=\sqrt{\vone^2+\vtwo^2},
  \label{SSB}
\end{align}
where \g{} and \gy{} denote the weak isospin and hypercharge gauge
couplings, and $\mw$ and $\mz$ the \PW- and \PZ-boson masses,
respectively.  The mixing angle $\beta$ is related to the ratio of
vevs according to $\tb \equiv \tan\beta = \vtwo/\vone$.  The angle
$\alpha$ is chosen such that it diagonalizes the symmetric
mass-squared matrix defined by
\begin{align}
  M_{ij}:= \left.\frac{\partial^2 V}{\partial \rho_i
  \partial\rho_j}\right|_{\varphi=0},
  \label{neutralmm}
\end{align}
where $V$ is the potential in Eq.~\eqref{eq:hpgb}. 
The solution reads \cite{Gunion:2002}
\begin{align}
  \stwoa = \frac{2 M_{12}}
  {\sqrt{({M}_{11} - M_{22})^2 + 4 M_{12}^2}}.
  \label{solalpha}
\end{align}

The parameters of the Higgs potential can then be substituted for
physical parameters after SSB and after diagonalizing the neutral
Higgs sector. The minimum conditions for the scalar potential,
$\langle \rho_i\rangle=0$, read
\begin{align}
        m_1^2&= \Msb^2 \sbeptwo-\frac{2 \mw^2}{g^2} \Bigl[
\lambda _1 \cbeptwo +
\left(\lambda _3+\lambda_4+\lambda _5\right) \sbeptwo
\Bigr],\notag\\
        m_2^2&= \Msb^2 \cbeptwo -\frac{2 \mw^2}{g^2} \Bigl[\lambda _2 \sbeptwo + 
\left(\lambda _3+\lambda _4+\lambda _5\right) \cbeptwo \Bigr],
  \label{eq:vacuumstability}
\end{align}
where we have defined the soft-breaking scale $\Msb$ as
\begin{align}
  \Msb^2=\frac{m_{12}^2}{\cbe\sbe}.
  \label{eq:softz2breakingterm}
\end{align}
The quartic coupling parameters $\lambda_i$ are expressed by the
masses of the physical particles, \ie the Higgs-boson masses \mhl,
\mhh, \mha, \mhc{} and the gauge-boson masses $\mw, \mz$, the
soft-breaking scale $\Msb$, and the mixing angles \al{} and \be{} as
\begin{align}
  \lambda _1&= \frac{g^2}{4 \mw^2\cbeptwo}
  \left[\catwo \mhh^2+ \satwo \mhl^2-\sbeptwo \Msb^2\right],\notag\\
  \lambda _2&= \frac{g^2}{4\mw^2\sbeptwo} \left[ 
   \satwo \mhh^2 + \catwo \mhl^2-\cbeptwo \Msb^2\right],\notag\\
  \lambda _3&= \frac{g^2}{4\mw^2}\left[\frac{\ca \sa}{\cbe \sbe }
  \left(\mhh^2-\mhl^2\right)+2 \mhc^2-\Msb^2\right],\notag\\
  \lambda _4&= \frac{g^2}{4\mw^2}\left(\mha^2-2
\mhc^2+\Msb^2\right),\notag\\
  \lambda _5&= \frac{g^2}{4\mw^2}\left(\Msb^2-\mha^2\right),
  \label{eq:lambdasol}
\end{align}
where the vev $v$ has been substituted using Eq.~\eqref{SSB}.

\subsection{Yukawa Lagrangian for the  type-II \THDM}
In the type-II \THDM, the up-type quarks couple to $\Phitwo$, while
the down-type quarks and leptons couple to $\Phione$. This corresponds
to the Higgs sector in the \MSSM, but here, it is realized by the
discrete $Z_2$ symmetry $\Phione\to-\Phione$,
$d_\mathrm{R}\to-d_\mathrm{R}$, $l_\mathrm{R}\to-l_\mathrm{R}$ and all
other fields unchanged. The corresponding Yukawa Lagrangian reads
\begin{align}
  \mathcal{L}_Y&=
  -\Gamma_\mathrm{d} \overline{Q}_\mathrm{L}\Phi_1 d_\mathrm{R}
  -\Gamma_\mathrm{u} \overline{Q}_\mathrm{L}\tilde{\Phi}_2 u_\mathrm{R}
  -\Gamma_\mathrm{l} \overline{L}_\mathrm{L}\Phi_1 l_\mathrm{R}
  +\mathrm{h.c.},
  \label{eq:Yukawa}
\end{align}
where $\tilde{\Phi}_2$ is the charge-conjugated Higgs doublet of
$\Phi_2$. Neglecting flavour mixing, the coefficients are expressed by
the fermion masses $m_\mathrm{d}$, $m_\mathrm{u}$ and $m_\mathrm{l}$,
and the mixing angle $\beta$,
\begin{align}
  \Gamma_\mathrm{d}=\frac{g\, m_\mathrm{d}}{\sqrt{2} \mw \cbe}, \qquad
  \Gamma_\mathrm{u}=\frac{g\, m_\mathrm{u}}{\sqrt{2} \mw \sbe}, \qquad
  \Gamma_\mathrm{l}=\frac{g\, m_\mathrm{l}}{\sqrt{2} \mw \cbe}.
\end{align}
Again, the vev $v$ has been substituted using Eq.~\eqref{SSB}.

\subsection{Physical parameters}
In the mass eigenbasis the physical parameters resulting from the
Higgs sector are identified with the Higgs-boson masses, \mhl\ (light
Higgs boson), \mhh\ (heavy Higgs boson), \mha\ (pseudoscalar Higgs
boson), \mhc\ (charged Higgs boson), the two mixing angles \al{} and
\be{}, the soft-$Z_2$-breaking scale \Msb, and the vacuum expectation
value $v$. The mass of the light Higgs boson is commonly identified
with the mass of the observed Higgs boson, and $v$ also keeps its SM
value being directly related to the W-boson mass \mw{}. This leaves a
total of six new parameters compared to the \SM. This identification
allows to translate the parameters in the symmetric basis to the mass
eigenbasis
\begin{align}
  \lambda_1,\lambda_2,\lambda_3,\lambda_4,&\lambda_5,m_1,m_2,m_{12}
   \;\to\;
  \mhl,\mhh,\mha,\mhc,\Msb,\alpha,\beta,\mw/g.
\end{align}
Note that the vevs $\vone$ and $\vtwo$ are no independent physical
parameters.  In the following, we choose a more natural representation
for the angles in view of the alignment limit \cite{Gunion:2002}
\begin{align}
  \alpha,\; \beta \quad\to \quad\cab:=\cos(\al-\be),\; \tb:= \tan \beta,
\end{align}
which can be achieved by using simple trigonometric identities.\footnote{
  $\cbe = \frac{1}{\sqrt{1+\tb^2}}$,
  $\sbe = \frac{\tb}{\sqrt{1+\tb^2}}$, 
  $\ca = \frac{\cab - \sab \tb}{\sqrt{1+\tb^2}}$,
  $\sa = \frac{\sab + \cab \tb}{\sqrt{1+\tb^2}}$.}
We have chosen the sign convention for the angles in such a way that the
alignment limit is reached by
\begin{align}
        \sab\to-1, \qquad
        \cab\to0.
\end{align}


\section{Renormalization conditions in the \THDM}
\label{sec:renoconditions}
As argued in \refse{sec:tadpoles}, the \THDM{} is an example of a
theory with new parameters, namely the mixing angles $\alpha$ and
$\beta$ and the soft-$Z_2$-breaking scale $\Msb$, whose on-shell
renormalization through vertex functions would introduce a process
dependence or would be plagued by IR singularities. This has been
discussed for the decays of heavy scalar and charged Higgs bosons in
\citere{Krause:2016oke} and for the renormalization of the mixing
angle $\beta$ in the context of the MSSM in \citere{Freitas:2002um}.
Therefore, an \msbar{} renormalization is advantageous. It requires,
however, care in the treatment of tadpoles to assure the gauge
independence of the bare physical parameters of the theory and thereby
of the \SMatrix{}.  This is guaranteed by the \ts{} presented in
\refse{sec:tadpoles}. In this section, this scheme is applied to the
\THDM{}.  The corresponding renormalization conditions are listed in
\refses{sec:ren-masses-fields}--\ref{sec:ren-ab-msb}, including the
tadpole counterterms which are essential for the gauge independence of
the expressions.

The bare parameters split into the finite, renormalized parameters and
counterterms,
\begin{align}
  \label{eq:2HDMrenorm}
        e_{\B}      &= e + \delta e,\notag\\
        \mwb^2    &= \mw^2 + \delta \mw^2,
        &\mzb^2    &= \mz^2 + \delta \mz^2,\notag\\
        \mhlb^2    &= \mhl^2 + \delta \mhl^2, 
        &\mhhb^2    &= \mhh^2 + \delta \mhh^2, \notag\\
        \mhab^2    &= \mha^2 + \delta \mha^2, 
        &\mhcb^2    &= \mhc^2 + \delta \mhc^2, \notag\\
        \alpha_{\B} &= \alpha + \delta\alpha,
        &\beta_{\B}  &= \beta + \delta\beta,\notag\\
        M_{\rm sb,\B}^2 &= \Msb^2 + \delta\Msb^2, \notag\\
        \mfb &= \mf + \dmf,
\end{align}
where $f$ stands for any fermion.  In the \SM{}, only the $\PZ$-boson
field $Z$ and the photon field $A$ mix and require the introduction of
renormalization matrices. In the \THDM{}, we introduce additional
renormalization matrices for the mixing of the two neutral scalars
$\Hl$ and $\Hh$, the pseudo scalars $\GZ$ and $\Ha$, and the charged
scalars $\Gpm$ and $\Hpm$. The complete field renormalization is given
by
\begin{align}
        \begin{split}
        W^{\pm}_{\B} &=\left(1+\frac{1}{2}\dzw\right)W^{\pm},\\
        \left(\begin{array}{c} Z_{\B} \\ A_{\B} \end{array}\right)
        &=
        \left(\begin{array}{cc} 1+\frac{1}{2}\dzz       & \frac{1}{2}\dzza \\[1.5ex]
                                                        \frac{1}{2}\dzaz                & 1+\frac{1}{2}\dza \end{array}\right)
        \left(\begin{array}{c} Z \\ A \end{array}\right),\\[1.5ex]
        \left(\begin{array}{c} S_{\B} \\ S'_{\B} \end{array}\right)
        &=
          \left(\begin{array}{cc} 1+\frac{1}{2}\dzss      & \frac{1}{2}\dzssp \\[1.5ex]
      \frac{1}{2}\dzsps  & 1+\frac{1}{2}\dzspsp \end{array}\right)
        \left(\begin{array}{c} S \\ S' \end{array}\right),\\[1.5ex]
        \end{split}
        \intertext{where $SS'=\{\GZ\Ha,\Gpm\Hpm,\Hh\Hl\}$. The fermion field renormalization is defined as}
        \begin{split}
        f^{\rm L}_{\B} &=\left(1+\frac{1}{2}\dzfl\right)f^{\rm L},\\
        f^{\rm R}_{\B} &=\left(1+\frac{1}{2}\dzfr\right)f^{\rm R},
        \end{split}
\end{align}
for left-handed (L) and right-handed (R) fermions, where we neglect
fermion mixing.
%
%
\subsection{The \bfts{} applied to the \THDM}\label{sec:tadpolesTHDM}
The \THDM{} as presented in the previous section contains two Higgs doublets
with the corresponding two vevs, such that 
Eq.~\eqref{eq:barelag} 
becomes
\begin{align}
  &\mL_{\rm H,\B}\left( \rhobone + \vbone + \dvone,\rhobtwo + \vbtwo + \dvtwo; \ldots \right).
  \label{eq:properbarelag2HDM}
\end{align}
As in the SM, we use Eq.~\eqref{eq:rentadpoleeq1} to obtain $\dvone$
and $\dvtwo$ expressed by the $L$-loop tadpole counterterms, but
instead of calculating the tadpoles in the generic basis, we define
the vevs in terms of the tadpole counterterms associated to the
physical Higgs fields \Hl{} and \Hh. Of course, the result does not
depend on the choice of parametrization.  In the \THDM{}, the tadpole
counterterms are defined as
\begin{align}
  \label{eq:tadpoleeq2HDM}
  \begin{split}
  \thl &=\Delta\mL_{\Hl}\left( \rhobone + \vbone + \dvone,\rhobtwo +
  \vbtwo + \dvtwo; \ldots \right),\\
  \thh &=\Delta\mL_{\Hh}\left( \rhobone + \vbone + \dvone,\rhobtwo +
\vbtwo + \dvtwo; \ldots \right).
  \end{split}
\end{align}
At tree level, the tadpole counterterms $\thl$ and $\thh$ vanish, such
that $\dvone=\dvtwo=0$. This provides the conditions
\eqref{eq:vacuumstability} for the potential minimum at tree level,
and the relations between the generic and the physical Higgs basis
\eqref{eq:lambdasol} for bare quantities. Thus, the bare parameters in
the symmetric basis are expressed by the bare parameters in the
physical basis.  Linearizing Eq.~\eqref{eq:tadpoleeq2HDM} by using the
expansion \eqref{eq:propervexp}, we can solve for $\dvoneone$ and
$\dvtwoone$. The results for \dvoneone and \dvtwoone simplify after
using the potential minimum conditions at lowest order and the
relation between the parameters in the generic and physical basis.

Evaluating the linearized versions of Eq.~\eqref{eq:tadpoleeq2HDM} for bare physical
parameters, we obtain the one-loop expressions
\begin{align}
  \begin{split}
  \dvoneone &=  \phantom{-}\frac{\thlone \sa}{\mhl^2}-\frac{\thhone\ca}{\mhh^2},\\
  \dvtwoone &=  -\frac{\thlone\ca}{\mhl^2}-\frac{\thhone\sa}{\mhh^2}.
  \end{split}
  \label{eq:tpcondtionsol2HDM1L}
\end{align}
Just as in the \SM{}, tadpole counterterms arise from all terms in the
Lagrangian that depend on the vevs. This results in tadpole
counterterms to two- and three-point functions involving scalars and
vector bosons as well as to fermionic two-point functions. 

In the following, we assume that the tadpole counterterms are fixed
according to Eq.~\eqref{eq:rentadpoleeq4}. Explicit results for the
tadpoles $\Thl$ and $\Thh$ in the \THDM\ in the \rxi-gauge are listed
in \refapp{sec:appA}.
%


%
\subsection{Renormalized two-point functions}\label{sec:two-point-ren}
Using Eq.~\eqref{eq:fullselfenergya} and the condition $\hat{T}_i=0$,
the renormalized self-energies $\hat{\Sigma}(q^2)$ for vector bosons
are given by
\begin{align}
        \gvvth &(q^2) = \gvvt(\qq^2) + (\qq^2-\mv^2)\dzvv - \dmv^2 - \tvv \quad
  \text{with}\quad \tvv^{\mu \nu} =: g^{\mu \nu} \tvv,
  \label{eq:seVec}\\
        \intertext{for $V=\{W,Z,A\}$, with the 1PI contributions $\gvvt$, and}
        \gazth &(q^2) = \gazt(\qq^2) + \frac 1 2(\qq^2-\mz^2)\dzza + \frac 1 2\qq^2 \dzaz \label{eq:seAZ}
        \intertext{for the mixing of photons and \PZ~bosons. The scalar sector works similarly, with}
        \gssh &(q^2) = \gss(\qq^2) + (\qq^2-\ms^2) \dzss - \dms^2 + \tss \label{eq:seSca}
        \intertext{for $S=\{G_0,G^{\pm},H_a,H^{\pm},H_l,H_h\}$ and}
        \gssph &(q^2)= \gssp(\qq^2) + \frac{1}{2}(\qq^2-\ms^2)\dzssp +
        \frac{1}{2}(\qq^2-\msp^2) \dzsps + \tssp \label{eq:seScaOff}
        \intertext{for the mixing of the scalar fields, where
          $SS'=\{\GZ\Ha,\Gpm\Hpm,\Hh\Hl\}$. The renormalized
          scalar--vector-boson mixing energy reads}
        \gvsh &(\qq) = \tvs^{\mu} + \gvs(\qq)
        \label{eq:seVS},
        \intertext{where $VS=\{W^{\pm}\Gmp,W^{\pm}\Hmp,Z\GZ,Z\Ha\}$. Defining the
    helicity projectors}
\Prl &= \frac{1 - \gamma_5}{2}, \qquad 
  \Prr = \frac{1 + \gamma_5}{2},
  \intertext{the renormalized fermionic self-energies can be decomposed into covariants}
        \gffh &(\qq) = \slq\Prl\gffLh(\qq^2) + \slq\Prr\gffRh(\qq^2)+\gffSh(\qq^2),
        \label{eq:seFerm}\\
        \intertext{which are given by}
        \label{eq:covFerm}
        \begin{split}
            \gffLh(\qq^2) & = \gffL(\qq^2)+\dzfl,\\
            \gffRh(\qq^2) & = \gffR(\qq^2)+\dzfr,\\
            \gffSh(\qq^2) & = \gffS(\qq^2)-\frac{1}{2} \mf(\dzfl+\dzfr)-\dmf+\tff.
        \end{split}
\end{align} 
We omit the renormalization of Faddeev--Popov ghosts which are not
needed for the discussion of the processes under consideration (see
\refse{sec:processes}) and in general not at one-loop order.  As a
result of using the \ts, all self-energies involving massive particles
receive tadpole contributions. These tadpole contributions assure the
gauge independence of the on-shell self-energies.  Results for the
tadpole contributions entering the renormalized self-energies in terms
of the tadpole counterterms $\thl$ and $\thh$ are provided in the
't~Hooft--Feynman gauge in \refapp{sec:appB}.

\subsection{Mass and field renormalization conditions}\label{sec:ren-masses-fields}
In the complex-mass scheme%
\footnote{In the usual on-shell scheme, the real part should be taken
  in all renormalization conditions \eqref{eq:conVec},
  \eqref{eq:conScaOff}, \eqref{eq:conFerm}, and \eqref{eq:renalpha0}.}
\cite{Denner:1999gp,Denner:2005fg,Denner:2006ic}, the scalar and
vector-boson mass and field renormalization constants are derived from
the conditions
\begin{alignat}{2}
  \gvvth(\mv^2) & = 0, \qquad  \left.\frac{\partial\gvvth(\qq^2)}{\partial\qq^2}\right|_{\qq^2=\mv^2} &&= 0
,\notag\\
  \gssh(\ms^2) & = 0, \qquad  \left.\frac{\partial\gssh(\qq^2)}{\partial\qq^2}\right|_{\qq^2=\ms^2} &&= 0.
        \label{eq:conVec}
\end{alignat}
The off-diagonal elements of the field renormalization matrices of
scalars and vector bosons are obtained from requiring
\begin{alignat}{2}
  \gazth(0) & = 0,\qquad\gazth(\mz^2) &&= 0,
\notag\\
  \gssph(\ms^2) & = 0,\qquad\gssph(\msp^2) &&= 0.
        \label{eq:conScaOff}
\end{alignat}
For fermions, the renormalization conditions are given by
\begin{align}
        \label{eq:conFerm}
                \mf\gffLh(\mf^2)+\gffSh(\mf^2) & = 0,\notag\\
                \mf\gffRh(\mf^2)+\gffSh(\mf^2) & = 0,\notag\\
                \gffRh(\mf^2)+\gffLh(\mf^2)
                \notag\\
                {}+2\frac{\partial}{\partial\qq^2}\bigg[\mf^2\Bigl(\gffRh(\qq^2)&+\gffLh(\qq^2)\Bigr)
                +2\mf\gffSh(\qq^2)\bigg]
                \Bigg|_{\qq^2=\mf^2}  = 0.
\end{align}
Inserting the expressions \eqref{eq:seVec}--\eqref{eq:covFerm} into the
renormalization conditions \eqref{eq:conVec}--\eqref{eq:conFerm}, we obtain the
mass and field renormalization constants in terms of the 1PI self-energy
and tadpole contributions.

\subsection{Renormalization of the electroweak coupling}

The electromagnetic coupling $e$ as well as the weak coupling $g$ can be
related to the fine-structure constant $\alpha$ (not to be confused
with the mixing angle of the neutral, scalar Higgs bosons)
\begin{align}
  e=g \sw=\sqrt{4\pi\alpha},
\end{align}
where we define the weak mixing angle in the on-shell scheme as
\begin{align}
\cw=\cos\theta_{\mathrm{w}}=\frac{\mw}{\mz}, \qquad \sw = \sqrt{1-\cw^2}.
\end{align}

Renormalizing the electromagnetic coupling in the Thomson limit and
using a Ward identity leads to \cite{Denner:1991,Bohm:2001yx}
\begin{align}
  \frac{\delta e}{e} &{}=
  \left.\frac{1}{2}\frac{\partial\change{\gaat}(\qq^2)}{\partial\qq^2}\right|_{\qq^2=0} 
-\frac{\sw}{\cw}\frac{\change{\gazt}(0)}{\mz^2}.
\label{eq:renalpha0}
\end{align}
\change{Since the counterterm $\delta e$ does not receive any tadpole
contributions, the 1PI self-energies can be replaced by the full
self-energies in Eq.~\refeq{eq:renalpha0}.}

In the $\Gf$ scheme, the fine-structure constant is expressed by the
Fermi coupling constant $\Gf$, using the well-known relation
\begin{align}
  \mw^2\left(1-\frac{\mw^2}{\mz^2}\right) = \frac{\pi\alpha}{\sqrt{2}\Gf}\left(1+\Delta r\right),
\end{align}
where $\Delta r$ contains the EW corrections to muon decay. The
correction term $\Delta r$ depends on the on-shell photon self-energy
\change{$\gaat(0)$, the on-shell Z-boson self-energy $\gzzt(\mz^2)$,
  the W-boson self-energy $\gwwt$ at $\qq^2=0$ and $\qq^2=\mw^2$, the
  photon--Z-boson mixing energy $\gazt(0)$}, and explicit vertex and
box contributions to the muon decay
\cite{Sirlin:1980nh,Marciano:1980pb,Sirlin:1981yz}.  \change{Since the
  tadpole terms cancel within $\Delta r$, this quantity
  takes the same form in terms of the full self-energies or their 1PI
  parts.}  Under the assumption that the couplings of the Higgs bosons
to electrons and muons are negligible, only the self-energies are
modified in the \THDM{}, while the vertex and box contributions to the
muon decay remain the same as in the \SM{}.

In the $\Gf$ scheme, the renormalization constant for the
electromagnetic coupling reads
\begin{align}
  \frac{\delta e}{e} &{}=
  \left.\frac{1}{2}\frac{\partial\change{\gaat}(\qq^2)}{\partial\qq^2}\right|_{\qq^2=0} 
-\frac{\sw}{\cw}\frac{\change{\gazt}(0)}{\mz^2} - \frac{1}{2} \Delta r,
\end{align}
using the conventions of \citere{Denner:1991}.

\subsection
{Renormalization of the parameters \boldmath{$\alpha$}, \boldmath{$\beta$},
  and \boldmath{$\Msb$}}
\label{sec:ren-ab-msb}

\subsubsection
{Mixing angle \boldmath{$\beta$}}

The angle $\beta$ is renormalized using \msbar{} subtraction for the
process $\Ha\to\tau^-\tau^+$,
\begin{align}
\left.\raisebox{-29pt}{\includegraphics{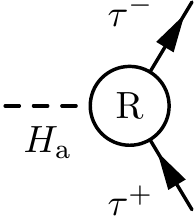}}\;\right|_\PP =
\left[\;\raisebox{-28pt}{\includegraphics{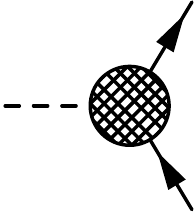}} +
\raisebox{-28pt}{\includegraphics{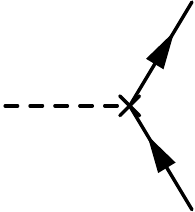}}\;\right]_\PP
\overset{!}{=} 0,
\label{eq:betarenocond}
\end{align}
where \PP\ denotes the projection onto the pole part including the
generic finite parts that are subtracted within the \msbar{} scheme,
\ie the terms proportional to
\begin{align}
\frac{2}{4-D} - \gamma_\mathrm{E} + \log (4 \pi) 
\label{msbarterms}
\end{align}
with the space--time dimension $D$.  The corresponding counterterm
explicitly reads
\begin{align}
\raisebox{-28pt}{\includegraphics{\pics/hatatact.pdf}} =
\frac{e m_\tau \tb}{2 \mw \sw} \biggl[ &
\frac{\delta m_{\tau}}{m_{\tau}} +
\frac{\delta e}{e} +
\frac{\cw^2 - \sw^2}{2 \sw^2} \frac{\delta \mw^2}{\mw^2}-
\frac{\cw^2}{2 \sw^2} \frac{\delta \mz^2}{\mz^2}+
\frac{1+\tb^2 }{\tb}\delta \beta
 -\frac{1}{\tb} \frac{\dzgha}{2} \biggr].
 \label{eq:dbct}
\end{align}
Since there are no explicit tadpoles for this vertex, it follows that
there are no tadpole counterterms in the \ts.
The renormalization condition \eqref{eq:betarenocond} determines $\delta \beta^{\msbar}$.

It has been remarked before (e.g.\ \citeres{Krause:2016oke,Kanemura:2015mxa})
that the relation
\begin{align}
  \delta \beta^{\msbar} = \frac{\dzgha^{\msbar} - \dzhag^{\msbar}}{4},
  \label{eq:betarel}
\end{align}
holds at one-loop order,
and we have explicitly verified this in the
general \rxi-gauge.
This relation has also been used as a
renormalization condition \cite{Krause:2016oke,Kanemura:2015mxa}.
It is particularly useful because it is valid in any of the schemes
which we consider in \refse{sec:gaugedep}.  We verified by explicit
calculation in the \rxi-gauge that $\delta \beta^{\msbar}$ is gauge
independent in the \ts. The gauge dependence of $\delta
\beta^{\msbar}$ in Schemes 1 and 2 is discussed in
\refapp{sec:gdepbeta}.

\subsubsection
{Mixing angle \boldmath{$\alpha$}}
The angle $\alpha$ is renormalized using \msbar{} subtraction in the
process $\Hl\to\tau^-\tau^+$,
\begin{align}
\left.\raisebox{-30pt}{\includegraphics{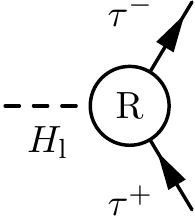}}\;\right|_\PP=
\left[\;\raisebox{-28pt}{\includegraphics{\pics/hatata.pdf}} +
\raisebox{-28pt}{\includegraphics{\pics/hatatact.pdf}}\;\right]_\PP
\overset{!}{=} 0.
\label{eq:alpharenocond} 
\end{align}
The corresponding counterterm is given by
\begin{align}
\raisebox{-28pt}{\includegraphics{\pics/hatatact.pdf}} = 
\notag
\frac{\ii e m_\tau}{2 \mw \sw} \biggl[ &\left(\sab + \cab \tb\right)
 \left(
 \frac{\delta m_{\tau}}{m_{\tau}} +
 \frac{\delta e}{e} +
\frac{\cw^2 - \sw^2}{2 \sw^2} \frac{\delta \mw^2}{\mw^2}-
\frac{\cw^2}{2 \sw^2} \frac{\delta \mz^2}{\mz^2} +
\tb \delta \beta \right) \\
&+\left(\cab - \sab \tb\right)\left( \delta \alpha - \frac{\dzhhhl}{2}\right)
 \biggr].
\end{align}
Again, there is no tadpole dependence, and the renormalization
condition \eqref{eq:alpharenocond} determines $\delta\alpha^{\msbar}$.
Similarly to Eq.~\eqref{eq:betarel} the
relation
\begin{align}
  \delta \alpha^{\msbar} = \frac{\dzhhhl^{\msbar} - \dzhlhh^{\msbar}}{4},
  \label{eq:alpharel}
\end{align}
is valid at one-loop order, which we have explicitly verified in the
\rxi-gauge. Moreover, we have checked by explicit calculation in the
\rxi-gauge that $\delta \alpha^{\msbar}$ is gauge independent in the
\ts{} but gauge dependent in Schemes 1 and 2.  In addition, we have
verified that the renormalized vertex $\Hl \tau^- \tau^+$ is gauge
independent in the \ts, while it is gauge dependent in Schemes 1 and
2.

\subsubsection
{Soft-breaking scale \boldmath{$\Msb$}}
The parameter $\Msb$ is renormalized using the \msbar{} subtraction of the process $\Hh\to\Hl\Hl$,
\begin{align}
\left.\raisebox{-30pt}{\includegraphics{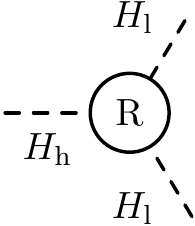}}\;\right|_\PP =
\left[\;\raisebox{-28pt}{\includegraphics{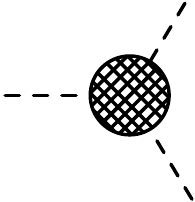}} +
\raisebox{-28pt}{\includegraphics{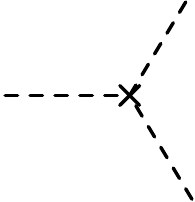}}\;\right]_\PP
\overset{!}{=} 0.
\end{align}
The dependence of this vertex on $\delta \Msb$ and the 
tadpole counterterm 
$t_{\Hh\Hl\Hl}$ reads
\begin{align}
  \raisebox{-28pt}{\includegraphics{\pics/hhhlhlct.pdf}} = 
  \notag
  \frac{e}{\mw\sw}\left[(\ldots)+\left(-2\cab+3\cab^3+\frac{3}{2}\sab\cab^2\left(\frac{1}{\tb}-\tb\right)\right)\delta\Msb^2\right]+t_{\Hh\Hl\Hl},
\end{align}
where $(...)$ stands for other counterterms.


\section{Discussion of gauge dependence\label{sec:gauge_dependence}}
\label{sec:gaugedep}
In this section, we discuss the gauge dependence of \SMatrix{}
elements assuming that the renormalization conditions listed in
\refse{sec:renoconditions} are employed. We show that tadpole
renormalization schemes that are commonly used in literature in
combination with \msbar\ renormalization lead to gauge-dependent
predictions, while the use of the \ts{} ensures gauge independence.

\subsection{Gauge-fixing Lagrangian \label{sec:gaugefixing}}

To verify gauge independence of \SMatrix{} elements and counterterms
of physical parameters in the \ts, we use a general $\rxi$-gauge. The
corresponding gauge-fixing Lagrangian is given by
\begin{align}
        \mL_{\mathrm{GF}}={}&-\frac{1}{\xiw} C^+C^- -\frac{1}{2\xiz}(C^Z)^2-\frac{1}{2\xia}(C^{A})^2\label{eq:GF}
        \intertext{with}
        C^{A} ={}& \partial^{\mu}A_{\mu},\qquad
        C^Z= \partial^{\mu} Z_{\mu} - \xiz \mz G_0,\qquad
        C^+= \partial^{\mu} W_{\mu}^\pm \mp\ii \xiw \mw G^\pm.
        \label{eq:GFRxi}
\end{align}
 We do not renormalize the gauge-fixing Lagrangian, \ie we write it directly in
 terms of renormalized fields, which is sufficient to assure that all \SMatrix{}
 elements are finite \cite{'tHooft:1972fi,Lee:1973fn}.  To compensate the
 unphysical components in $\mL_{\mathrm{GF}}$, Faddeev--Popov ghosts are
 introduced as usual.

\subsection{Characterizing different schemes}\label{sec:charac-schemes}

In the literature different tadpole renormalization schemes are
employed.  In order to efficiently generate the tadpole counterterms
we follow the recipe presented at the end of \refse{sec:gaugeinvsm}
for the SM.  \change{We start from the tree-level Lagrangian
  \eqref{eq:barelag} in the symmetric basis in terms of the
  theory-defining parameters $m^2_{i,\B}$, $i=1,2$, $m^2_{12,\B}$,
  $\lambda^2_{j,\B}$, $j=1,\ldots,5$, where the fields have been
  shifted by independent parameters $v_{i,\B}$. Then we perform the
  shifts of the parameters as defined below. Thereafter, the vevs
  $v_{i,\B}$ are determined at leading order, and the bare physical
  basis is introduced by using the tree-level relations
  \eqref{eq:vacuumstability}--\eqref{eq:lambdasol}, \ie for
  $\thl=0=\thh$.  Finally, the bare parameters are renormalized
  according to Eq.~\eqref{eq:2HDMrenorm}.} 

The tadpole renormalization in the different schemes can be generated
by shifting the corresponding bare parameters as follows:

\begin{description}
  
\item [Scheme 1] \strut\\{} A commonly used renormalization scheme for
  the SM was proposed in \citere{Denner:1991}. There, the bare
  physical masses are defined as the coefficients of the quadratic
  terms in the fields, and the tadpoles are 
the coefficients of the terms
  linear in the fields. Applying this definition to the \THDM, we can
  construct the corresponding Lagrangian by a shift in the bare
  parameters as
\begin{align}
  \loneb &\to \loneb - \frac{1}{\vone^3} \kone,\notag\\
  \ltwob &\to \ltwob + \frac{1}{\vtwo^3} \ktwo,\notag\\
  \lthrb &\to \lthrb - \frac{2\vtwo^2}{\vone v^4} \kone 
                     + \frac{2\vone^2}{\vtwo v^4} \ktwo,\notag\\
  \lfoub &\to \lfoub + \frac{\vtwo^2}{\vone v^4} \kone 
                     - \frac{\vone^2}{\vtwo v^4} \ktwo,\notag\\
  \lfivb &\to \lfivb + \frac{\vtwo^2}{\vone v^4} \kone 
                     - \frac{\vone^2}{\vtwo v^4} \ktwo,\notag\\
   \moneb &\to \moneb + \frac{3}{2 \vone} \kone,\notag\\
   \mtwob &\to \mtwob - \frac{3}{2 \vtwo} \ktwo.
 \label{eq:dennerscheme}
\end{align}
One can verify that the prescription \eqref{eq:dennerscheme} leads to
the tadpole equations \eqref{eq:tadpoleeq2HDM} in the \THDM{}.  Note
that in the alignment limit, the \SM{} tadpoles (see App.~A in
\citere{Denner:1991}) are reproduced.
\item [Scheme 2]\strut\\
  In the $\beta_h$ scheme of \citere{Actis:2006}, the mass parameters
  in the Higgs potential are eliminated in favour of explicit
  tadpoles, while the quartic Higgs couplings $\lambda_i$ are kept
  fixed. Thus, no tadpole counterterm contributions appear in the
  triple and quartic vertices between scalars, but the mass parameters
  of the Higgs potential and thus the two-point functions are shifted
  by tadpole counterterms,
\begin{align}
  \lib &\to \lib,\notag\\
  \moneb &\to \moneb + \frac{\kone}{\vone},\notag\\
  \mtwob &\to \mtwob - \frac{\ktwo }{\vtwo}.
  \label{eq:bhscheme}
\end{align}
For explicit computations, Scheme 2 is very simple
because tadpole counterterms appear only in two-point functions.  This
scheme is widely used, \eg in the
\THDM~\cite{Santos:1996vt,Kanemura:2004mg,LopezVal:2009qy,LopezVal:2012zb,Kanemura:2015mxa}
and in the \MSSM~\cite{Pierce:1992hg,Freitas:2002um,Baro:2008bg}.  In
contrast, in Scheme~1 two-point functions do not receive tadpole
counterterms due to the definition of the bare masses in that scheme.
  
\item[Scheme 3] \strut\\ As described in detail in
  \refse{sec:tadpoles-gen}, in the \ts{}, the 
vevs 
are replaced by
  \vtone{} and \vttwo{}, which corresponds to the following shift
\begin{align}
  \vbone &\to \vbone   +\frac{\thl\sa}{\mhl^2}
                       -\frac{\thh\ca}{\mhh^2}
                     ,\notag\\
  \vbtwo &\to \vbtwo  -\frac{\thl\ca}{\mhl^2} 
                      -\frac{\thh\sa}{\mhh^2}
                      .
  \label{eq:tadpolescheme}
\end{align}
This prescription has to be applied to the full Lagrangian and is not
restricted to the Higgs potential.
\end{description}
We stress again, as we have shown in \refse{sec:tadpolesSM}, that the
bare parameters of the theory are shifted by (gauge-dependent) tadpole
contributions in Schemes~1 and 2, as opposed to the prescription of
the \ts{} \eqref{eq:tadpolescheme}, where only the unphysical vevs
receive a shift.

\subsection{Differences of counterterms in different renormalization schemes}
\label{se:differences}

Employing different schemes leads to different expressions for the
counterterms. Since we are mainly interested in the changes of
amplitudes between different tadpole renormalization schemes, we
compare counterterms in the different schemes. We name the schemes as
in the previous section, \ie Scheme~1 for the scheme employed in
\citere{{Denner:1991}} and Scheme~2 for the $\beta_h$ scheme of
\citere{Actis:2006}. The \ts\ is referred to as Scheme~3. We
generically label the difference in the schemes for a counterterm
$\delta c_i$ by
\begin{align}
  \Delta_{i} \delta c &= \delta c_i - \delta c_3, \quad i=1,2,
\end{align}
where the $\Delta_i$ describe the difference of Scheme~$i$ with
respect to the Scheme 3.

In the following, we list the results for the counterterm parameters.
Thereby, we make use of results for tadpoles listed in
\refapps{sec:appA} and \ref{sec:appB}.
As a first result, we note that the counterterms of couplings in the
\SM{} are not affected by the choice of the tadpole renormalization
scheme, \ie
\begin{align}
  \label{eq:ctschemedep}
  \Delta_{i} \delta e &= \Delta_{i} \delta \cw = 0, \quad i=1,2.
\end{align}
However, the masses of all fermions and gauge bosons change equally
for $i=1,2$ as
\begin{align}
  \Delta_{i} \dmv^2 &= \frac{\g}{\mw} \mv^2 \left(\frac{\thl}{\mhl^2}\sab  -
                            \frac{\thh}{\mhh^2}\cab\right),\notag\\
  \Delta_{i} \dmfD &= \frac{\g}{2 \mw} \mfD
                            \left(
                            \frac{\thl}{\mhl^2} \left(\cab \tb+\sab\right)
                            +\frac{\thh}{\mhh^2} \left(\sab\tb-\cab\right)
                            \right), \notag\\
  \Delta_{i} \dmfU &= \frac{\g}{2 \mw \tb} \mfU 
                            \left(
                            \frac{\thl}{\mhl^2} \left(\sab \tb-\cab\right)
                            -\frac{\thh}{\mhh^2} \left(\cab\tb+\sab\right)
                            \right), \notag\\
  \Delta_{i} \dmfL &= \frac{\g}{2 \mw} \mfL
                            \left(
                            \frac{\thl}{\mhl^2} \left(\cab \tb+\sab\right)
                            +\frac{\thh}{\mhh^2} \left(\sab\tb-\cab\right)
                            \right),
  \label{eq:masstadpoledep}
\end{align}
which is easily derived because neither in Scheme~1 nor in Scheme~2
there are tadpole contributions to two-point functions of fermions and
gauge bosons.  Therefore, the difference is the full tadpole
dependence of these two-point functions in the \ts{} obtained from
Eq.~\eqref{eq:tadpolescheme}. The results for the scalar fields are
more complicated but not needed in the following.  For Scheme~1, the
difference is again given by the full tadpole dependence in the \ts{},
which can be found in \refapp{sec:appA}.

In the \ts{}, the mass counterterms are gauge independent by
definition, which we have verified in a general \rxi-gauge.
Consequently, the mass counterterms in Schemes~1 and 2 are gauge
dependent, and their gauge dependence is given by the gauge dependence
of the corresponding tadpole counterterms.

Next, we give the results for the new parameters in the \THDM. Since
those parameters are renormalized in the \msbar~scheme, we only need
to study the UV-divergent parts of vertex functions and can use the
Eqs.~\eqref{eq:betarel} and \eqref{eq:alpharel}, which hold in any of
the presented schemes. For $\beta$, we obtain
\begin{align}
  \Delta_{i} \delta \beta^\msbar &= \Delta_{i} \frac{\dzgha^\msbar-\dzhag^\msbar}{4}
  = -\frac{\Delta_{i} \tgha^\msbar}{\mha^2},  \quad
  i=1,2,
  \label{eq:gaugeinvtadpolecomb2}
\end{align}
where ``$\msbar$'' denotes the UV-divergent part of the corresponding
expression together with the finite terms in the \msbar-scheme
according to \eqref{msbarterms}.  In the first step, we use
Eq.~\eqref{eq:betarel}. The second step can be derived by solving the
renormalization conditions for the relevant mixing energies
\begin{align}
  \left(p^2-\mha^2\right) \frac{\dzhag}{2} + p^2 \frac{\dzgha}{2} +\tgha + \text{self-energy
  diagrams} \overset{!}{=} \text{finite},
\end{align}
where we omitted any explicit tadpoles because of $\hat{T}_i=0$. Since
the self-energy diagrams do not depend on the scheme, the
scheme-dependent divergence of the tadpole counterterms has to cancel
against the scheme-dependent divergence of the non-diagonal field
renormalization constants, and for $i=1,2$ we obtain
\begin{align}
  \Delta_{i}\left(\left(p^2-\mha^2\right) \dzhag^\msbar + p^2 \dzgha^\msbar +
  2 \tgha^\msbar\right) \overset{!}{=} 0,
\end{align}
which implies
\begin{align}
  \Delta_{i}\dzhag^\msbar = 2\frac{\Delta_{i}\tgha^\msbar}{\mha^2}, \qquad
   \Delta_{i}\dzgha^\msbar = -2\frac{\Delta_{i}\tgha^\msbar}{\mha^2}.
\end{align}
Therefore, the scheme dependence of $\delta\beta^\msbar$ is given by
the one of the tadpole contribution $\tgha$ of
Eq.~\eqref{eq:gaugeinvtadpolecomb2}.  The explicit results for $\tgha$
in the \ts\ are listed in \refapp{sec:appB}, and those for Schemes~1
and 2 are given by
\begin{align}
\label{tZHa12}
t_{\GZ\Ha,1} = t_{\GZ\Ha,2} = \frac{g}{2\mw}\left(\thl\cab +\thh\sab\right),
\end{align}
and hence 
\begin{align}
    \frac{\Delta_{i} \tgha}{\mha^2} = \frac{g}{2\mw}\left(\cab
    \frac{\thl}{\mhl^2}+ \sab \frac{\thh}{\mhh^2}\right), 
    \quad i=1,2
    \label{eq:gaugeinvtadpolecomb}.
\end{align}
While the change in $\delta\beta^\msbar$ at one-loop order is
independent of the gauge parameters in the usual \rxi-gauge and in
their generalizations to non-linear gauges, we show in
\refapp{sec:appA} that it is nevertheless already gauge dependent at
the one-loop level in the \THDM.  We expect that this applies as well
to the MSSM, where it is known that $\delta \beta^\msbar$ becomes
gauge dependent at two loops \cite{Yamada:2001ck}.

For the difference in the counterterms to the mixing angle $\alpha$, we obtain
\begin{align}
  \Delta_1 \delta \alpha^\msbar &= 
  -\frac{\Delta_1 \dzhlhh^\msbar}{2} =
                             \frac{\Delta_1
                             \thhhl^\msbar}{\mhh^2-\mhl^2} =
                             -\frac{
                             \thhhl^\msbar}{\mhh^2-\mhl^2}
    \label{eq:delta1alpha}
  \intertext{and}
  \Delta_2 \delta \alpha^\msbar &=  
  -\frac{\Delta_2 \dzhlhh^\msbar}{2} =
                             \frac{\Delta_2
                             \thhhl^\msbar}{\mhh^2-\mhl^2} =-
                             \frac{\thhhl^\msbar-t_{\Hh\Hl,2}^\msbar}
                             {\mhh^2-\mhl^2},
    \label{eq:delta2alpha}
\end{align}
with $\thhhl$ defined in 
\refapp{sec:appB} and   
\begin{align}
 t_{\Hh\Hl,2} = - \frac{g}{2\mw\tb}\left(\thl\cab+\thh\sab\right).
\label{tHhHl2}
\end{align}
Here we used Eq.~\eqref{eq:alpharel} and the antisymmetry
of $\Delta_1 \dzhlhh^\msbar$, 
which can be derived similarly as the one of
$\Delta_{1,2}\dzhag^\msbar$ above.
The result for 
$\Delta_1 \delta \alpha^\msbar$ 
can be
expressed by the tadpole counterterm in the \ts{} only, because 
Scheme~1 [see Eq.~\eqref{eq:dennerscheme}]
does not induce tadpole counterterms for
two-point functions that do not involve external would-be Goldstone bosons.

The differences $\Delta_1\delta\alpha^\msbar$ and
$\Delta_2\delta\alpha^\msbar$ are both gauge dependent at one-loop
order, which can be seen by inserting the explicit expressions for the
tadpoles from \refapp{sec:appA} in the \rxi-gauge. This result is used
in the next section to demonstrate the gauge dependence of \SMatrix{}
elements in Schemes~1 and 2.

\subsection
{The $\Hl \tau^+ \tau^-$ vertex}
\label{se:hltata}

We study the renormalized $\Hl \tau^+ \tau^-$ vertex for the different
tadpole renormalization schemes defined above.  In this section, we
assume that the renormalized tadpole terms vanish, $\hThlone = 0$ and
$\hThhone = 0$. Then, the shifts in the bare parameters that originate
from tadpole counterterms \thlone\ and \thhone\ can be expressed in
terms of the one-loop tadpole contributions \Thlone\ and \Thhone.

As the tadpole renormalization schemes do not modify the bare
Lagrangian, the bare loop amplitudes are not altered. However, the
finite parts of the counterterms are affected, and thus receive gauge
dependencies, as we demonstrate in the following. In particular, we
show that the change in the renormalized vertex function is gauge
dependent in the \rxi-gauge, \ie
\begin{align}
\partial_\xi \Delta_i \raisebox{-30pt}{\includegraphics{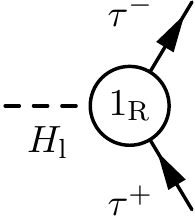}} \ne 0,
\label{eq:vertexhltata}
\end{align}
which has been verified by direct computation. 

The relevant Feynman rules read
\begin{align}
  \raisebox{-28pt}{\includegraphics{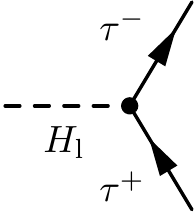}}\ = \frac{\ii e m_\tau}{2 \mw
  \sw} \left(\cab \tb + \sab\right)
  , \qquad
  \raisebox{-28pt}{\includegraphics{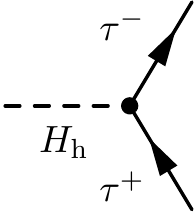}}\ = -\frac{\ii e m_\tau}{2 \mw
  \sw} \left(\cab - \sab \tb\right).
\end{align}
Computing the difference of the renormalized 
vertex function in different tadpole
schemes yields
\begin{align}
  \Delta_i \raisebox{-29.4pt}{\includegraphics{\pics/hltataR1}} ={}&
  \Delta_i \raisebox{-27.5pt}{\includegraphics{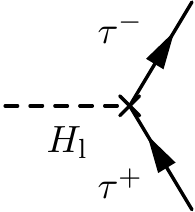}}\notag\\ 
  ={}&
  \frac{\ii e m_\tau}{\mw \sw}
  \Biggl[\left(\cab \tb + \sab\right) \left( \Delta_i \delta \beta^\msbar +
    \frac{\Delta_i \delta m_\tau}{m_\tau}
    - \frac{\Delta_i \delta \mw}{\mw}
  \right) \notag\\
  &-\left(\cab - \sab\tb \right) \left( \Delta_i\frac{\dzhhhl}{2} -
\Delta_i\delta \alpha^\msbar \right)
  \Biggr].
\end{align}
The terms can be split into two parts which are separately UV finite,
thus allowing for a simple interpretation
\begin{align}
  \Delta_i \delta \beta^\msbar + \frac{\Delta_i \delta m_\tau}{m_\tau} -
  \frac{\Delta_i \delta \mw}{\mw}
  &{}= \Delta_i\frac{\dzhag^\mathrm{fin}}{2}, \notag\\
  \Delta_i\frac{\dzhhhl}{2} - \Delta_i\delta \alpha^\msbar &{}= 
      -\Delta_i \frac{\dzhlhh^\mathrm{fin}}{2},
\end{align}
where we used Eqs.~\eqref{eq:masstadpoledep}, \eqref{eq:gaugeinvtadpolecomb2},
\eqref{eq:delta1alpha}, and
\eqref{eq:delta2alpha} 
and ``fin'' denotes the UV-finite part, \ie the remnant after \msbar{}
subtraction.
The final result reads
\begin{align}
  \Delta_i \raisebox{-29.5pt}{\includegraphics{\pics/hltataR1}} &=
  \raisebox{-27.5pt}{\includegraphics{\pics/hltata}}
  \times
  \Delta_i\frac{\dzhag^\mathrm{fin}}{2} \;-\;
  \raisebox{-27.5pt}{\includegraphics{\pics/hhtata}}
  \times
  \Delta_i\frac{\dzhlhh^\mathrm{fin}}{2},
\label{eq:gaugedephltata}
\end{align}
with 
\begin{align}
  \Delta_{1} \dzhag^\mathrm{fin} &= -2\frac{\tgha^\mathrm{fin}-t_{\GZ\Ha,1}^\mathrm{fin}}{\mha^2}, \qquad
  \Delta_{2} \dzhag^\mathrm{fin} = -2\frac{\tgha^\mathrm{fin}-t_{\GZ\Ha,2}^\mathrm{fin}}{\mha^2},
  \notag\\
  \Delta_1 \dzhlhh^\mathrm{fin} &= 2\frac{\thhhl^\mathrm{fin}}{\mhh^2-\mhl^2}, \qquad
\Delta_2 \dzhlhh^\mathrm{fin} =
                             2\frac{\thhhl^\mathrm{fin}-t_{\Hh\Hl,2}^\mathrm{fin}}
                             {\mhh^2-\mhl^2},
\end{align}
where $\tgha$ and $\thhhl$ are defined in \refapp{sec:appB}, and
$t_{\GZ\Ha,1,2}$ and
$t_{\Hh\Hl,2}$ in Eqs.~\eqref{tZHa12} and \eqref{tHhHl2}, respectively.  
The first contribution in
Eq.~\eqref{eq:gaugedephltata} appears owing to the differences in the
definition of $\beta$, the second one is a consequence of the
definition of $\alpha$.  Both are gauge dependent at one-loop order as
discussed in \refse{se:differences}.

The \ts{} yields gauge-independent predictions for the decay rate
$\Hl\to\tau^+\tau^-$, whereas in Schemes~1 and 2 the prediction is
gauge dependent. This has been confirmed via explicit calculation of
the \SMatrix{} element in the \rxi-gauge.

At one-loop order, the results for the \ts{} can be obtained from
Schemes~1 and 2 via the mapping
\begin{align}
  \left(\delta \beta^\msbar\right)_i \to \left(\delta \beta^\msbar\right)_i -
  \Delta_i\frac{\dzhag^\mathrm{fin}}{2}, \quad 
  \left(\delta \alpha^\msbar\right)_i \to \left(\delta \alpha^\msbar\right)_i -
  \Delta_i\frac{\dzhlhh^\mathrm{fin}}{2}, \quad i=1,2.
  \label{eq:schemecorr}
\end{align}

\change{It is interesting to mention that the ``Tadpole scheme''
  introduced in \citere{Freitas:2002um} for the renormalization of
  \tb\ in the \MSSM\ is equivalent to the \ts\ applied to the \MSSM\ 
  combined with $\msbar$~subtraction for \tb. Indeed for the \MSSM\ the
  finite shift $\delta\tb^{\mathrm{fin}}$ defined in Eq.~(43) of
  \citere{Freitas:2002um} corresponds exactly to the shift of
  $\bigl(\delta \beta^\msbar\bigr)_2$ in Eq.~\refeq{eq:schemecorr},
  which translates the popular Scheme~2 to the \ts. While in the \ts\
  the $\msbar$~subtracted \tb\ is directly gauge independent, an
  additional finite renormalization is required in Scheme~2 to restore
  the gauge independence after  $\msbar$~subtraction.}

\subsection
{The $Z Z \Hh$ vertex}
In this section, we present the finite correction of the $Z Z \Hh$
vertex due to the tadpole scheme.  We obtain formally analogous
results as in the previous section.  The following Feynman rules were
used
\begin{align}
  \raisebox{-23pt}{\includegraphics{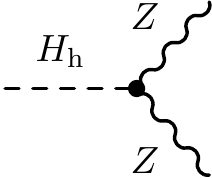}} = \frac{\ii e \cab}{\sw \cw}
  \frac{\mw}{\cw} g^{\mu\nu}, \qquad
  \raisebox{-23pt}{\includegraphics{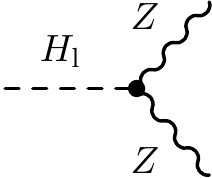}} = -\frac{\ii e \sab}{\sw \cw}
  \frac{\mw}{\cw} g^{\mu\nu}.
\end{align}
The calculation proceeds as in \refse{se:hltata} except that one has
to take into account a tadpole contribution which is given by
\begin{align}
  \raisebox{-23pt}{\includegraphics{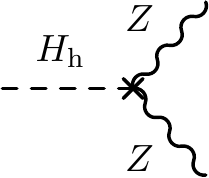}} &\supset t_{ZZ\Hh}=
  -\frac{\ii e^2}{2 \sw^2 \cw^2}
  \frac{\thh}{\mhh^2} g^{\mu\nu}.
\label{eq:frZZH}
\end{align}
With the same line of arguments as in the previous section we obtain
the difference of the renormalized vertex in different tadpole schemes
as
\begin{align}
  \Delta_i \raisebox{-26pt}{\includegraphics{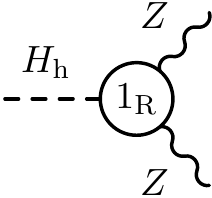}} &=
  \raisebox{-23pt}{\includegraphics{\pics/zzh1}}
  \times
  \Delta_i\frac{-\dzhag^\mathrm{fin} + \dzhlhh^\mathrm{fin} }{2}.
\label{eq:gaugedepZZHh}
\end{align}
The gauge independence of the renormalized $ZZ\Hh$ vertex has been
verified in the \ts{} by explicit computation in the \rxi-gauge.  In
this way it has also been confirmed that this vertex is gauge
dependent in Scheme~1.  Schemes~1 and 2 can be mapped to the \ts{} by
a redefinition of $\alpha$ and $\beta$ via Eq.~\eqref{eq:schemecorr}.
It is expected that the same is true for other vertices which are
sensitive to the renormalization of $\alpha$, $\beta$, and \SM{}
parameters, but not to $\Msb$.  Since the mapping
\eqref{eq:schemecorr} is gauge dependent, the renormalized $ZZ\Hh$
vertex becomes gauge dependent in Schemes~1 and 2.

\section{Electroweak NLO corrections to Higgs-boson production processes in the
\THDM{} \label{sec:processes}}
In this section, we analyze the EW NLO corrections for two Higgs-boson
production channels in the \THDM{}. First, in Section
\ref{sec:gluon-fusion}, we discuss results for the production of a
light \SM{}-like Higgs boson produced through gluon fusion for
scenarios in the alignment limit.  A more detailed description of the
implementation of this process and results for the production of a
light or a heavy neutral Higgs boson for the case of non-alignment
will be presented elsewhere. In Section \ref{sec:vb-fusion}, we
provide results for the production of a light \SM{}-like Higgs boson
in vector-boson fusion at NLO. Also here, a more detailed study
including the description of the implementation of the process will be
published separately.

In both processes all external particles are \SM{} particles such that
the new Higgs bosons only appear as virtual particles in the loops. In
both cases, we apply the renormalization scheme defined in
\refse{sec:renoconditions}%
\footnote{Since both considered processes do not depend on the
  soft-breaking scale $\Msb$ at LO, this parameter does not require
  renormalization.}  and discuss the size of the EW corrections. We
study the dependence on the renormalization scale that appears owing
to the \msbar{} renormalization of the mixing angles $\alpha$ and
$\beta$ and analyze the decoupling of the new (heavy) Higgs particles.
Besides investigating scenarios close to the decoupling limit, we
provide results for selected benchmark points from
\citeres{HXSWG2016,Baglio:2014nea}.  The benchmark points BP21A--D,
BP22A, and BP43--5 were originally designed for the study of exotic
Higgs-boson decays, the points BP3A1 and BP3B1--2 for a successful EW
baryogenesis and the points a-1 and b-1 for Higgs-boson pair
production.  All benchmark points fulfil theoretical constraints from
vacuum stability and perturbativity as well as experimental
constraints in flavour physics, EW precision measurements, and direct
searches.
In \refta{tab:BP} we list the benchmark points in the alignment limit, which we
study in gluon fusion and Higgs strahlung. In \refta{tab:BPNA} we provide
benchmark scenarios that are not in the alignment limit and which we study in
Higgs strahlung only.
\begin{table}
        \centering
        \begin{tabular}{|c|c|c|c|c|c||c|}
        \hline     & $\mhh$    & $\mha$    & $\mhc$    & $m_{12}$  & $\tb$ & $\Msb$\\
        \hline BP21A  & $200\GeV$ & $500\GeV$ & $200\GeV$ & $135\GeV$ & $1.5$ & $198.7\GeV$ \\
        \hline BP21B & $200\GeV$ & $500\GeV$ & $500\GeV$ & $135\GeV$   & $1.5$   & $198.7\GeV$  \\
        \hline BP21C & $400\GeV$ & $225\GeV$ & $225\GeV$ & $0\GeV$   & $1.5$   & $0\GeV$  \\
        \hline BP21D & $400\GeV$ & $100\GeV$  & $400\GeV$ & $0\GeV$ & $1.5$   & $0\GeV$ \\
        \hline BP3A1 & $180\GeV$ & $420\GeV$  & $420\GeV$ & $70.71\GeV$ & $3$  & $129.1\GeV$\\
        \hline
        \end{tabular}
        \caption{\THDM{} benchmark points in the alignment
          limit, \ie $\sab\to-1$, $\cab\to0$, taken from
          \citere{HXSWG2016}. The parameter $\Msb$
          depends on the other parameters and is given for
          convenience. \label{tab:BP}}
\end{table}
\begin{table}
\begin{tabular}{|c|c|c|c|c|c|c||c|}
\hline     & $\mhh$    & $\mha$    & $\mhc$    & $m_{12}$    & $\tb$  & $\cab$ & $\Msb$\\
\hline a-1 & $700\GeV$ & $700\GeV$ & $670\GeV$ & $424.3\GeV$ & $1.5$  & $-0.0910$ & $624.5\GeV$\\
\hline b-1 & $200\GeV$ & $383\GeV$ & $383\GeV$ & $100\GeV$   & $2.52$ & $-0.0346$ & $204.2\GeV$\\
\hline BP22A & $500\GeV$ & $500\GeV$ & $500\GeV$ & $187.08\GeV$   & $7$ & $0.28 $ & $500\GeV$\\
\hline BP3B1 & $200\GeV$ & $420\GeV$ & $420\GeV$ & $77.78\GeV$   & $3$ & $0.3 $
  & $142.0\GeV$\\
\hline BP3B2 & $200\GeV$ & $420\GeV$ & $420\GeV$ & $77.78\GeV$   & $3$ & $0.5 $
  & $142.0\GeV$\\
\hline BP43 & $263.7\GeV$ & $6.3\GeV$ & $308.3\GeV$ & $52.32\GeV$   & $1.9$ & $0.14107 $ & $81.5\GeV$\\
  \hline BP44 & $227.1\GeV$ & $24.7\GeV$  & $226.8\GeV$ & $58.37\GeV$ & $1.8$ &
  $0.14107$ & $89.6\GeV$\\
\hline BP45 & $210.2\GeV$ & $63.06\GeV$ & $333.5\GeV$ & $69.2\GeV$   & $2.4$ &
  $0.71414 $ & $116.2\GeV$\\
\hline
\end{tabular}
\caption{\THDM{} benchmark points outside the alignment limit taken from \citere{Baglio:2014nea} (a-1, b-1)
  and \citere{HXSWG2016}. The parameter $\Msb$ depends on the other parameters and is given for
convenience. \label{tab:BPNA}}
\end{table}

For the numerical evaluation of the two Higgs-boson production
processes we use the following values for the \SM{} input
parameters~\cite{Agashe:2014kda}:
\begin{align}
 \begin{split}
  \Gf=1.16638\cdot10^{-5}\GeV^{-2},\quad
  \mt=173.21\GeV, \quad
  \mh=125.09\GeV=\mhl, \\
  \mw=80.385\GeV, \quad \Gamma_{\PW}=2.0850\GeV, \quad
  \mz=91.1876\GeV, \quad \Gamma_{\PZ}=2.4952\GeV.
 \end{split}
\end{align}
The numerical results presented in the following have been obtained in
the 't~Hooft--Feynman gauge.


\subsection{Higgs-boson production in gluon fusion\label{sec:gluon-fusion}}
Higgs-boson production through gluon fusion is a loop-induced process,
\ie its LO contribution appears at the one-loop level.  Despite its
loop suppression, it is the dominant Higgs-boson production mechanism
in the \SM{} at the LHC.  Since the Yukawa couplings of the Higgs
boson to fermions are proportional to the fermion mass, the dominant
contribution arises from top-quark loops.  Treating all other fermions
as massless, the LO partonic cross section $\hat{\sigma}$ for \SM{}
Higgs-boson production is generated only via a top-quark loop.

In the \SM{}, the QCD corrections to Higgs-boson production in gluon
fusion are known up to N$^3$LO and are
large~\cite{Anastasiou:2014vaa,Anastasiou:2014lda,Duhr:2014nda,Anastasiou:2015ema,Anastasiou:2016cez}.
The complete NLO EW corrections have been calculated in
\citeres{Actis:2008ug,Actis:2008ts} and are also sizable.  EW
radiative corrections may significantly change a process, if BSM
particles propagate in the loop. In
\citeres{Passarino:2011kv,Denner:2011vt}, for example, the influence
of a fourth generation of heavy fermions on the EW corrections to
Higgs-boson production in gluon fusion has been discussed, and the EW
corrections turned out to be large.  In the following, we present the
behaviour of the NLO EW corrections to this Higgs-boson production
channel in the alignment limit of the \THDM{} of type II, where the
light, neutral Higgs boson $\Hl$ becomes \SM{}-like.
All results are calculated in the \ts{} with the renormalization
conditions given in \refse{sec:renoconditions} with the exception of the
top-quark mass which has been renormalized in the on-shell scheme for 
gluon fusion.

The coupling of the light neutral Higgs boson to top quarks in the
\THDM{} of type II is given by the $\Hl t\bar{t}$ vertex:
\begin{align}
  \raisebox{-28pt}{\includegraphics{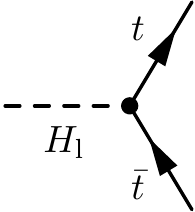}}\ =
  -\frac{\ii e \mt}{2 \mw \sw} \left( \frac{\cab}{\tb} - \sab \right).
\end{align}
In the alignment limit ($\cab=0, \sab=-1$) the coupling of the light
neutral Higgs boson to up-type fermions equals the one in the \SM{}.
Therefore, the LO production cross section of the light neutral Higgs
boson through gluon fusion in the \THDM{} is the same as in the \SM{},
and the QCD corrections do not change. For small $\tb$ the alignment
limit is reached slower and one speaks of a delayed
decoupling~\cite{Gunion:2002}. Without alignment, the LO cross section
changes only by the factor $(\cab/\tb-\sab)^2$, such that the relative
QCD corrections stay the same.  In the alignment limit, the $\tb$
dependence of the process disappears at LO, but survives in the NLO EW
corrections. The derivation of the counterterms for the NLO
calculation requires special care. The fact that the alignment limit
implies $\cab=0$ (and consequently $\sab=-1$), but does not affect
$\tb$, leads to an explicit dependence of the $\Hl t\bar{t}$
counterterm on $\tb$, $\delta \alpha$, and $\delta \beta$:
\begin{align}
  \left.\raisebox{-28pt}{\includegraphics{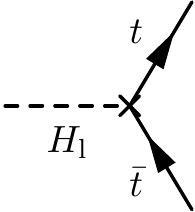}}\;\right|_{\cab=0, \sab=-1} \supset
  -\frac{\ii e  m_t}{2 \mw \sw} \frac{\delta \alpha - \delta \beta}{\tb}.
  \label{eq:hlttct}
\end{align}
As a result, the NLO corrections to this process are still scale
dependent despite of the alignment limit. The scale dependence
originates from the renormalization of $\alpha$ and $\beta$ in the
$\msbar$ scheme.

The calculation of the NLO EW corrections proceeds in several steps.
First, the \THDM{} Feynman rules are derived using
{\tt{FeynRules}}~\cite{Alloul:2013bka}.  These are used in the code
{\tt{QGS}} in order to construct the amplitudes based on Feynman
diagrams generated with {\tt{QGRAF}}~\cite{Nogueira:1991ex}.  The
program {\tt{QGS}} is an extension of {\tt{GraphShot}}~\cite{Actis},
which has been used to accomplish the corresponding \SM{} calculation
and performs the algebraic manipulations of the amplitudes with
{\tt{Form}}~\cite{Vermaseren:2000nd}. The reducible scalar products
are removed, the symmetries are taken into account in order to reduce
the number of loop integrals, the UV renormalization as well as the
cancellation of collinear logarithms is performed analytically, and
finally the remaining finite integrals are mapped onto form factors.
The latter are evaluated numerically with Fortran routines.  In the
alignment limit of the \THDM{} the same types of Feynman integrals
arise as in the \SM{} calculation of
\citeres{Actis:2008ts,Actis:2008ug} such that we can employ the same
Fortran library for their numerical evaluation. For the numerical
evaluation of the two-loop massive diagrams the library uses the
methods of \citeres{Passarino:2001wv,Passarino:2001jd} for
self-energies and of
\citeres{Ferroglia:2003yj,Passarino:2006gv,Actis:2004bp,Actis:2008ts}
for vertex functions.

The NLO EW corrections to the partonic cross section are expressed as
percentage correction
$\delta^{\mbox{\scriptsize{NLO}}}_{\mbox{\scriptsize{EW}}}$ relative
to the LO result,
\begin{align}
  \hat{\sigma} = \hat{\sigma}^{\rm LO}+\hat{\sigma}^{\rm NLO}=\hat{\sigma}^{\rm LO}(1+\delta^{\mbox{\scriptsize{NLO}}}_{\mbox{\scriptsize{EW}}}).
\end{align}
In order to study the scale dependence we consider the following
scenario: We choose $\tan\beta=2$ and $\mscale=700\GeV$ as a typical
mass scale for the new degrees of freedom.  The soft-breaking scale
$M_{\rm sb}$ and the masses of all heavy Higgs bosons, except for the
heavy, neutral one, are set equal to $\mscale$.  By allowing for a
different mass value for the heavy, neutral Higgs boson, we find
enhanced scale-dependent logarithms in the alignment limit.  This can
be seen from the analytical expression for the scale-dependent part of
the relative correction to the LO matrix-element squared
\begin{align}
\delta^{\mbox{\scriptsize{NLO}},\mu\mathrm{{-}dep.}}_{\mbox{\scriptsize EW}} &=
\frac{\Gf\sqrt{2}}{8\pi^2\tb^2\mhh^2(\mhh^2-\mhl^2)}\ln\frac{\mu^2}{\mhl^2}\notag\\
&\quad\times\biggl[(1-\tb^2)(\mhh^2-\Msb^2) \Bigl[3\mhh^2\mhl^2 +
\Msb^2(\mha^2+2\mhc^2-3\mhh^2)\Bigr] \notag\\
&\quad{}+6\mt^2(\mhh^2\mhl^2-4\Msb^2\mt^2) \biggr].
\label{eq:scaledep_ggh}
\end{align}
This expression is proportional to
$(\delta\alpha-\delta\beta)/t_\beta$ in the \msbar{} scheme as
expected from Eq.~\eqref{eq:hlttct}.  If the mass of the heavy,
neutral Higgs boson differs from the soft-breaking scale and the
masses of the other heavy Higgs bosons, the terms in the second line
dominate the scale dependence.  If also the heavy, neutral Higgs-boson
mass is chosen to be equal to the typical mass scale~$\mscale$ only
the top-mass-dependent terms in the last line contribute.  In order to
see the effect of the enhanced logarithms, we thus vary the mass of
the heavy, neutral Higgs boson.  The variation of the renormalization
scale $\mu$ can be used in order to estimate the uncertainty due to
unknown higher-order corrections.  To this end, we evaluate the NLO
corrections for different values of the renormalization scale
$\mu=\mscale,\mscale/2,\mscale/4$.

\begin{figure}
  \begin{center}
  \includegraphics[width=0.7\textwidth]{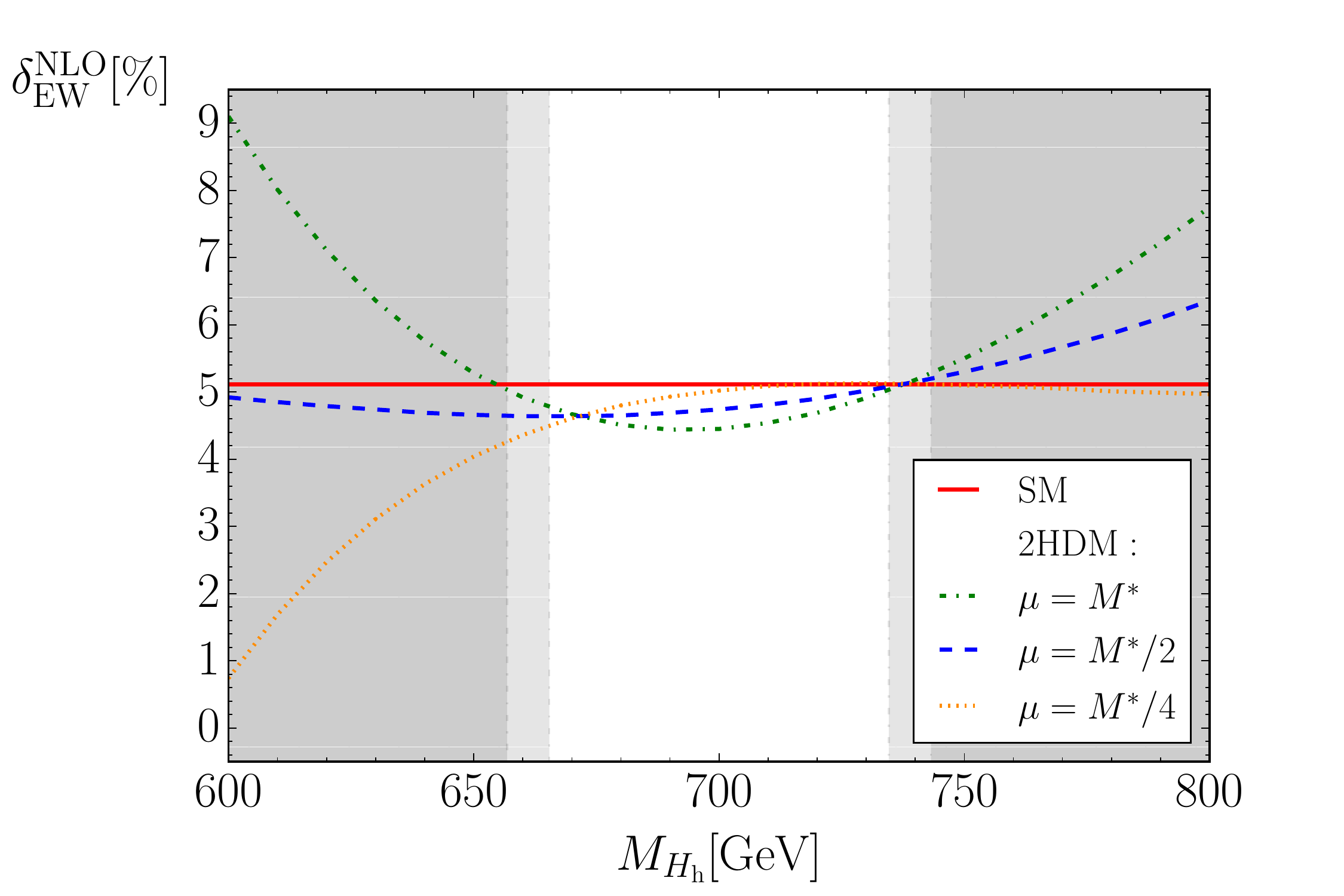}
  \caption{EW NLO corrections to Higgs-boson production in gluon fusion. The
    {\red{solid}} line indicates the \SM{} result. The
    {\green{dashed-dotted}}, {\blue{dashed}} and {\orange{dotted}} lines are
    the percentage corrections in the \THDM{} as a function of the
    heavy, neutral Higgs-boson mass $\mhh$ for different values of the
    renormalization scale $\mu=\mscale,\,\mscale/2,\,\mscale/4$. The heavy
    Higgs-boson masses $\mha=\mhc=M_{\rm sb}=\mscale=700\GeV$ are kept
    constant and $\tb=2$.\label{fig:Hgg-scale-dep}}
  \end{center}
\end{figure}

In \reffi{fig:Hgg-scale-dep} we show the percentage correction
$\delta^{\mbox{\scriptsize{NLO}}}_{\mbox{\scriptsize{EW}}}$ as a
function of the heavy, neutral Higgs-boson mass $\mhh$ for the three
different values of the renormalization scale $\mu$ and compare the EW
corrections of the \THDM{} to those in the \SM{}.
For a small mass splitting, \ie for a heavy, neutral Higgs-boson mass
$\mhh$ in the vicinity of $\mscale=700\GeV$, the size of the
corrections is comparable to the one in the \SM{}, and the scale
dependence is small, \ie perturbation theory is well-behaved and
higher-order EW corrections can be expected to be small. For large
mass splittings, \eg for a heavy, neutral Higgs-boson mass $\mhh$ that
deviates significantly from the mass values of the other heavy Higgs
bosons and thus from $\mscale=700\GeV$, the scale dependence becomes
large. This indicates large uncertainties owing to unknown
higher-order corrections, which signals the breakdown of the
perturbative expansion and the onset of a non-perturbative regime.
This behaviour is expected, since the mass splitting of the heavy
Higgs bosons is restricted by perturbativity. The parameters
$\lambda_i$ in Eq.~\eqref{eq:hpgb} have to fulfil the condition
$|\lambda_i|\lesssim\mathcal{O}\left(1\right)$ in order to ensure that
the Higgs-boson sector does not become strongly
coupled~\cite{Gunion:2002}. This is important to maintain tree-level
unitarity~\cite{Arhrib:2000is}. The requirement
$|\lambda_i|\lesssim\mathcal{O}\left(1\right)$ leads to a bound on the
mass splitting~\cite{Gunion:2002} of
\begin{align}
|\mhh-\mscale|,\;\;|\mha-\mscale|,\;\;|\mhc - \mscale|\;\;\lesssim\mathcal{O}\left(\frac{v^2}{\mscale}\right),
  \label{eq:masssplit}
\end{align}
where $v\approx246\GeV$ is the \SM{} vev. Using this information as
well as the knowledge of the NLO EW corrections, we study the
region of $\mhh$ in \reffi{fig:Hgg-scale-dep} for which perturbativity
or non-perturbativity can be expected.  For this purpose, we constrain
the allowed region of $\mhh$ around the typical mass scale $\mscale$
via
\begin{align}
\mscale-f\frac{v^2}{\mscale}<\mhh<\mscale+f\frac{v^2}{\mscale}.
\label{eq:heavyhiggsmasssplit}
\end{align}
Perturbativity should be realized if the parameter $f$ is sufficiently
smaller than one.  In \reffi{fig:Hgg-scale-dep} the $\mhh$ region
corresponding to $f>0.4$ is marked in light grey, and the region
$f>0.5$ in dark grey.  The scale dependence can be used to estimate
uncertainties from unknown higher-order corrections and provides
useful means to determine the onset of the non-perturbative regime.
This is supported by \reffi{fig:Hgg-scale-dep}, where the scale
dependence becomes stronger when $f$ becomes large. We conclude that
the scale dependence introduced by the \msbar{} renormalization of the
mixing angles $\alpha$ and $\beta$ is less than a percent, as long as
perturbativity is not violated, \ie $f<0.5$.  Thus, the \msbar{}
renormalization of $\alpha$ and $\beta$ provides stable results for
sound scenarios in the perturbative regime.

\begin{figure}
  \begin{center}
  \includegraphics[width=0.7\textwidth]{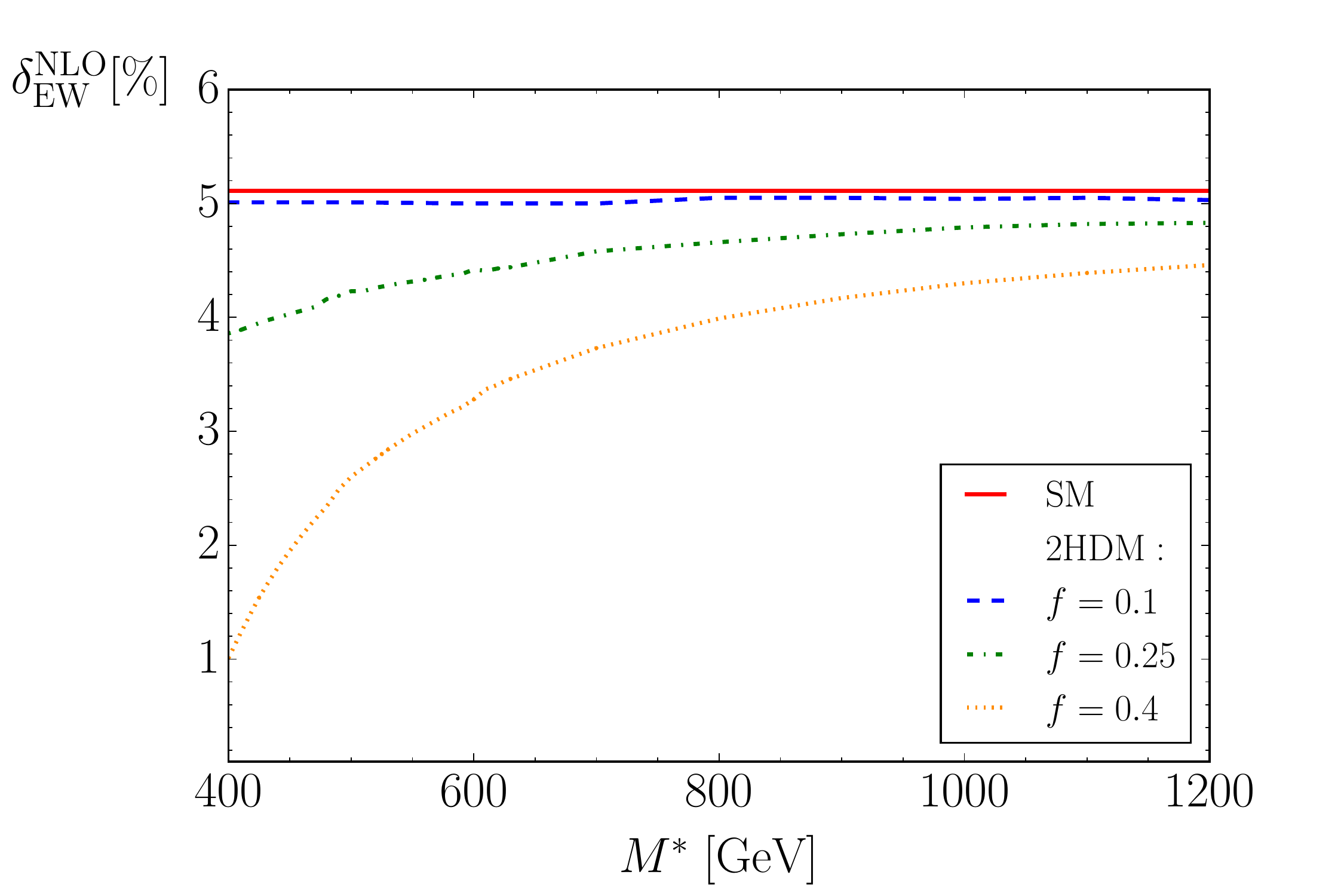}
  \caption{EW NLO corrections  to Higgs-boson production in gluon
    fusion as a function of the scale $\mscale$.  The values of
    $\mhh$, $\mha$, $\mhc$ and $\Msb$ are chosen according to
    Eq.~\eqref{eq:show_decoup}.  The solid line represents the \SM.
    The {\blue{dashed}}, {\green{dashed-dotted}}, and {\orange{dotted}}
    lines correspond to $f=0.1$, $f=0.25$ and
    $f=0.4$.\label{fig:Hgg-decoup}}
  \end{center}
\end{figure}
In \reffi{fig:Hgg-decoup}, we analyze the decoupling of the heavy
Higgs-boson sector for different values of~$f$. We vary the mass scale
$\mscale$ and choose the masses of the heavy Higgs bosons as
\begin{align}
        \mhh = \mscale-f\frac{v^2}{\mscale},\qquad
        \mha = M_{\rm sb}=\mscale, \qquad
        \mhc = \mscale+f\frac{v^2}{\mscale}.
        \label{eq:show_decoup}
\end{align}
For the rather large value $f=0.4$, the NLO corrections in the \THDM{}
approach the \SM{} value of $5.1\%$ only slowly; at a typical mass
scale $\mscale=1200\GeV$ the decoupling limit is almost reached.  For
smaller values of $f$ decoupling is approached considerably faster.
For $f=0.1$ the decoupling of the heavy Higgs-boson sector already
occurs at about $400\GeV$.

Finally, in \refta{tab:BP_ggH} we present the relative NLO corrections
for the benchmark points of \refta{tab:BP} which fulfil
perturbativity in the sense that $|\lambda_i|<4\pi$~\cite{HXSWG2016}.
\begin{table}
        \begin{center}
\begin{tabular}{|c|c|c|c|}
\hline $\mu$        & $\mhl$ & $2\mhl$ & $4\mhl$ \\
\hline BP21A        & $ 8.5\%$ & $-1.3\%$  & $-11.2\%$ \\
\hline BP21B        & $ 7.3\%$ & $-2.7\%$  & $-12.7$\% \\
\hline BP21C        & $13.2\%$ & $12.6\%$  & $ 12.0$\% \\
\hline BP21D        & $15.1\%$ & $14.6\%$  & $ 14.0$\% \\
\hline BP3A1        & $21.3\%$ & $13.2\%$  & $  5.1$\% \\
\hline
\end{tabular}
        \caption{Relative NLO corrections
          $\delta^{\mbox{\scriptsize{NLO}}}_{\mbox{\scriptsize{EW}}}$
           to Higgs-boson production in gluon fusion for the benchmark points of \refta{tab:BP}.
The scale $\mu$ is varied as a function of $\mhl$.
\label{tab:BP_ggH}}
        \end{center}
\end{table}
For all scenarios, we observe large NLO EW corrections on top of the
\SM{} value of $5.1\%$.  Owing to the large corrections, it should be
possible to exclude these scenarios at the LHC, as soon as
computations for the relevant decay channels
like $\Hl\to\gamma\gamma$ 
are available at the same order in the weak coupling.

The scale uncertainty turns out to be at the level of $\pm10\%$ for
the benchmark points BP21A, BP21B and BP3A1, but small for BP21C and
BP21D.  The large scale dependencies are due to rather large values of
the $\lambda_i$ in these benchmark scenarios, the largest values for
$|\lambda_i|$ ranging between 3.7 and 7.7.  The results from
\refta{tab:BP_ggH} can be understood from the analytic expression for
the scale dependence in Eq.~\eqref{eq:scaledep_ggh}.  For BP21A and
BP21B, $\mhh\approx\Msb$ and, thus, basically only the $\mt$-dependent
terms in the last line of Eq.~\eqref{eq:scaledep_ggh} contribute.  The
scale dependence is enhanced by the factor
$(\mhl^2-4\mt^2)/(\mhh^2-\mhl^2)$. For BP21C and BP21D all terms
involving $\Msb$ vanish, and the scale dependence is suppressed by the
ratio $\mhl^2/\mhh^2\sim 0.1$. For BP3A1, the $\mt$-independent terms
dominate the scale dependences, the leading term being proportional to
$\mhc^2/\mhh^2\sim 5.4$.

\subsection{Higgs production in association with a weak boson
\label{sec:vb-fusion}}

Besides the gluon-fusion channel and the vector-boson fusion
channel, the associated Higgs production with a vector boson, also
called Higgs strahlung, is used to study the properties of the Higgs
boson.  In this section, we focus on this process which allows in
particular to measure the decay mode $\PH \to \Pb \bar\Pb$ and to
study BSM physics in the $V V H$ vertex.

There has been enormous progress in higher-order calculations to Higgs
strahlung in the \SM. The QCD corrections are known up to NNLO for the
inclusive cross section
\cite{Hamberg:1990np,Brein:2003wg,Brein:2011vx} as well as for
differential cross sections \cite{Ferrera:2011bk,Ferrera:2014lca}. The
NLO EW corrections were first computed in \citere{Ciccolini:2003jy}
for stable vector bosons.  Meanwhile public codes are available
including the vector-boson decays, \eg V2HV \cite{Spira2015}, MCFM
\cite{Campbell2015}, HAWK2.0 \cite{Denner:2014cla} and vh@nnlo
\cite{Brein:2012ne}, allowing to study any final state in this process
class at NLO QCD and EW and also partially at NNLO QCD.  Higgs
strahlung has also been investigated in the \THDM\ 
\cite{Harlander:2013mla}, where the ratio of inclusive $\PW\PH$ and
$\PZ\PH$ production for light and heavy Higgs bosons has been studied
and the impact of type-I and type-II Yukawa couplings to the \SM{}
Higgs production has been analyzed including all available and
numerically relevant contributions.

The following analysis is restricted to the case of two charged
leptons in the final state, $\Pp\Pp \to \PH l^+ l^- + X$. For massless
leptons one has to be careful with final-state collinear radiation,
which requires special treatment (see \eg \citere{Denner:2011id}). We do
not recombine collinear photons and leptons and assume that the
leptons can be perfectly isolated, which is justified for a pair of
muons in the final state. We employ the cuts used in the analysis
of \citere{Chatrchyan:2013zna}, \ie we require the muons to
\begin{itemize}
  \item have transverse momentum $p_{\mathrm{T}}^l> 20\GeV$ for $l=\mu^+, \mu^-$,
  \item be central with rapidity $\left|\eta_l\right| < 2.4$ for $l=\mu^+, \mu^-$,
  \item have 
a pair invariant mass 
$m_{ll}$ of  $75\GeV < m_{ll} < 105\GeV$.
\end{itemize}
In addition, we demand a boosted Z boson with
\begin{itemize}
  \item transverse momentum  $p_{\mathrm{T}}^\PZ > 160\GeV$.
\end{itemize}
All predictions are for the hadronic cross section at the 
center-of-mass energy of $13\TeV$ using the NLO
PDF set NNPDF2.3 with QED corrections \cite{Ball:2013hta}. 

The numerical results were produced using an extended version of
RECOLA \cite{Actis:2016mpe} and HAWK 2.0 \cite{Denner:2014cla}. RECOLA
has been used to calculate all needed one-loop \SMatrix{} elements in
the \THDM{}, and HAWK 2.0 served as integrator for Higgs strahlung.

As in the case of gluon fusion, we discuss the scale dependence in the
decoupling limit.  In the alignment limit the \THDM{} leaves its marks
in Higgs strahlung only at NLO, but, in contrast to gluon fusion,
Higgs strahlung is scale independent in the alignment limit because
the tree-level vertices $V V H$ do not depend on $\tb$ but only on
$\cab$ and $\sab$ (see discussion in \refse{sec:gluon-fusion}). For
this reason the analysis has been extended to the decoupling limit
with only approximate alignment compatible with perturbative
unitarity, \ie the requirement
$|\lambda_i|\lesssim\mathcal{O}\left(1\right)$ is extended
\cite{Gunion:2002} by
\begin{align}
  |\cab| \lesssim\mathcal{O}\left(\frac{v^2}{\mscale^2}\right).
  \label{eq:cabpert}
\end{align}
We perform an analysis for Higgs strahlung similar to the one
for gluon fusion in \reffi{fig:Hgg-scale-dep}. 
\begin{figure}
  \centering
  \subfigure[$k=0.4$]{\includegraphics[width=0.49\textwidth]{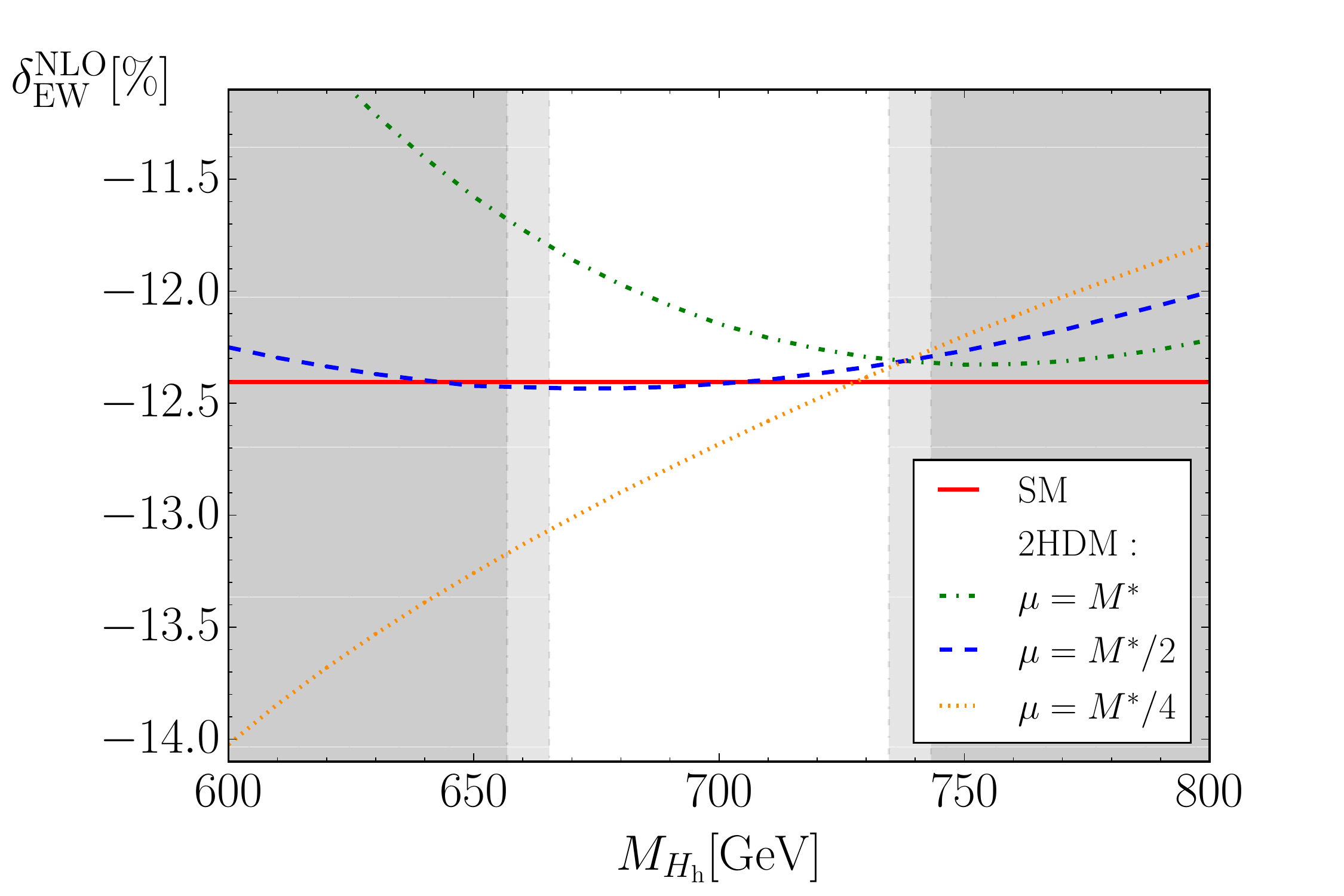}}
  \subfigure[$k=0.25$]{\includegraphics[width=0.49\textwidth]{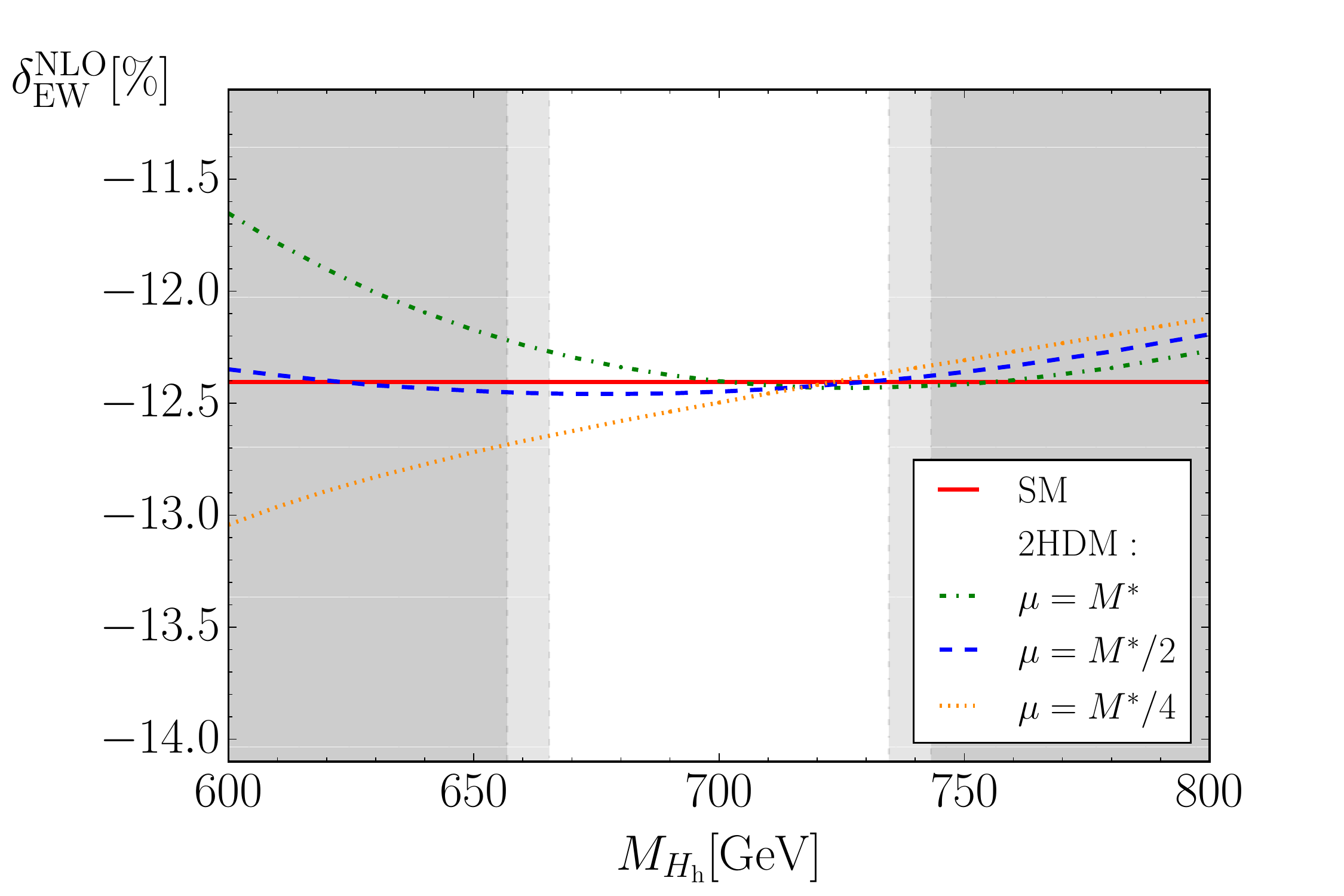}}
  \subfigure[$k=0.1$]{\includegraphics[width=0.70\textwidth]{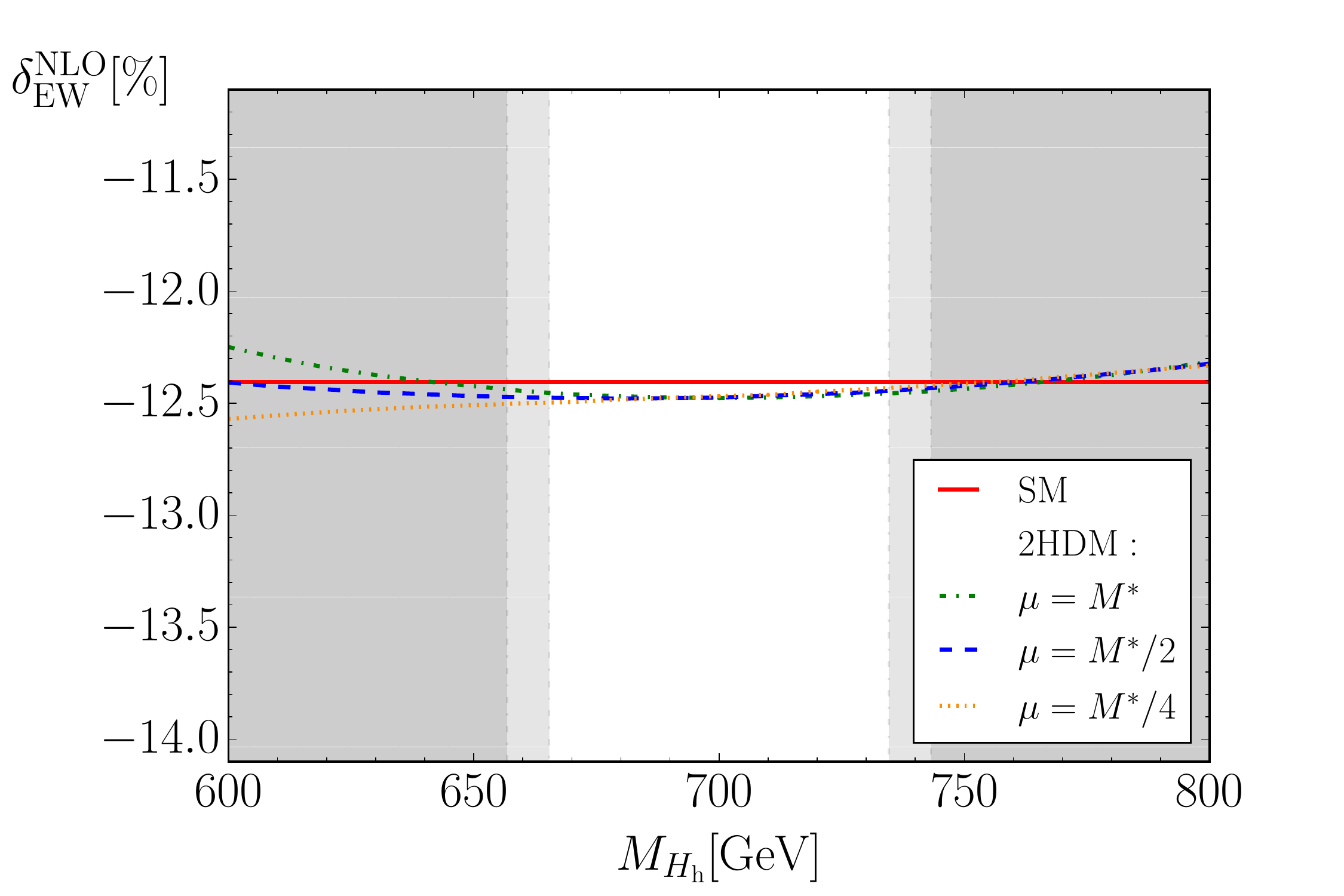}}
  \caption{
    NLO percentage EW corrections 
          to the integrated cross section of $\Pp\Pp \to \PH l^+ l^- + X$
as a function of the heavy
    Higgs-boson mass \mhh\ for three different scenarios, which differ
    by the value of $\cab$ parametrized according to
    Eq.~\eqref{eq:cabpertf}. The solid line indicates the SM result.
    The \green{dashed-dotted}, \blue{dashed}, and \orange{dotted}
    lines show the percentage correction in the \THDM\ (normalized to
    the \THDM\ Born) for different values of the renormalization scale
    $\mu = \mscale , \mscale/2 , \mscale/4$.  The heavy masses
    $\mha=\mhc=M_{\rm sb}=\mscale=700\GeV$ and $\tb=2$ are kept
    constant. The bright grey band represents the region $0.4 \le f
    \le 0.5$ [see Eq.~\eqref{eq:heavyhiggsmasssplit}] and the dark
    grey band the region $f>0.5$.}
  \label{fig:ppHll-scale}
\end{figure}
In \reffi{fig:ppHll-scale} we present the percentage EW correction
$\delta^{\mbox{\scriptsize{NLO}}}_{\mbox{\scriptsize{EW}}}$ as a
function of the heavy, neutral Higgs-boson mass \mhh\ for three
different scales $\mu$ centered around $\mscale/2$. All other
parameters are kept fixed, the masses are set to the decoupling scale
$\mscale=700\GeV$, and we choose $\tb=2$. The results are presented
for three different scenarios where we investigate the decoupling in
terms of $\cab$, parametrizing
\begin{align}
  \cab = k \frac{v^2}{\mscale^2}
  \label{eq:cabpertf}
\end{align}
with $k=0.1, 0.25, 0.4$.  The expected non-perturbative region is
shown in dark grey for $f> 0.5$ and the transition region in bright
grey defined by $0.4 \le f \le 0.5$. The first plot in
\reffi{fig:ppHll-scale} shows the case $k=0.4$, which is at the border
of perturbative unitarity independently of $f$ because $\cab$ is close
to its upper limit [see Eq.~\eqref{eq:cabpert}]. This scenario
exhibits moderate scale uncertainties of the order of one percent in
the perturbative regime. Decreasing $k$ to $0.25$ reduces the scale
dependence to below one percent in the perturbative region $f\le 0.4$,
and for $k=0.1$ almost no scale dependence is left. While the
considered scenario is not in the decoupling limit, the resulting
corrections are nevertheless comparable to those in the \SM. The
decoupling limit is reached by setting $\cab=0$, where the corrections
coincide with those in the \SM.

In \reftas{tab:BP_Hll_alignment} and \ref{tab:BP_Hll_noalignment} we
present the results for the benchmark points of \reftas{tab:BP} and
\ref{tab:BPNA}.
\begin{table}
        \begin{center}
        \begin{tabular}{|c|c|c|c|}
        \hline $\mu$            & $\mhh$ & $2\mhh$ & $4\mhh$ \\
        \hline BP21A            & $-11.8$ \%& $-11.8$ \%  & $-11.8$ \% \\
        \hline BP21B            & $-13.1$ \%& $-13.1$ \%  & $-13.1$ \% \\
        \hline BP21C            & $-13.2$ \%& $-13.2$ \%  & $-13.2$ \% \\
        \hline BP21D            & $-13.6$ \%& $-13.6$ \%  & $-13.6$ \% \\
        \hline BP3A1            & $-13.3$ \%& $-13.3$ \%  & $-13.3$ \% \\
        \hline
        \end{tabular}
        \caption{Relative NLO correction
          $\delta^{\mbox{\scriptsize{NLO}}}_{\mbox{\scriptsize{EW}}}$
          to the integrated cross section of $\Pp\Pp \to \PH l^+ l^- + X$
          for the benchmark points in the alignment limit of
          \refta{tab:BP}.
        \label{tab:BP_Hll_alignment} 
        The
        \SM\ correction is $-12.4\%$.}
        \end{center}
\end{table}
\begin{table}
        \begin{center}
        \begin{tabular}{|c|c|c|c|}
        \hline $\mu$            & $\mhh$ & $2\mhh$ & $4\mhh$ \\
        \hline a-1              & $-7.6 $ \%& $-10.5$ \%  & $-13.3$ \% \\
        \hline b-1              & $-12.5$ \%& $-12.5$ \%  & $-12.4$ \% \\
        \hline BP22A            & $-239 $ \%& $-54.8$ \%  & $130$ \% \\
        \hline BP3B1            & $-23.2$ \%& $-20.0$ \%  & $-16.9$ \% \\
        \hline BP3B2            & $-56.0$ \%& $-39.5$ \%  & $-23.0$ \% \\
        \hline BP43             & $-11.9 $ \%& $-10.6$  \%  &  $-9.3$\% \\
        \hline BP44             & $-11.1 $ \%& $-11.2$ \%  & $-11.3$ \% \\
        \hline BP45             & $-50.6$ \% &  $-14.3$    \%  & $21.9$\% \\
        \hline
        \end{tabular}
        \caption{Relative NLO correction
          $\delta^{\mbox{\scriptsize{NLO}}}_{\mbox{\scriptsize{EW}}}$
          to the integrated cross section of $\Pp\Pp \to \PH l^+ l^- + X$
          for the benchmark points outside the alignment limit of
          \refta{tab:BPNA}.\label{tab:BP_Hll_noalignment} The 
        \SM\ correction is $-12.4\%$}
        \end{center}
\end{table}
For scenarios in the alignment limit compiled in
\refta{tab:BP_Hll_alignment} there is no scale dependence, and the
differences between EW corrections in the \THDM{} and in the \SM{},
where they amount to $-12.4\%$, are typically at the level of one
percent.

The scenarios outside the alignment limit shown in
\refta{tab:BP_Hll_noalignment} are more interesting.  In the scenarios
a-1, b-1, BP43, and BP44, which are close to the alignment limit, the
scale variations are small of the order of 0.5\%. The corrections are
comparable to those in the \SM{}, differing typically at the level of
one percent.  The scenarios BP3B1, BP3B2, BP45 significantly deviate
from the alignment limit with a mass splitting of more than $200\GeV$
and exhibit scale uncertainties up to $35 \%$. For all these scenarios
we find absolute values of $\lambda$ of the order of
$\left|\lambda_i\right|/(4 \pi) \approx 0.3$.  The scenario BP22A is
in the decoupling limit, but does not fulfil condition
\eqref{eq:cabpert}. We observe large scale uncertainties of the order
of $180\%$ which raises the question of perturbativity of this
scenario. In fact, we find $\lambda_2/(4 \pi) \approx 1.1$ and
$\lambda_3/(4 \pi) \approx 0.7$ for BP22A. Thus, large scale
uncertainties signal a breakdown of the perturbative expansion.

In conclusion, as in gluon fusion, violation of perturbative unitarity
and large scale dependence are connected. We observe small scale
uncertainties of the EW corrections in the \THDM{} both in the
decoupling limit and for benchmark points that are close to the
alignment limit or involve small mass splittings, while respecting
perturbative unitarity.


\section{Conclusion\label{sec:conclusion}}

The precise study of theories with extended Higgs sectors is of utmost
importance for the investigation of the Higgs sector at the LHC.  To
this end, NLO corrections of QCD and electroweak origin have to be
calculated.

We have proposed a consistent gauge-independent renormalization scheme
for the CP-conserving \THDM\ of type II. While masses are renormalized
in the on-shell scheme, the mixing angles of the Higgs sector and the
soft-$Z_2$-symmetry-breaking
scale are renormalized in the \msbar{} scheme. To
render this approach gauge independent, a consistent treatment of
tadpoles is crucial. This is provided by the method proposed by
Fleischer and Jegerlehner many years ago for the Standard Model.

We have generalized this method specifically to the 2-Higgs-Doublet
Model of type II. We have investigated the difference to popular
renormalization schemes used in the literature and clarified their
range of applicability.  We showed in particular that an $\msbar$
renormalization of the mixing angles in the extended Higgs sector
within popular schemes leads to gauge-dependent predictions in the
2-Higgs-Doublet Model of type II. We expect that this is also the case
in the Minimal Supersymmetric Standard Model.
 
The proposed extension of the Fleischer--Jegerlehner tadpole scheme
can be straightforwardly applied to more general theories.  This opens
the way for consistent renormalization prescriptions of theories with
more complicated extended Higgs sectors.

We have applied the renormalization scheme to the calculation of NLO
EW corrections for Higgs production in gluon fusion and Higgs
strahlung and have, in particular, investigated the scale dependence
of the corrections and the decoupling of the heavy Higgs bosons within
this scheme.


\section{Acknowledgements\label{Acknowledgements}}

We thank S.\ Uccirati for valuable discussions and providing an early
version of the code QGS.  A.D. and J.-N.L. acknowledge support from
the German Research Foundation (DFG) via grants DE~623/2-1 and
DE~623/4-1. The work of L.J. and C.S. was supported by the DFG under
contract STU~615/1-1 and the work of J.-N.L. by the Studienstiftung des
Deutschen Volkes.


\appendix

\begin{appendices}

\section{Results for tadpoles in the \THDM}
\label{sec:appA}
We give the results for the tadpoles $\thl$ and $\thh$ corresponding
to the Higgs bosons \Hl\ and \Hh\ in the \THDM\ in the \rxi-gauge
defined in \refse{sec:gaugefixing},
\begin{align}
\label{eq:explThl}
&\thl = - \Thl =\;\frac{\sab\, g}{8\pi^2\mw}\Bigg\{ - 3\mt^2 \Az\left(\mt\right)\notag \\
&+ \frac{\mhl^2}{8} \left( \Az\left(\sqrt{\xiz}\mz\right) + 2\Az\left(\sqrt{\xiw}\mw\right)\right)
 + \frac{(D-1)}{4}\left(\mz^2\Az\left(\mz\right) + 2\mw^2\Az\left(\mw\right)\right)\notag \\
&+ \frac{3}{8}\left( \mhl^2 \left(1 + 2 \cab^2\right) - 2\cab^2\Msb^2
\right)\Az\left(\mhl\right)\notag\\
&+ \frac{1}{8}\bigg( \left(1-2\cab^2\right) \left(\mhl^2+2\mhh^2\right) -
2\Msb^2
        \left(1-3\cab^2\right)\bigg)\Az\left(\mhh\right)\notag\\ 
&+\frac{1}{8}  \left( 2\mha^2 + \mhl^2 - 2\Msb^2 \right)\Az\left(\mha\right)
 + \frac{1}{4} \left( 2\mhc^2 + \mhl^2 - 2\Msb^2 \right)\Az\left(\mhc\right)
 \Bigg\}\notag\\ 
&+\frac{\cab\, g}{8\pi^2 \mw \tb}\Bigg\{
        3\mt^2\Az\left(\mt\right)\notag \\
&+\frac{\tb^2-1}{8}\bigg( 3\cab^2\left(\mhl^2 - \Msb^2\right)\Az\left(\mhl\right)
 +\sab^2 \left(  2 \mhh^2 +\mhl^2 - 3\Msb^2\right)\Az\left(\mhh\right)\notag \\
&+\left(\mhl^2 - \Msb^2\right)\Az\left(\mha\right)  + 
        2\left(\mhl^2 - \Msb^2\right)\Az\left(\mhc\right)\bigg)\Bigg\},
\end{align}
\begin{align}
\label{eq:explThh}
&\thh= - \Thh = \frac{\cab\, g}{8 \pi^2 \mw} \Bigg\{ 3\mt^2
\Az\left(\mt\right)\notag \\
&- \frac{\mhh^2}{8} \left( \Az\left(\sqrt{\xiz}\mz\right) +
2\Az\left(\sqrt{\xiw}\mw\right)\right)
 -
 \frac{(D-1)}{4}\left(\mz^2\Az\left(\mz\right) + 2\mw^2\Az\left(\mw\right)\right)\notag \\
&- \frac{3}{8}\left( \mhh^2 \left(1+ 2 \sab^2\right) - 2\sab^2\Msb^2
          \right)\Az\left(\mhh\right)\notag\\
&- \frac{1}{8} \bigg( \left(1-2\sab^2\right) \left(\mhh^2+2\mhl^2\right) -
          2\Msb^2
          \left(1-3\sab^2\right)\bigg)\Az\left(\mhl\right)\notag\\ 
&-\frac{1}{8}  \left( 2\mha^2 + \mhh^2 - 2\Msb^2 \right)\Az\left(\mha\right) -
        \frac{1}{4} \left( 2\mhc^2 + \mhh^2 - 2\Msb^2 \right)\Az\left(\mhc\right)
      \Bigg\}\notag\\
&+\frac{\sab g}{8\pi^2 \mw \tb}\Bigg\{ 3\mt^2\Az\left(\mt\right)\notag \\
&+\frac{\tb^2-1}{8}\bigg( 3\sab^2 \left(\mhh^2-\Msb^2\right)\Az\left(\mhh\right)
          +\cab^2\left(\mhh^2+2\mhl^2 - 3\Msb^2\right)\Az\left(\mhl\right)\notag\\
&+\left(\mhh^2 - \Msb^2\right)\Az\left(\mha\right)
          +2\left(\mhh^2 - \Msb^2\right)\Az\left(\mhc\right)\bigg)\Bigg\}.
\end{align}
The scalar integral $A_0$ is defined in $D$ dimensions by
\begin{align}
  A_0(m) = \frac{\left(2 \pi \mu\right)^{4-D}}{\ii \pi^2} \int \mathrm{d}^Dq\;
  \frac{1}{q^2-m^2+\ii \epsilon}.
\end{align}
Note that by the transformation
\begin{align}
  \sab \to \cab, \; \cab \to  -\sab, \; \mhl\leftrightarrow \mhh
\end{align}
the tadpoles turn into each other in the following way
\begin{align}
\thl \to -\thh,\; \thh \to \thl.
\end{align}

\section{Results for 2-point tadpole counterterms in the \bfts{} in the \THDM}
\label{sec:appB}

In this section, we list the tadpole counterterms for the two-point
functions derived according to the definition 
\eqref{eq:rentadpoleeq2}. Using the abbreviations 
\begin{align}
\ftone(a,b)&= a\, \tb +b \left(1-\tb^2\right),\\
\tab&= \left(\sab+\cab\tb\right)\left(\cab-\sab\tb\right),\\
\tfp &= \frac{g}{2\mw}\left[-\frac{\thl}{\mhl^2}\left(\sab+\tb\cab\right)
      +\frac{\thh}{\mhh^2}\left(\cab-\sab\tb\right)\right]\label{eq:tfp},\\
\tfm &= \frac{g}{2\mw\tb}\left[\frac{\thl}{\mhl^2}\left(\cab-\sab\tb\right)
    +\frac{\thh}{\mhh^2}\left(\sab+\cab\tb\right)\right] \label{eq:tfm},
\end{align}
the expressions read:
\begin{align}
\thlhl ={}& -\thl\frac{3 g}{2\mw\tb}\left[
           \sab \tab  +
           \ftone(2\sab, -\cab)\left( 1 - 
           \frac{\Msb^2 \cab^2}{\mhl^2}
           \right) \right]\notag\\
        & -\thh\frac{g\,\cab}{2\mw\tb}\left[
          -\left(1+\frac{2\mhl^2}{\mhh^2}\right)\tab      
          +\frac{\Msb^2}{\mhh^2} 
          \ftone(2(\cab^2-2\sab^2),3\sab \cab)
               \right],\\
\thhhh ={}& -\thh\frac{3g}{2\mw\tb} \left[
             \cab \tab -
             \ftone(2\cab,\sab) \left(1-
             \frac{\Msb^2}{\mhh^2}  \sab^2
             \right)
             \right]\notag\\*
        & -\thl\frac{g\sab}{2\mw\tb} \left[
            -\left(1+\frac{2\mhl^2}{\mhh^2}\right)\tab      
            -\frac{\Msb^2}{\mhl^2} 
            \ftone(2(\sab^2-2\cab^2), -3\sab \cab)
             \right],\\
\label{eq:thhhl}
\thhhl ={}& \thlhh = \frac{g}{2 \mw \tb} \thl \cab \left[ 
                      \left(2+\frac{\mhh^2}{\mhl^2}\right) \tab +
                      \frac{\Msb^2}{\mhl^2} \ftone(2(2\sab^2-\cab^2),-3\sab \cab)
           \right] \notag\\
         & {}+\frac{g}{2 \mw \tb} \thh \sab \left[ 
              \left(2+\frac{\mhl^2}{\mhh^2}\right) \tab -
              \frac{\Msb^2}{\mhh^2} \ftone(2(2\cab^2-\sab^2),3\sab \cab)
           \right],\\
\tha ={}& \thl\frac{g}{2\mw\tb}\left[\frac{\Msb^2}{\mhl^2}\ftone(2\sab,-\cab)
        -\frac{\mha^2}{\mhl^2}2\sab\tb-\ftone(\sab,-\cab)\right]\notag\\
      & + \thh\frac{g}{2\mw\tb}\left[-\frac{\Msb^2}{\mhh^2}\ftone(2\cab,\sab)
        +\frac{\mha^2}{\mhh^2}2\cab\tb+\ftone(\cab,\sab)\right]\label{eq:tha},\\
\thc ={}& \tha\left(\mha\to\mhc\right) \label{eq:thc},\\
\tgzero ={}& \tgpm = \frac{g}{2\mw}\left(-\thl \sab +
           \thh\cab\right) \label{eq:tg0},\\
\tgha ={}& \thag = \thl\frac{g\,\cab}{2\mw}\left(1-\frac{\mha^2}{\mhl^2}\right)
          +\thh\frac{g\,\sab}{2\mw}\left(1-\frac{\mha^2}{\mhh^2}\right)
          \label{eq:tgha},\\
\tghc ={}& \thcg =\tgha\left(\mha\to\mhc\right) \label{eq:tghc},\\
\tw ={}& g^{\mu\nu} g\change{\mw}\left(\frac{\thl}{\mhl^2}\sab -
      \frac{\thh}{\mhh^2}\cab\right) \label{eq:tw},\\
\tz ={}& \frac{\tw}{\ct} \label{eq:tz},\\
\twhc ={}& \ii q^\mu \frac{g}{2}\left(\frac{\thl}{\mhl^2}\cab + \frac{\thh}{\mhh^2} \sab\right)\label{eq:twhc},\\
\twg ={}& \pm q^{\mu} \change{\frac{g}{2}}\left(\frac{\thl}{\mhl^2}\sab -
           \frac{\thh }{\mhh^2}\cab\right) \label{eq:twg},\\
\tzha ={}& -\ii\frac{\twhc}{\ct} \label{eq:tzha},\\
\tzg ={}& -\ii \frac{\twmgp}{\cw} \label{eq:tzg},\\
\tff ={}&\left\{\begin{array}{ll}
m_f \tfp &\mbox{if $f$ couples to $\Phi_1$ in Eq.~\eqref{eq:Yukawa}}\\[1.5ex]
m_f \tfm &\mbox{if $f$ couples to $\Phi_2$ in Eq.~\eqref{eq:Yukawa}}
\end{array},\right.
\end{align}
with $q^\mu$ being the incoming momentum of the corresponding vector
boson.  The Feynman rules 
for the tadpole counterterms 
are obtained by
multiplying the tadpole expression with the imaginary unit $\ii$.

\section{Tadpoles in  the two-loop Higgs-boson self-energy
  in the \bfts{}}  
\label{sec:2lse}
In this appendix, we relate the renormalized two-loop self-energy of
the Higgs boson in the \SM{} in the two schemes based on $\hat T_h=0$ 
and
$\dv =0$.  Analogously to Eq.~\eqref{eq:tadpoleinterp} at one
loop, we show that the two-loop tadpole contributions to the
self-energy, which are generated by $\dv$ in the scheme with $\hat
T_h=0$, reproduce the self-energy in the scheme with $\dv = 0$, where
unrenormalized tadpoles are explicitly taken into account.  The latter
situation is given in Eq.~\eqref{eq:fullselfenergy2loop} if the
renormalized tadpoles are replaced by unrenormalized ones.  

To start with, we note that the 1PI two-loop self-energy and tadpole
contributions depend on the tadpole renormalization scheme, more
precisely, they differ by $\dv$-dependent terms. Having performed the
renormalization at one-loop, the 1PI two-loop self-energy diagrams in
both schemes are related via
\begin{align}
  \left.\raisebox{-7pt}{\includegraphics{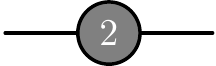}}\right|_{\hat T_h=0} &=
  \left.\raisebox{-7pt}{\includegraphics{\pics/2p22B}}\right|_{\dv =0} +
  \raisebox{-7pt}{\includegraphics{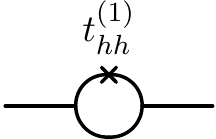}},
  \label{eq:relselfenergyts}
\end{align}
where the second diagram on the right-hand side schematically denotes
all one-loop self-energy diagrams with an additional insertion of the
one-loop two-point tadpole counterterm $t^{(1)}_{hh}$.  Using the
one-loop result \eqref{eq:tadpoleinterp} which relates $t^{(1)}_{hh}$
with the bare one-loop tadpole $t^{(1)}_{h}$, this can be written as
\begin{align}
  \left.\raisebox{-7pt}{\includegraphics{\pics/2p22B}}\right|_{\hat T_h=0} &=
  \left.\raisebox{-7pt}{\includegraphics{\pics/2p22B}} \right|_{\dv =0}
  + \raisebox{-6pt}{\includegraphics{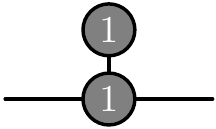}}.
  \label{eq:relselfenergyts2}
\end{align}
Next, we consider the two-loop 1PI tadpole which fixes
$t^{(2)}_h$ in the scheme where $\hat T_h=0$,
\begin{align}
  -t^{(2)}_h = \left.\raisebox{-14pt}{\includegraphics{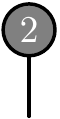}}\right|_{\hat
  T_h=0} =
\left.\raisebox{-14pt}{\includegraphics{\pics/t2}}\right|_{\dv=0} +
  \raisebox{-14pt}{\includegraphics{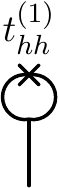}}
  =
  \left.\raisebox{-14pt}{\includegraphics{\pics/t2}}\right|_{\dv=0} +
  \raisebox{-14pt}{\includegraphics{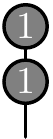}}.
  \label{eq:reltadpolets}
\end{align}
The first equality is the renormalization condition. In the second
equality we separate the $\dv$-dependent terms, where the second
diagram schematically represents all tadpole one-loop diagrams with an
additional insertion of $t^{(1)}_{hh}$. In the third equality we use
again the one-loop result \eqref{eq:tadpoleinterp}.  In the scheme
where the tadpoles are renormalized according to $\hat{T}_{h}=0$ the
renormalized two-loop self-energy can be expressed exclusively by 1PI
contributions and is given by
\begin{align}
\left.\hat{\Sigma}_{hh}^{(2)}\left(\qq^2\right)\right|_{\hat T_h=0} =&
  \left.\left[\raisebox{-7pt}{\includegraphics{\pics/2p22B}} + 
  \raisebox{-1pt}{\includegraphics{\pics/2pct}}\right]
\right|_{\hat T_h=0}.
\label{eq:renselfenergy2}
\end{align}
The \ts{} for $\hat T_h=0$ includes tadpoles via the $\dv$-dependent
counterterms.  In addition to the $\dv^{(1)}$-dependent one-loop
counterterms appearing in Eqs.~\eqref{eq:relselfenergyts} and
\eqref{eq:reltadpolets}, the two-loop counterterm induces a further
dependence on $\dv^{(1)}$ and $\dv^{(2)}$. In the \SM{} the additional
two-loop tadpole counterterms are derived from
Eq.~\eqref{eq:twopointtadpole}, which can be written as
\begin{align}
t_{hh} \hb^2 = 
  \left(\lambda_{hhh, \mathrm{B}} \dv + \frac{\lambda_{hhhh, \mathrm{B}}}{2}
(\dv)^2\right) \hb^2
\label{eq:twopointtadpole2}
\end{align}
upon identifying $\lambda_{hhh, \mathrm{B}}$ and $\lambda_{hhhh,
  \mathrm{B}}$ as the bare triple and quartic Higgs-boson couplings.
The dependence on the two-loop tadpole counterterm $t_h^{(2)}$
originates from the term proportional to $\lambda_{hhh,\mathrm{B}}
\dv^{(2)}$. Using Eqs.~\eqref{eq:dvhtwosol} and
\eqref{eq:reltadpolets} the contribution of $t_{h}^{(2)}$ can be
written as
\begin{align}
  \left.
  \raisebox{-15.5pt}{\includegraphics{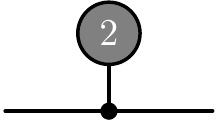}}
  \right|_{\hat T_h=0}
  =
  \left.
  \raisebox{-15.5pt}{\includegraphics{\pics/2p2B}}
  \right|_{\dv=0}
  +
  \raisebox{-15.5pt}{\includegraphics{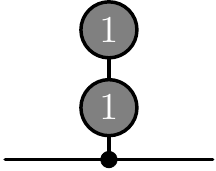}}.
  \label{eq:t1pi2}
\end{align}
Next, we consider the quadratic 1PI one-loop tadpole contributions
which are included in $\dvhtwo$ and $\left(\dvhone\right)^2$ being
proportional to $\lambda_{hhh}$ and $\lambda_{hhhh}$, respectively. We
identify the two contributions with
\begin{align}
  \raisebox{-0.5pt}{\includegraphics{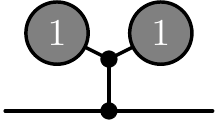}},\quad
  \raisebox{-0.5pt}{\includegraphics{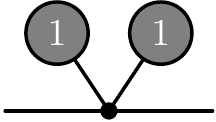}}. 
  \label{eq:t1pi11}
\end{align}
The last two-loop tadpole counterterms result from products of the
one-loop tadpole $t^{(1)}_h$ with the counterterms to $\lambda_{hhh}$,
the Higgs-boson mass [entering via Eq.~\eqref{dvhonesol}], and the
Higgs-boson field-renormalization constant and can be represented as
follows:
\begin{align}
 \raisebox{-1.5pt}{\includegraphics{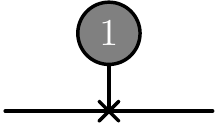}}, \quad
  \raisebox{-0.5pt}{\includegraphics{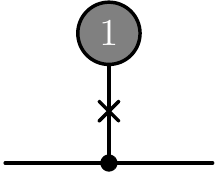}},
  \label{eq:t1pi1ct}
\end{align}
where in the counterterms $\Delta v=0$ is understood.  Finally, we
separate the $\dv$ dependence from the renormalized two-loop
self-energy \eqref{eq:renselfenergy2}. For the bare 1PI two-loop
self-energy we use the result \eqref{eq:relselfenergyts2}. The
two-loop $\dv$-dependent counterterms are given by the sum of the
diagrams in Eqs.~\eqref{eq:t1pi2}, \eqref{eq:t1pi11}, and
\eqref{eq:t1pi1ct}. The result reads
\begin{align}
\left.\hat{\Sigma}_{hh}^{(2)}\left(\qq^2\right)\right|_{\hat T_h=0} =&
  \left[\raisebox{-7pt}{\includegraphics{\pics/2p22B}} +
  \raisebox{-1pt}{\includegraphics{\pics/2pct}}
  \right]_{\hat T_h=0} \notag \\
  =&\left[\raisebox{-23.9pt}{\includegraphics{\pics/2p22R}} +
  \raisebox{-17pt}{\includegraphics{\pics/2p2B}}
  +\raisebox{-23.9pt}{\includegraphics{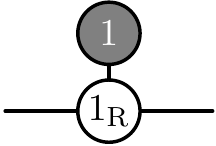}} +
  \raisebox{-17pt}{\includegraphics{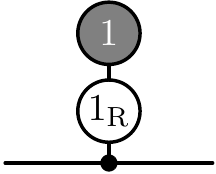}}
  \right]_{\dv=0}
   \notag \\
  &+\raisebox{-0.5pt}{\includegraphics{\pics/2p3TB}} +
  \raisebox{-0.5pt}{\includegraphics{\pics/2p4TB}} \notag \\[1.5ex]
  =& \left.\hat{\Sigma}_{hh}^{(2)}\left(\qq^2\right)\right|_{\dv=0}.
\end{align}
In the second equation we combine counterterms, evaluated for $\Delta
v = 0$, and bare-loop topologies to renormalized objects.  The result
matches the renormalized two-loop self-energy in
Eq.~\eqref{eq:fullselfenergy2loop} when tadpoles are not renormalized,
\ie for $\dv=0$.

\section
{Gauge dependence of \boldmath{$\beta$} in popular tadpole schemes}
\label{sec:gdepbeta}

In the MSSM, $\delta \beta^\msbar$ as obtained in Scheme~2 is gauge
dependent at two loops \cite{Yamada:2001ck}, while it does not
dependent on the gauge parameters in the \rxi-gauge at one-loop order.
The latter result translates to the \THDM{} of type II.

In the \THDM{} the apparent gauge independence of $\delta
\beta^\msbar$ in the \rxi-gauge at the one-loop level can be
understood as follows: Consider the linear combination of $\Phione$
and $\Phitwo$ that does not have a vev.  Using the explicit rotations
in Eq.~\eqref{eq:rotationfields}, this Higgs doublet is identified as
\begin{align}
  \cbe \varPhi_2 - \sbe \varPhi_1 = 
  \left(
  \begin{array}{c}
    \Hpm\\
    \frac{1}{\sqrt{2}}\left(\cbe \vtwo - \sbe \vone + \Hl \cab  + \Hh
      \sab +\ii \Ha\right)
  \end{array}
  \right),
  \label{eq:gbindepdoublet}
\end{align}
with $\cbe \vtwo - \sbe \vone=0$.  Performing the shift in the vevs
according to Eq.~\eqref{eq:tadpoleeq2HDM} and using
Eq.~\eqref{eq:tpcondtionsol2HDM1L}, we obtain the tadpole
corresponding to the neutral Higgs field $\Hnov=\Hl \cab+\Hh \sab$,
\begin{align}
  \sbe \Delta\vone - \cbe \Delta\vtwo =
  \cab \frac{\thl}{\mhl^2}+ \sab \frac{\thh}{\mhh^2},
  \label{eq:rxigaugeidtadpole}
\end{align}
which enters the shift of $\Delta_{i}\beta^\msbar$ between the \ts{}
and the popular Schemes~1 and 2 found in
Eqs.~\eqref{eq:gaugeinvtadpolecomb2} and
\eqref{eq:gaugeinvtadpolecomb}.  The Higgs field $\Hnov$ does neither
couple to two gauge bosons nor to two would-be Goldstone bosons, and,
moreover, it does not enter the gauge fixing in the \rxi-gauge and
thus does not couple to Faddeev--Popov ghost fields.  Consequently,
there are no gauge-dependent Feynman diagrams for the \Hnov\ tadpole
at one-loop order and thus $\Delta_{i} \delta \beta^\msbar$ does not
depend on the gauge-parameter in the \rxi-gauge.  Since $\delta
\beta^\msbar$ is gauge independent in the \ts{}, this translates to
Schemes~1 and 2. This argument can be generalized to non-linear
\rxi-gauges at one-loop order.\footnote{Specifically, we verified the
  independence of $\delta \beta^\msbar$ of the gauge parameters for a
  non-linear gauge-fixing function $C^Z = \partial_{\mu} Z_{\mu} -
  \xiphiz \mz G_0 \left(1+\xihh \Hh\right)$ for general $\xiphiz$ and
  $\xihh$.}

Nevertheless, it is possible to demonstrate the gauge dependence of
$\delta \beta^\msbar$ in Schemes~1 and 2 at one-loop order in a
suitably chosen gauge.  Since $\Hnov$ couples to one gauge boson or
would-be Goldstone boson and $\Ha$, we can generate a gauge-dependent
contribution to its tadpole by allowing for mixing propagators induced
by the gauge fixing.  Here, we provide an appropriate gauge-fixing
function and prove the gauge dependence of $\delta \beta^\msbar$ via
two different approaches. In addition, we show that also in this class
of gauges the gauge independence of $\delta \beta^\msbar$ is preserved
in the \ts.

The appropriate choice of the gauge-fixing function can be motivated
as follows.  From the point of view of the \ts{} the gauge dependence
appears in the Schemes~1 [Eq. \eqref{eq:dennerscheme}] or 2
[Eq.~\eqref{eq:bhscheme}] if it is possible to generate a
gauge-dependent tadpole contribution of the form of
Eq.~\eqref{eq:rxigaugeidtadpole}.  For a gauge-fixing function $C$
linear in the gauge fields the infinitesimal variation of Green's
functions under a change in the gauge-fixing function, $\Delta C$,
with respect to some parameter can be derived (see \eg Section 2.5.4.4
of \citere{Bohm:2001yx}, Section 12.4 of \citere{Collins:1984xc}, or
\citere{Lee:1973fn}). For the one-point function of a field $\varphi$
this reads
\begin{align}
  \delta_{\Delta C}\left\langle T\varphi(x)\right\rangle:&=
  \delta_{\Delta C} \raisebox{-21pt}{\includegraphics{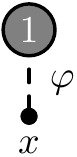}}=
  \left(\raisebox{-20pt}{\includegraphics{\pics/tadpoleGF.pdf}}\right)_{C+\Delta C} -
  \left(\raisebox{-20pt}{\includegraphics{\pics/tadpoleGF.pdf}}\right)_{C}
  =\ii \langle T(\dBRS \varphi(x))\int\mathrm{d}^4y\,\bar u(y)\Delta C(y)\rangle\notag\\
  &=\ii \int \mathrm{d}^4y\;\left[
    \raisebox{-14pt}{\includegraphics{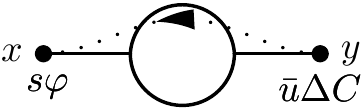}}
\right],
\label{eq:gaugedepTadpole}
\end{align}
where $\dBRS\varphi$ represents the BRST transformation
of the field $\varphi$ at the space--time point $x$ and $\bar u$ is
the anti-ghost field associated to the gauge-fixing function $C$, both
at the space--time point $y$.
For 
\begin{align}
  \varphi = \Hnov =\cab \Hl + \sab \Hh,
\end{align}
the required BRST transformation in
Eq.~\eqref{eq:gaugedepTadpole} reads
\begin{align}
  \dBRS \Hnov = \frac{e}{2 \sw \cw}  u^\PZ \Ha + \frac{\ii e}{2 \sw} \left(
  u^- H^+ - u^+ H^-\right). 
\end{align}
We note that $\dBRS \Hnov$ does neither induce would-be Goldstone bosons
nor vevs.  Hence, one can easily read off the condition for a gauge
dependence of Eq.~\eqref{eq:rxigaugeidtadpole}. We modify the
gauge-fixing function in Eq.~\eqref{eq:GFRxi} by setting $\xiw=\xia=\xiz=1$ and
adding a term proportional to $\Ha$ to $C^Z$,
\begin{align}
  C^Z = \partial^\mu Z_\mu - \mz G_0 - \xib \mha \Ha, 
  \label{eq:xibetagauge}
\end{align}
which is required to obtain non-vanishing contributions to
Eq.~\eqref{eq:gaugedepTadpole}. The resulting gauge-fixing function
\eqref{eq:xibetagauge} in the $\xib$-gauge looks simple, but gives
rise to a non-diagonal propagator matrix (see \refapp{sec:FRxi} for
the Feynman rules).  An infinitesimal change in the gauge-fixing
function is obtained by performing an expansion for small $\xib$, \ie
we identify $\Delta C$ with $-\xib \mha \Ha$, defining
\begin{align}
  \dxib X := \left.\frac{\partial}{\partial \xib} X
  \right|_{\xib=0}\xib .
\end{align}
While we work only to leading order in $\xib$, an exact calculation is
possible and straightforward in the gauge of
Eq.~\eqref{eq:xibetagauge}.  At one-loop order we find after Fourier
transformation to momentum space
\begin{align}
   \ft\int \mathrm{d}^4y\;\left[
    \raisebox{-14pt}{\includegraphics{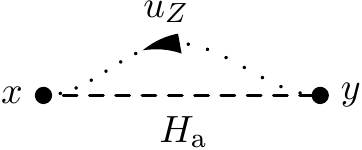}}
  \right] = \int \frac{\mathrm{d}^4q}{\left(2 \pi\right)^4}\;
  \frac{\ii}{q^2-\mz^2} \frac{\ii}{q^2-\mha^2},
\end{align}
and hence using Eq.~\eqref{eq:gaugedepTadpole}
\begin{align}
  \left \langle \Hnov\right \rangle_{\xib}:=\ft\, \dxib \left\langle T\Hnov(x)\right\rangle ={}&
  \ft\left[\cab\ \dxib\raisebox{-20pt}{\includegraphics{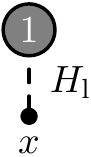}}+
           \sab\ \dxib\raisebox{-20pt}{\includegraphics{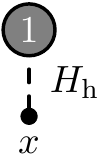}}
     \right]
    \label{eq:gaugedepTadpoleNLO}
   \\
={}&-\frac{\ii e \xib \mha}{2 \sw \cw} \int
  \frac{\mathrm{d}^4q}{\left(2 \pi\right)^4}\; \frac{\ii}{q^2-\mz^2}
  \frac{\ii}{q^2-\mha^2}.
\label{eq:gaugedepTadpoleResNLO}
\end{align}
Consequently, there is a non-zero gauge-dependent and UV-divergent
contribution to the tadpole in Eq.~\eqref{eq:rxigaugeidtadpole}, which
proves the gauge dependence in the popular schemes, where tadpole
contributions are absorbed in bare parameters.  Note that this
argument can be carried over to the supersymmetric case, where $\left
  \langle \Hnov\right \rangle_{\xib}$ does not change if the same
gauge is used.  This result is used below to derive the $\xib$
dependence of $\delta \beta^\msbar$ in Schemes~1 and 2 [see
Eq.~\eqref{eq:xibetagaugedep}].

We validate Eq.~\eqref{eq:gaugedepTadpoleResNLO} in the \THDM\ using
an explicit Feynman-diagrammatic calculation of the tadpole $\langle
\Hnov\rangle$.  Inspecting the Feynman rules listed in
\refapp{sec:FRxi}, we find three sources that can induce a linear
$\xib$ dependence of the tadpole $\langle \Hnov\rangle$. These are
provided by the mixing propagators $Z\Ha$ and $G_0 \Ha$ and the
coupling of the neutral Higgs bosons $\Hl$ and $\Hh$ to Faddeev--Popov
ghosts $\bar{u}^Z$ and $u^Z$. For the $\xib$-dependent tadpole
contributions corresponding to Eq.~\eqref{eq:gaugedepTadpoleNLO} in
momentum space we obtain
\begin{align}
  \dxib\raisebox{-13pt}{\includegraphics{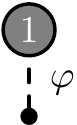}}
  =
  \raisebox{-13pt}{\includegraphics{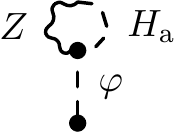}}
  +
  \raisebox{-13pt}{\includegraphics{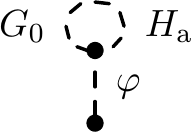}}
  +
  \raisebox{-13pt}{\includegraphics{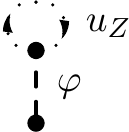}}
  \quad \text{for} \quad \varphi= \Hl, \Hh,
\end{align}
and the sum of the contributions yields
\begin{align}
  \dxib \raisebox{-13pt}{\includegraphics{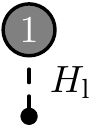}}\;
  = -\frac{\ii e \xib \mha \cab}{2 \sw \cw} \int
  \frac{\mathrm{d}^4q}{\left(2 \pi\right)^4}\; \frac{\ii}{q^2-\mz^2}
  \frac{\ii}{q^2-\mha^2}= \cab \left \langle \Hnov\right \rangle_{\xib}, \notag\\
  \dxib \raisebox{-13pt}{\includegraphics{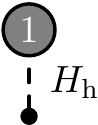}}
  = -\frac{\ii e \xib \mha \sab}{2 \sw \cw} \int
  \frac{\mathrm{d}^4q}{\left(2 \pi\right)^4}\; \frac{\ii}{q^2-\mz^2}
  \frac{\ii}{q^2-\mha^2} = \sab \left \langle \Hnov\right \rangle_{\xib}.
  \label{eq:gaugedepTadpoleResNLO2}
\end{align}
Thus, we reproduce the result in Eq.~\eqref{eq:gaugedepTadpoleResNLO}.

Finally, we show explicitly that $\delta \beta^\msbar$ remains gauge
independent in the \ts{} at one-loop order but depends explicitly on
\xib{} in the gauge of Eq.~\eqref{eq:xibetagauge} in Schemes~1 and 2.
We cannot make use of Eq.~\eqref{eq:betarel} because it does not hold
in the $\xib$-gauge for Schemes~1 and 2, but instead we derive the
gauge dependence directly from the renormalized vertex function in
Eq.~\eqref{eq:betarenocond}.  We consider only the terms linear in
$\xib$.

In the \ts{} it is enough to verify that all counterterm parameters
that enter the renormalization of $\beta$ are gauge independent and
that no gauge dependence is introduced by the bare vertex function in
Eq.~\eqref{eq:betarenocond}.  The renormalization constant $\delta
Z_e$ is independent of $\xib$ since no Higgs-boson couplings enter
this quantity. For $\delta \mw^2$ and $\delta \mz^2$, the tadpole
contributions to the $WW$ and $ZZ$ two-point functions are
proportional to (see \refapp{sec:appB})
\begin{align}
  \sab \frac{\thl}{\mhl^2} - \cab \frac{\thh}{\mhh^2},
\end{align}
which is not sensitive to our choice of gauge-fixing function. The
\PW-boson self-energy receives no other contributions linear in
$\xib$.  The linear $\xib$-dependent contribution induced in the
\PZ-boson self-energy contributes only to its longitudinal part and
does not influence $\delta \mz^2$.  This implies the $\xib$
independence of $\delta \mw^2$ and $\delta \mz^2$ which we have also
verified via explicit calculation in the $\xib$-gauge.  For the vertex
$\Ha\tau^+\tau^-$ there is no $\xib$-dependent and at the same time
UV-divergent term. This is consistent with the fact that there is no
tadpole contribution to the bare vertex function which could cancel a
would-be gauge dependence.  For $\dzgha$ and $\delta m_{\tau}$ no such
argument can be given, and a cancellation of $\xib$-dependent terms
between self-energy diagrams and tadpoles takes place. We explicitly
show this cancellation starting with $\dzgha$.

The terms linear in $\xib$ contributing to the
$G_0\Ha$ mixing energy are given by%
\footnote{In the alignment limit, the second line in Eq.~\eqref{eq:seGFcontr}
vanishes.}  
\begin{align}
  \dxib\raisebox{-13.5pt}{\includegraphics{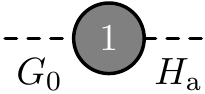}}
  \;= \;\;&{}
  \raisebox{-23.7pt}{\includegraphics{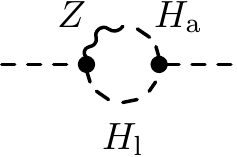}}+
  \raisebox{-23.7pt}{\includegraphics{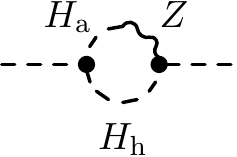}}+ 
  \;{}\raisebox{-23.7pt}{\includegraphics{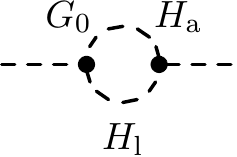}}+
  \raisebox{-23.7pt}{\includegraphics{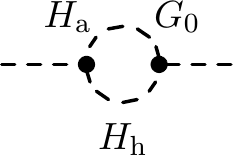}}
  \notag\\
  &{}+\;\raisebox{-23.7pt}{\includegraphics{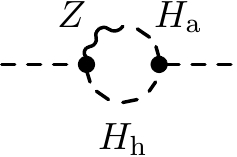}}
  +\raisebox{-23.7pt}{\includegraphics{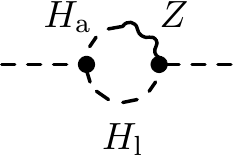}}
  +\raisebox{-23.7pt}{\includegraphics{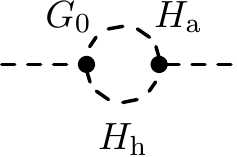}}
  +\raisebox{-23.7pt}{\includegraphics{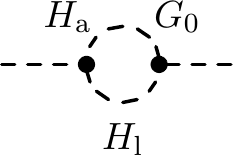}}
  \notag\\
  &+\;\raisebox{0.0pt}{\includegraphics{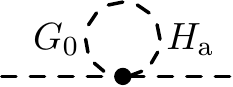}}
  \;.
  \label{eq:seGFcontr}
\end{align}
Note that each diagram contains one mixing propagator.
The diagrams involving a neutral Higgs boson propagator and a mixing
propagator of a pseudoscalar Higgs boson and a would-be
Goldstone boson 
do not contribute to the $\msbar$ renormalization of
$\beta$ because they are UV finite. The other self-energy
diagrams are UV divergent, and we obtain for the combined
contributions to the $G_0\Ha$ mixing energy
%

\begin{align}
  \sum_{\varphi=\Hl,\Hh}\raisebox{-21.7pt}
  {\includegraphics{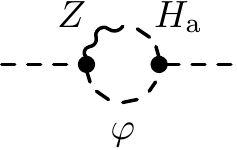}} ={} &
  -\frac{\ii e}{2 \sw\mw}
  \Biggl[\sab^2 \mhl^2 +\cab^2 \mhh^2 + 2 \mha^2 - 2\Msb^2 \notag\\
  &+ \cab \sab \frac{1-\tb^2}{\tb}\left(\mhh^2-\mhl^2\right)\Biggr] \left
    \langle \Hnov\right \rangle_{\xib} + \text{UV-finite terms},
    \label{eq:Zleftcontr}
    \\
  \sum_{\varphi=\Hl,\Hh}\raisebox{-21.7pt}{\includegraphics{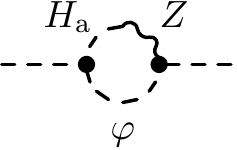}} ={}&
  \frac{\ii e}{2 \sw\mw} \left[\mha^2 - \sab^2\mhh^2 -\cab^2 \mhl^2\right]
  \left \langle \Hnov\right \rangle_{\xib} + \text{UV-finite terms},
  \label{eq:Zrightcontr}
  \\
  \raisebox{0.0pt}{\includegraphics{\pics/selfenergyGFMS5.pdf}} ={}& 
  \frac{\ii e}{2 \sw\mw} \Biggl[\mhl^2 + 2 \mhh^2 - 2 \Msb^2 \notag\\
    &+\cab \left(-\cab + \sab
  \frac{1-\tb^2}{\tb}\right)\left(\mhh^2-\mhl^2\right)\Biggr]
  \left \langle \Hnov\right \rangle_{\xib},
\end{align}
where for arriving at the Eqs.~\eqref{eq:Zleftcontr} and
\eqref{eq:Zrightcontr} the numerator structure has been cancelled
against one of the neutral Higgs-boson propagators. 
Adding all contributions leads to
\begin{align}
  \dxib\raisebox{-13.5pt}{\includegraphics{\pics/selfenergyGFMS.pdf}}
  =  -\frac{\ii e}{2 \sw \mw} \left[ \mha^2- \sab^2 \mhh^2 - \cab^2 \mhl^2\right]
      \left \langle \Hnov\right \rangle_{\xib} + \text{UV-finite terms}
  \label{eq:xibdepselfenergy}
\end{align}
for the linear dependence of self-energy diagrams on $\xib$.
The tadpole contributions to the $G_0\Ha$ mixing energy are derived
using the results in Eq.~\eqref{eq:gaugedepTadpoleResNLO2} 
leading to
\begin{align}
  \sum_{\varphi=\Hl,\Hh}
  \raisebox{0.0pt}{\includegraphics{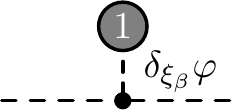}} =
  \frac{\ii e}{2 \sw \mw} \left[ \mha^2- \sab^2 \mhh^2 - \cab^2 \mhl^2\right]
      \left \langle \Hnov\right \rangle_{\xib},
\end{align}
which cancels against the $\xib$ dependent terms in
Eq.~\eqref{eq:xibdepselfenergy} contributing to the renormalization of
$\beta$.  Thus, we have proven that
\begin{align}
  \left(\dxib \dzgha^\msbar\right)_3 = 0.
\end{align}
For the on-shell renormalization of $\delta m_{\tau}$ we pursue the
same strategy. The $\xib$-dependent contributions to the $\tau$
self-energy are given by
\begin{align}
  \dxib\raisebox{-12pt}{\includegraphics{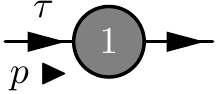}}
  = 
  \raisebox{-19.4pt}{\includegraphics{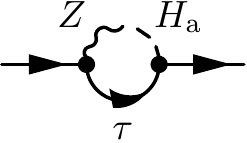}}
  +\raisebox{-19.4pt}{\includegraphics{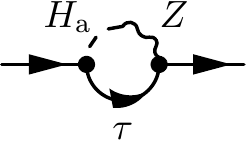}}
  +\raisebox{-19.4pt}{\includegraphics{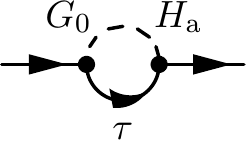}}
  +\raisebox{-19.4pt}{\includegraphics{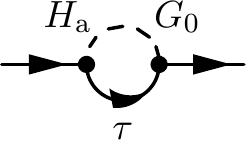}}\;.
\end{align}
Projecting the $\tau$ self-energy onto a Dirac spinor, putting the
momentum on shell, using the Dirac equation, and considering only the
scalar and vector part that is relevant for the mass counterterm, we
find
\begin{align}
  \left.\dxib\raisebox{-12pt}{\includegraphics{\pics/selfenergyGFMStau.pdf}}
  u(p) \right|_{p^2= m_\tau^2}
  ={}& \frac{\ii e^2 \tb m_{\tau} \xib \mha}{4 \mw \sw^2 \cw}  u(p) \times \\
   &\left.\int\frac{\mathrm{d}^4q}{\left(2 \pi\right)^4}
    \frac{\ii}{q^2-\mz^2}
    \frac{\ii}{q^2-\mha^2} 
    \frac{\ii}{\left(p+q\right)^2-m_{\tau}^2}
    \left( \left(p+q\right)^2 -m_\tau^2 \right)\right|_{p^2=m_\tau^2} \notag\\
    ={}&-\left.\frac{\ii e m_\tau \tb}{2 \sw \mw} \left \langle \Hnov \right
    \rangle_{\xib} u(p)\right|_{p^2=m_\tau^2}.
    \label{eq:taugaugedepxib}
\end{align}
The tadpole contribution to the $\tau$ self-energy is derived using
the results in Eq.~\eqref{eq:gaugedepTadpoleResNLO2}
leading to
\begin{align}
  \sum_{\varphi=\Hl,\Hh}
  \raisebox{0.0pt}{\includegraphics{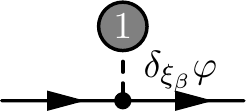}} =
  \frac{\ii e m_\tau \tb}{2 \sw \mw} \left \langle \Hnov \right \rangle_{\xib}.
  \label{eq:tadpoletauxib}
\end{align}
The tadpole contribution in Eq.~\eqref{eq:tadpoletauxib} cancels against the
self-energy contribution in Eq.~\eqref{eq:taugaugedepxib} and we conclude that
$\delta m_\tau$ is 
independent of $\xib$
in on-shell renormalization,
\begin{align}
  \left(\dxib \delta m_\tau\right)_3 = 0.
\end{align}
Altogether, we have proven the gauge independence of
$\delta\beta^\msbar$ in the $\xib$-gauge in the \ts,
\begin{align}
  \left(\dxib \delta\beta^\msbar\right)_3 = 0.
\end{align}
Finally, we can give the precise \xib{} dependence of $\delta \beta$
for Schemes~1 and 2 originating from the gauge dependence of the
tadpoles in $\delta m_\tau$ and $\dzgha$.  Using Eq.~\eqref{eq:dbct}
for the counterterm, the full $\xib$ dependence is obtained as
\begin{align}
  (\dxib \delta \beta^\msbar)_i = \left.\frac{e}{2 \sw \mw}
  \left[ 1- \frac{\sab^2\mhh^2 + \cab^2 \mhl^2}{\mha^2
\left(1+\tb^2\right)}\right]
  \left \langle \Hnov \right \rangle_{\xib}\right|_{\PP},
  \quad i=1,2,
  \label{eq:xibetagaugedep}
\end{align}
which is evidently nonzero.

\section
{Feynman rules in the \boldmath{$\xib$}-gauge \label{sec:FRxi}}

In this section, we list the Feynman rules used to derive the gauge
dependence of $\delta \beta^\msbar$ in \refapp{sec:gdepbeta}.  The
gauge-fixing function \eqref{eq:xibetagauge} gives rise to mixing of
propagators, which is required to actually observe the gauge
dependence at one-loop order.  The gauge-fixing Lagrangian includes
the following mixing terms
\begin{align}
  \mL_{\mathrm{GF}} \supset  \xib \mha \left(\partial_\mu Z^\mu +\mz
  G_0\right) \Ha.
\end{align}
The corresponding 2-point vertex function in the basis $(Z_\mu, G_0,
\Ha)$ reads
\begin{align}
  \Gamma = \left(
  \begin{array}{c@{\quad}c@{\quad}c}
    -\left(p^2-\mz^2\right) g^{\mu \nu} & 0        & \ii \xib \mha p^\mu \\
    0      & p^2-\mz^2   & \xib \mha \mz     \\
    -\ii \xib\mha p^\mu  & \xib \mha \mz & p^2-\mha^2
  \left(1+\xib^2\right)
  \end{array}
  \right).
\end{align}
By inverting the vertex function to linear order in $\xib$ we obtain
the propagators as
\begin{align}
  \raisebox{-0pt}{\includegraphics{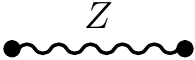}} ={}& \frac{-\ii g^{\mu
  \nu}}{p^2-\mz^2} +\mathcal{O}\left(\xib^2\right), \\
  \raisebox{-0pt}{\includegraphics{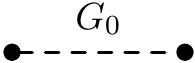}} ={}&
  \frac{\ii}{p^2-\mz^2}+\mathcal{O}\left(\xib^2\right), \\
  \raisebox{-0pt}{\includegraphics{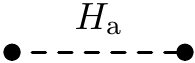}} ={}&
  \frac{\ii}{p^2-\mha^2},\\
  \raisebox{-0pt}{\includegraphics{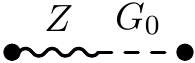}} ={}&
  \mathcal{O}\left(\xib^2\right),\\ 
  \raisebox{-12pt}{\includegraphics{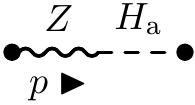}} ={}&
  - \xib \mha p^\mu \frac{\ii}{p^2-\mz^2} \frac{\ii}{p^2-\mha^2},\\ 
  \raisebox{-0pt}{\includegraphics{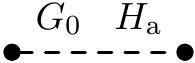}} ={}&
  \ii \xib \mha \mz \frac{\ii}{p^2-\mz^2} \frac{\ii}{p^2-\mha^2},
\end{align}
where the momentum flows from left to right.  We identify mixing
propagators by two particle labels.  The Faddeev--Popov-ghost
Lagrangian is derived by the standard methods which requires for the
$\xib$-gauge the BRST variation of $\Ha$,
\begin{align}
  \dBRS \Ha = -\frac{e}{2 \sw \cw}  u^Z \left(\cab \Hl + \sab
  \Hh\right) + \frac{e}{2 \sw} \left(u^+ H^- + u^- H^+\right).
\end{align}
The additional contribution to the ghost Lagrangian involving $\xib$ is then given by
\begin{align}
\mL_{\mathrm{gh}} \supset  \xib \mha \frac{e}{2 \sw \cw} \bar{u}^Z u^Z
  \left(\cab \Hl + \sab \Hh\right),
\end{align}
yielding the following gauge-dependent Feynman rules
\begin{align}
  \raisebox{-23pt}{\includegraphics{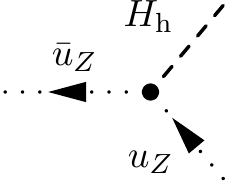}} = \frac{\ii e \sab}{2 \sw \cw}
  \xib \mha, \qquad
  \raisebox{-23pt}{\includegraphics{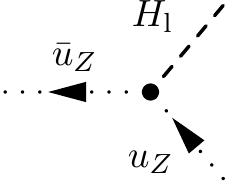}} = \frac{\ii e \cab}{2 \sw \cw}
  \xib \mha.
\end{align} 

Finally, we list all other vertices needed in the calculation of
\refapp{sec:gdepbeta} with the convention that all particles and
momenta are incoming:
\begin{alignat}{2}
  \raisebox{-24pt}{\includegraphics{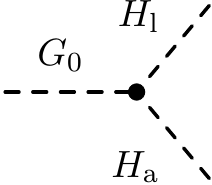}} &= 
 \frac{\ii \cab e}{2 \mw \sw} \left(\mha^2 - \mhl^2\right), \qquad
 &\raisebox{-24pt}{\includegraphics{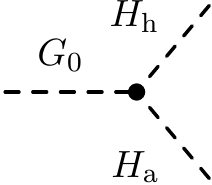}} &=
 \frac{\ii \sab e}{2 \mw \sw} \left(\mha^2 - \mhh^2\right), \\
 \raisebox{-24pt}{\includegraphics{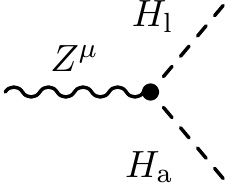}} &=
 \frac{\cab e}{2 \sw \cw} \left(p_{\Ha}^\mu - p_{\Hl}^\mu\right), \qquad
 &\raisebox{-24pt}{\includegraphics{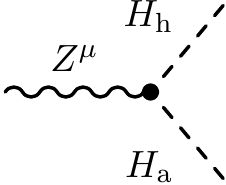}} &=
 \frac{\sab e}{2 \sw \cw}\left(p_{\Ha}^\mu - p_{\Hh}^\mu\right),\\
 \raisebox{-24pt}{\includegraphics{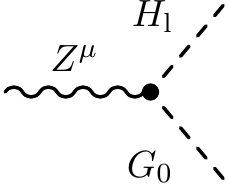}} &=  -\frac{e \sab}{2 \sw
 \cw} \left( p_{G_0}^\mu - p_{\Hl}^\mu \right)
 ,\qquad
 &\raisebox{-24pt}{\includegraphics{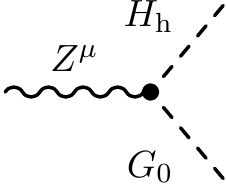}} &= 
 \frac{e \cab}{2 \sw \cw} \left( p_{G_0}^\mu - p_{\Hh}^\mu \right),
\end{alignat}
\begin{align}
 \raisebox{-24pt}{\includegraphics{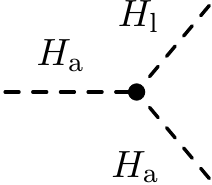}} &=
 \frac{\ii e}{2 \mw \sw} \left(\sab \left( \mhl^2+2\mha^2-2\Msb^2\right) - \cab
 \frac{1-\tb^2}{\tb} \left(\mhl^2-\Msb^2\right)\right),\\ 
  \raisebox{-24pt}{\includegraphics{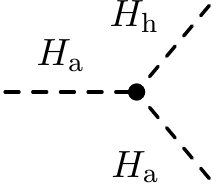}} &=
 \frac{\ii e}{2 \mw \sw} \left(-\cab \left( \mhh^2+2\mha^2-2\Msb^2\right) - \sab
 \frac{1-\tb^2}{\tb} \left(\mhh^2-\Msb^2\right)\right),\\
 \raisebox{-24pt}{\includegraphics{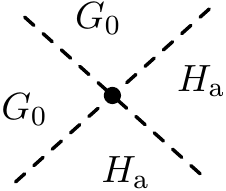}} &=
 -\frac{\ii e^2}{2 \mw^2 \sw^2} \left(\left( \mhl^2+2\mhh^2-2\Msb^2\right) -
 \cab \left(\cab - \sab \frac{1-\tb^2}{\tb}\right) \left(\mhh^2-\mhl^2\right)\right).
\end{align}

\end{appendices}


\bibliographystyle{JHEPmod}
\bibliography{2hdmrbib}

\end{document}